\documentclass[a4paper,12pt]{article}
\usepackage[english]{babel} 
\usepackage[T1]{fontenc}
\usepackage{float}
\usepackage{setspace}
\usepackage{appendix}
\usepackage{parcolumns}
\usepackage{cite}
\usepackage{times}
\usepackage[OT1]{fontenc}
\usepackage{type1cm}
\usepackage{url}
\usepackage{subfigure}
\usepackage{braket}
\usepackage{graphicx}
\usepackage{sidecap}
\usepackage{xspace}
\usepackage{tikz}
\usepackage{indentfirst}
\usepackage[mathscr]{eucal}
\usepackage{geometry}
\usepackage{booktabs}
\usepackage{indentfirst}
\usepackage{bm}
\usepackage{multirow}
\usepackage{dcolumn}
\usepackage{graphicx}
\usepackage{mathrsfs}  
\usepackage{enumerate}
\usepackage[utf8]{inputenc}
\usepackage{amsmath}
\usepackage{amssymb}
\usepackage{amsfonts}
\usepackage{amsthm}
\usepackage{mathrsfs}
\usepackage{graphicx}
\usepackage{multirow}
\usepackage{ulem}	
\usepackage{amsmath,mathtools}
\usepackage{comment}
\usepackage{amsbsy}
\usepackage{hyperref}
\usepackage{bookmark}

\makeatletter
\@addtoreset{equation}{section}

\makeatother

\def\Tr{{\rm Tr}}

\def\atanh{{\rm atanh}}

\newcommand{\be}{\begin{equation}}
\newcommand{\ee}{\end{equation}}
\newcommand{\bea}{\begin{eqnarray}}
\newcommand{\eea}{\end{eqnarray}}
\newcommand{\vs}[1]{\vspace{#1 mm}}

\newcommand{\EAAe}{\Gamma_E^k}

\newcommand{\dd}{\mathrm{d}}
\newcommand{\Eqref}[1]{Eq.~(\ref{#1})}
\newcommand{\MSbar}{$\overline{\text{MS}}\,$}

\begin{document}

\vs{3}
\begin{center}
{\Large\bf Functional Renormalization 
and \MSbar}
\vs{8}

{\large
Alessio Baldazzi${}^*$\footnote{e-mail address: abaldazz@sissa.it}\ 
Roberto Percacci${}^*$\footnote{e-mail address: percacci@sissa.it}\ 
and Luca Zambelli${}^\dagger$\footnote{e-mail address: luca.zambelli@uni-jena.de}
\vs{8}
}

${}^*${International School for Advanced Studies, via Bonomea 265, I-34136 Trieste, Italy}
{and INFN, Sezione di Trieste, Italy}

${}^\dagger${\mbox{ Theoretisch-Physikalisches Institut, Friedrich-Schiller-Universit{\"a}t Jena,}
\mbox{ D-07743 Jena, Germany}}

\vs{5}
{\bf Abstract}
\end{center}
Working with scalar field theories,
we discuss choices of regulator that, inserted in the functional
renormalization group equation, reproduce the results
of dimensional regularization at one and two loops.
The resulting flow equations can be seen as nonperturbative
extensions of the \MSbar scheme.
We support this claim by recovering all the multicritical
models in two dimensions.
We discuss a possible generalization to any dimension.
Finally, we show that the method also preserves
nonlinearly realized symmetries,
which is a definite advantage with respect to other regulators.

\section{Introduction}

Dimensional regularization (dimreg),
together with modified minimal subtraction (\MSbar),
\footnote{In the FRG one follows the flow of renormalized
quantities. Therefore, for a meaningful comparison,
we have to supplement dimreg by a renormalization prescription.}
is the most widely used regularization and renormalization 
method in particle physics.
It owes its popularity mainly to its simplicity and to
the fact that it respects gauge invariance,
one of the cornerstones of particle physics models.
It is also remarkably selective: in the language of momentum cutoffs,
it extracts only the logarithmic divergences, which for most
applications turn out to contain the important information
(in particular, the beta functions of the marginal couplings).
However, in its standard implementation, dimreg 
it is a purely perturbative device,
and it works only in even dimensions.

On the other hand, the Functional Renormalization Group (FRG)
Equation (FRGE)
is a convenient way of implementing Wilson's idea of
integrating out modes one momentum shell at the time.
At its core lies a choice of a ``regulator'' function $R_k$
that suppresses the contribution of low momentum modes
to the path integral.
The regulator depends on a scale parameter $k$ 
with dimension of mass, and the derivative
with respect to $k$ gives the contribution to the effective action
of an infinitesimal momentum shell.
The contribution to the functional integral of a momentum
shell of thickness $\Delta k$ can be written as a loop expansion.
The $\ell$-loop term is of order $(\Delta k/k)^\ell$,
so that the continuous FRGE ($\Delta k/k\to 0$) 
looks like a one-loop equation \cite{Wegner:1972ih}.
In the 1PI formulation, the FRGE reads~\cite{wett1,Morris:1993qb,Bonini:1992vh,Ellwanger:1993mw}
\be
\frac{d\Gamma_k}{dt} = \frac{\hbar}{2} \Tr \left( \frac{\delta^2 \Gamma_k}{\delta \phi \delta \phi} + R_k \right)^{-1} \frac{d R_k}{dt}
\label{FRGE}
\ee
where the functional $\Gamma_k$ is a scale-dependent
version of the Effective Action (EA),
generally called the Effective Average Action (EAA)
and $t=\log(k/k_0)$.
We refer to \cite{Bagnuls:2000ae,Pawlowski:2005xe,Gies:2006wv,Delamotte:2007pf,Dupuis:2020fhh} for reviews of this
equation and its applications.
The one-loop nature of the FRGE is manifest in the
presence of a single trace (momentum integration).
In fact, the FRGE can be represented graphically as
\be
\partial_t \Gamma_k =\frac12\ 
\begin{tikzpicture}
[>=stealth,scale=0.6,baseline=5mm]
\draw[double, very thick] (0,0) arc (270:-90:1cm);
\fill[white] (0,2) circle (0.2cm);	%
\draw (0,2) circle (0.2cm);			%
\draw (0,2) -- (0.141,2.141);		%
\draw (0,2) -- (-0.141,2.141);		%
\draw (0,2) -- (0.141,1.859);		%
\draw (0,2) -- (-0.141,1.859);		%
\end{tikzpicture} \ ,
\ee
where the double line represents the full propagators
and the crossed circle represents the insertion of the 
regulator $\partial_t R_k$.
The counterpart of this simplicity is that the equation
is only exact if one takes into account {\it all}
possible terms in the action.
Since it is practically impossible to solve the exact equation,
its effectiveness hinges crucially
on a good choice of approximation.
There are three main systematic expansion schemes.
We briefly recall their definition,
and then discuss the relation among them, and
to standard perturbation theory.
\smallskip

{\sl Loop expansion.}
This is an expansion in powers of $\hbar$.
We write for the EAA:
\begin{equation}
\Gamma_{k}[\phi]=S_{\Lambda}[\phi]+\sum_{L=1}^{n}\hbar^{L}\Gamma_{L,k}[\phi]\ .\label{PT_2}
\end{equation}
Inserting (\ref{PT_2}) in the the flow equation (\ref{FRGE})
one can reproduce the usual beta functions of perturbation theory.
First, introducing $S$ in the right-hand side (r.h.s.) of (\ref{FRGE}),
one calculates the one-loop beta functional $\partial_t\Gamma_{1,k}$.
Integrating over $k$ from $\Lambda$ to $k'$
gives the one-loop EAA $\Gamma_{1,k'}$,
and using this in the r.h.s.~of (\ref{FRGE}) 
one calculates the two-loop beta functional $\partial_t\Gamma_{2,k}$.
The procedure can be iterated.
Since in many cases the loop expansion coincides with the
expansion in the marginal coupling constant,
this approximation scheme is very close to standard 
weak-coupling perturbation theory.
\smallskip

{\sl Vertex expansion.}
The EAA can be Taylor-expanded in powers of the field:
\be
\Gamma_k[\phi]=\sum_n \int_{p_1}\ldots\int_{p_n}
\Gamma_k^{(n)}(p_1,\ldots,p_n)
\phi(p_1)\ldots\phi(p_n)
\label{verexp}
\ee
where $p_n$ are the external momenta.
By functionally differentiating \Eqref{FRGE}
one obtains an infinite sequence of flow equations
for the $n$-point functions $\Gamma^{(n)}$.
The vertex expansion consists in truncating this sequence
at some finite order.
The first three equations of the sequence for a
$\mathbb{Z}_2$-invariant scalar theory 
can be represented graphically as follows:
\begin{subequations}
\begin{align}
\partial_t \Gamma_k^{(2)} &=-\frac12\ 
\begin{tikzpicture}
[>=stealth,scale=0.6,baseline=0.2cm]
\draw [double, very thick] (-1.7,0) -- (0,0.0) node[anchor=east] {};
\draw[double, very thick] (0,0) -- (1.7,0.0) node[anchor=east] {};
\draw[double, very thick] (0,0) arc (270:-90:1cm);
\fill[white] (0,2) circle (0.2cm);	%
\draw (0,2) circle (0.2cm);			%
\draw (0,2) -- (0.141,2.141);		%
\draw (0,2) -- (-0.141,2.141);		%
\draw (0,2) -- (0.141,1.859);		%
\draw (0,2) -- (-0.141,1.859);		%
\draw[fill] (0,0) circle [radius=0.2];
\end{tikzpicture}\ ,\\
\partial_t \Gamma_k^{(4)} &=3\
\begin{tikzpicture}
[>=stealth,scale=0.6,baseline=-0.1cm]
\draw[double, very thick] (-1,0) arc (180:0:1cm);
\draw[double, very thick] (-1,0) arc (180:360:1cm);
\draw[double, very thick] (2,1) node[anchor=west] {} -- (1,0);
\draw[double, very thick] (2,-1) node[anchor=west] {} -- (1,0);
\draw[double, very thick] (-1,0) -- (-2,1) node[anchor=east] {};
\draw[double, very thick] (-1,0) -- (-2,-1) node[anchor=east] {};
\draw[fill] (-1,0) circle [radius=0.2];
\draw[fill] (1,0) circle [radius=0.2];
\fill[white] (0,1) circle (0.2cm);	
\draw (0,1) circle (0.2cm);		%
\draw (0,1) -- (0.141,1.141);		%
\draw (0,1) -- (-0.141,1.141);		%
\draw (0,1) -- (0.141,0.859);		%
\draw (0,1) -- (-0.141,0.859);
\end{tikzpicture}
-\frac12\ 
\begin{tikzpicture}
[>=stealth,scale=0.6,baseline=0cm]
\draw [double, very thick] (-1.7,0) -- (0,0.0) node[anchor=east] {};
\draw[double, very thick] (0,0) -- (1.7,0.0) node[anchor=east] {};
\draw[double, very thick, rotate=-45] (0,0) -- (1.7,0.0) node[anchor=east] {};
\draw [double, very thick, rotate=45] (-1.7,0) -- (0,0.0) node[anchor=east] {};
\draw[double, very thick] (0,0) arc (270:-90:1cm);
\fill[white] (0,2) circle (0.2cm);	%
\draw (0,2) circle (0.2cm);			%
\draw (0,2) -- (0.141,2.141);		%
\draw (0,2) -- (-0.141,2.141);		%
\draw (0,2) -- (0.141,1.859);		%
\draw (0,2) -- (-0.141,1.859);		%
\draw[fill] (0,0) circle [radius=0.2];
\end{tikzpicture} \ , \\
\partial_t \Gamma_k^{(6)} &=-45\
\begin{tikzpicture}
[>=stealth,scale=0.6,baseline=-0.2cm]
\draw[double, very thick] (-1,0) arc (180:0:1cm);
\draw[double, very thick] (-1,0) arc (180:360:1cm);
\draw[double, very thick] (2,1)  -- (1,0);
\draw[double, very thick] (2,-1) -- (1,0);
\draw[double, very thick] (-1,0) -- (-2,1);
\draw[double, very thick] (-1,0) -- (-2,-1);
\draw[double, very thick] (0,-1)  -- (-1,-2);
\draw[double, very thick] (0,-1)  -- (1,-2);
\draw[fill] (-1,0) circle [radius=0.2];
\draw[fill] (1,0) circle [radius=0.2];
\draw[fill] (0,-1) circle [radius=0.2];
\fill[white] (0,1) circle (0.2cm);	
\draw (0,1) circle (0.2cm);		%
\draw (0,1) -- (0.141,1.141);		%
\draw (0,1) -- (-0.141,1.141);		%
\draw (0,1) -- (0.141,0.859);		%
\draw (0,1) -- (-0.141,0.859);
\end{tikzpicture}
\ +15\
\begin{tikzpicture}
[>=stealth,scale=0.6,baseline=-0.1cm]
\draw[double, very thick] (-1,0) arc (180:0:1cm);
\draw[double, very thick] (-1,0) arc (180:360:1cm);
\draw[double, very thick] (1,0) -- (2.34, 0.437);
\draw[double, very thick] (1,0) -- (2.34, -0.437);
\draw[double, very thick] (1,0) -- (1.83, 1.14);
\draw[double, very thick] (1,0) -- (1.83, -1.14);
\draw[double, very thick] (-1,0) -- (-2,1) node[anchor=east] {};
\draw[double, very thick] (-1,0) -- (-2,-1) node[anchor=east] {};
\draw[fill] (-1,0) circle [radius=0.2];
\draw[fill] (1,0) circle [radius=0.2];
\fill[white] (0,1) circle (0.2cm);	
\draw (0,1) circle (0.2cm);		%
\draw (0,1) -- (0.141,1.141);		%
\draw (0,1) -- (-0.141,1.141);		%
\draw (0,1) -- (0.141,0.859);		%
\draw (0,1) -- (-0.141,0.859);
\end{tikzpicture}
\ -\frac12\
\begin{tikzpicture}
[>=stealth,scale=0.6,baseline=0cm]
\draw [double, very thick] (-1.7,0) -- (0,0.0) node[anchor=east] {};
\draw[double, very thick] (0,0) -- (1.7,0.0) node[anchor=east] {};
\draw[double, very thick, rotate=-36] (0,0) -- (1.7,0.0) node[anchor=east] {};
\draw [double, very thick, rotate=36] (-1.7,0) -- (0,0.0) node[anchor=east] {};
\draw[double, very thick, rotate=-72] (0,0) -- (1.7,0.0) node[anchor=east] {};
\draw [double, very thick, rotate=72] (-1.7,0) -- (0,0.0); 
\draw[double, very thick] (0,0) arc (270:-90:1cm);
\fill[white] (0,2) circle (0.2cm);	%
\draw (0,2) circle (0.2cm);			%
\draw (0,2) -- (0.141,2.141);		%
\draw (0,2) -- (-0.141,2.141);		%
\draw (0,2) -- (0.141,1.859);		%
\draw (0,2) -- (-0.141,1.859);		%
\draw[fill] (0,0) circle [radius=0.2];
\end{tikzpicture}\ .
\end{align}
\label{eqs:vertexexp}
\end{subequations}
Here the 
black dots represent full vertices.
The vertex expansion is clearly a good approximation in weak field situations,
and is widely used in particle physics,
where one generally deals with just a few quanta of the field.
In this approximation one retains the full momentum dependence.
\smallskip

{\sl Derivative expansion.}
When one is interested in low energy phenomena,
one can expand the action in powers of derivatives.
This is close to many applications of 
the effective field theory approach.
For a single scalar field the expansion starts with
\be
\Gamma_k[\phi]=
\int dx\left(
V_k(\phi)
+\frac{1}{2}Z_k(\phi)(\partial\phi)^2
+O(\partial^4)\right)
\label{derexp}
\ee
where $V_k$ and $Z_k$ are arbitrary functions of the field.
Inserting in \Eqref{FRGE} one obtains flow equations
for $V_k$, $Z_k$ etc.
This is complementary to the vertex expansion,
because one retains the full field dependence,
but only the lowest powers of momentum.
\smallskip

These expansions give rise to different forms of perturbation theory,
where different parameters are assumed to be small,
and a statement that is perturbative in one expansion
is generally nonperturbative in the others.
Thus for example, the leading order of the derivative
expansion, which is called the Local Potential Approximation (LPA),
consists in retaining in (\ref{derexp})
only the running potential $V_k$ and to put $Z_k=1$.
The beta function of the potential that can be obtained in this way
from the FRGE contains information about
infinitely many orders of the vertex expansion,
and to all loop orders.
If furthermore the potential is assumed to be a finite polynomial,
then one is working simultaneously in the derivative and vertex expansion.
Similarly, truncating the vertex expansion to a finite order
gives $n$-point functions that include all orders of the
derivative expansion and of the loop expansion,
and the EA calculated at a given order of the loop expansion 
contains information
that includes all orders of the derivative and vertex expansion.

In practice, in applications of perturbative quantum field theory
to particle physics, one generally considers two-, three- and four-point functions, at a finite order of the loop expansion,
and therefore one is working simultaneously in the vertex
and in the loop expansion.
This is what we shall refer to as ``standard perturbation theory''.
A different implementation of the 
weak coupling expansion is possible 
in presence of background fields,
since perturbation theory can then account for
the full dependence of the vertices on the latter
variables.
This is what might be called ``functional
perturbation theory''.
The application of dimreg to such functional
methods \cite{Jack:1982hf,Ford:1992pn,Martin:2001vx}
has been recently revived in the study of
conformal field theories \cite{Codello:2017hhh}.
In these respects, our study could be interpreted
as an attempt to extend these methods to the
nonperturbative domain.

Focusing again on a scalar field and
starting from a bare action that contains only
a quartic interaction $\lambda\phi^4$,
one can compute the two-, three- and four-point functions
at some finite loop order $L$,
by means of a functional integral, Feynman diagrams etc..
How does one obtain such higher-loop information from the FRGE,
which is a one-loop equation?
One has to recall that the FRGE is only exact when one uses
the full propagator and the full vertices.
The propagators and vertices that appear in the vertex expansion 
can be expanded in loops, and this gives rise to
the higher-loop effects in the beta function.
Integrating the flow from a bare action $S$ at
some UV cutoff scale $\Lambda$ down to $k=0$
gives the desired terms in the EA.
We shall discuss this in some detail in Section \ref{sec:twoloops}.

In summary, one of the most interesting
features of the FRGE is the availability of various approximation
schemes that sometimes allow us to follow the
flow of infinitely many couplings in a single stroke
and to go beyond standard perturbation theory.
On the other hand, the arbitrariness in the choice of the regulator
means that much of the information contained in the flow is
unphysical. 
One has to learn to extract physical information
from it.
\smallskip

Since the strengths and weaknesses of the FRGE
and of dimreg/\MSbar are quite complementary, 
it would be useful to transfer 
some of the strengths of one method to the other,
or at least to use them in a complementary way,
so as to overcome the respective weaknesses.
This paper is a first attempt in this direction.
The main question that we shall address is the following:
is there a choice of regulator that reproduces 
the beta functions of the \MSbar scheme in the 
standard perturbative domain?
We provide here a positive answer to this question:
we show that by bending the standard rules and procedures of the FRG 
it is possible to reproduce the results of
dimreg/\MSbar, at least up to two loops.
In this way, we will make manifest the kind of unphysical
features that one has to impose on a regulator so 
as to reproduce the results of dimreg.
For this reason we will talk about a ``pseudo-regulator'' that, 
upon use in the FRG equation, reproduces the
beta functions of \MSbar.

More importantly, having shown that the \MSbar pseudo-regulator
reproduces the results of dimreg in the perturbative domain,
we have a tool that potentially provides a nonperturbative 
extension of dimreg/\MSbar.
We will indeed show that with the pseudo-regulator
one can find and study all multicritical fixed points
in two dimensions, as well as the critical Sine-Gordon theory.
It is remarkable that in this way one can even write 
the potentials of these models in closed form.
Furthermore, the use of this tool is not limited to
even dimensions, as we shall show by considering the
Wilson-Fisher fixed point in three dimensions.

The use of dimreg is really of great advantage when one
deals with gauge theories.
We will not attempt here to use the \MSbar pseudo-regulator
in the FRGE for gauge theories,
but we will show that it has definite advantages
in the treatment of nonlinearly realized symmetries.

The paper is organized as follows.
In Section \ref{sec:statement} we state the problem in a precise way,
in the most straightforward and simplified
setting: the case of a linear scalar field theory in the LPA.
The solution of the problem and our pseudo-regulator
are given in Section \ref{sec:LPA}.
We also explore some of the intrinsic
freedom in the construction of the pseudo-regulator,
and we exhibit a one-parameter family of
regulators that continuously connects  the results
of standard FRG regulators with those of the \MSbar
pseudo-regulator.

In Section \ref{sec:beyondLPA}, we account for
the inclusion of the field's anomalous dimension.
This transition only requires minor generalizations
of the pseudo-regulator, allowing for some more
free parameters, which come along with 
corresponding forms of ``RG improvement'' in
the one-loop flow equations.
Section \ref{sec:DE} further shows that the same
pseudo-regulator is appropriate
for the $O(\partial^2)$ of the
derivative expansion.
This discussion offers us 
the chance to address two exploratory
applications of the \MSbar functional
RG equations.
The first is the description of
nonperturbative critical phenomena,
namely two-dimensional multicritical
scalar theories.
We perform this study with the main goal
to test the physical
content of the ``RG improvement'',
which is the imprint of the FRG
origin of our \MSbar equations.
The second application is
provided by nonlinear $O(N)$ models
in two dimensions, whose interest
in this context lies in the 
interplay between nonlinearly realized
symmetries and the FRG equations.
 
An even more general truncation is needed to
reproduce the two-loop \MSbar beta functions in
massive four-dimensional $\phi^4$ theory (the perturbatively
renormalizable linear $O(N)$ model).
This is discussed in Section \ref{sec:twoloops}.
This exercise serves as a proof
that by means of the FRG and our pseudo-regulator
one can, by considering large-enough truncations,
obtain \MSbar flow equations which are beyond 
a one-loop form.
 
Finally, in Section \ref{sec:epsilonexp}
we explore the role
of dimensionality in our construction.
In fact, while dimreg/\MSbar is usually
at work in an even number of dimensions $d$,
the FRG equations can be obtained and
applied for continuous $d$. We show
that the latter feature can be preserved
while taking the limit from the FRG to \MSbar. 

Section \ref{sec:conclusions} 
contains some concluding remarks and
an outlook on possible future developments.
Several appendices account for the details
and the subtleties of the computations presented.

%
%
%
%
%
%
%
%
%
%


\section{Statement of the problem}
\label{sec:statement}

In order to make our idea more precise, let us begin by stating
the conditions that are generally imposed on a regulator for 
the FRG equation.
A regulator is an additive modification of the 
inverse two-point function, and is therefore a function
of a single momentum $q$, or rather its modulus $z=q^2$, 
depending on an scale $k$. 
The regulator, which is denoted $R_k(z)$,
is typically assumed to satisfy the following conditions:
\begin{itemize}
\item to be positive (must suppress modes);
\item to be monotonically increasing with $k$, for all $z$;
\item to be monotonically decreasing with $z$, for all $k$;
\item $\lim_{k\to 0}R_k(z)=0$ for all $z$;
\item for $z>k^2$, $R_k$ goes to zero sufficiently fast, 
e.g. as an exponential;
\item $R_k(0)=k^2$.
\end{itemize}
The first three conditions are obvious properties of a cutoff.
The fourth guarantees that the path integral reproduces the
standard partition function for $k=0$.
The fifth condition ensures that high momentum modes are integrated out unsuppressed and guarantees the UV convergence of the r.h.s of the flow equation.
The sixth and last condition provides a sort of normalization.
For certain purposes, one may sometimes forgo the last two conditions and consider cutoffs that either do not decrease 
very fast for large momenta or even diverge when $z\to 0$.
These six conditions are useful in that
they provide a clear physical interpretation for the
coarse graining implemented by the regulator, 
and they ensure control on the UV and IR endpoints of the
momentum integrals. However, they are not needed in the derivation
of the FRG equation, which would keep its exact one-loop
form for any regulator choice.

Both $z$ and the function $R_k(z)$ have dimension of mass squared,
so we can write
\be
\label{berna}
R_k(z)=k^2 r(y)\ ;\qquad y=z/k^2\ ,
\ee
where $r$ is a dimensionless ``cutoff profile''.
The following are typical choices:
\begin{eqnarray}
r(y) & = & \frac{y}{e^y-1} \ ,
\label{cutoffs1}\\
r(y) & = & \frac{y^2}{e^{y^2}-1} \ ,
\label{cutoffs2}\\
r(y) & = & (1-y)\theta(1-y)\ .
\label{opt}
\end{eqnarray}
The third choice has been argued to provide ``optimized'' results,
in a certain class of models and truncations
\cite{Litim:2001up,Litim:2002cf}.
For certain purposes its non-differentiability is an issue,
but it has the great advantage of allowing an analytic evaluation
of momentum integrals.
Note that $k$ plays the role of an {\it infrared} cutoff:
its effect is to give a mass of order $k$ to the modes with $\sqrt{z}<k$,
and no mass to the modes with $\sqrt{z}>k$.

Introducing the cutoff in the functional integral and then
performing the Legendre transform leads to the 
FRGE (\ref{FRGE}).
We note that
the trace on the r.h.s.~is IR and UV-finite, and that
the equation contains no reference to a bare action 
or UV physics. 

In order to extract useful information from the exact equation
one has to approximate it in some way.
For definiteness, let us focus on a single scalar field in the LPA 
\be
\Gamma_k(\phi) =  \int d^dx \left( \frac{1}{2}\left( \partial_\mu \phi \right)^2 + V_k(\phi) \right)\ .
\label{LPA}
\ee

Inserting in the FRGE we obtain the ``beta functional''
\be
\partial_t V_k = \frac{1}{2(4\pi)^{d/2}} Q_{d/2} \left[ \frac{\partial_t R_k}{P_k+ V_k'' } \right]\ ,
\label{betaV}
\ee
where
\be
Q_n \left[ W \right]= \frac{1}{\Gamma(n)} \int_0^\infty dz \, z^{n-1} W(z)
\label{Qf}
\ee
is the momentum integral.
Assuming $\mathbb{Z}_2$-symmetry and Taylor-expanding the potential 
\be
V_k(\phi) = \sum_n \frac{\lambda_{2n}(k)}{(2n)!} \phi^{2n}\ ,
\label{potexp}
\ee
we can derive infinitely many beta functions
$\beta_{2n}=k\frac{\partial \lambda_{2n}}{\partial k}$.
These are
obtained by expanding both sides of (\ref{betaV}) in powers
of the field and equating the coefficients.
For arbitrary regulator, and in any dimension,
for the first few couplings
this leads to
\begin{subequations}
\begin{align}
\beta_2 &=
-\frac{1}{2(4\pi)^{d/2}} \lambda_4 Q_{d/2}\left[ \frac{\partial_t R_k}{(P_k+\lambda_2)^2} \right],
\\
\beta_4 &= \frac{1}{2(4\pi)^{d/2}}  \left(
6\lambda_4^2 Q_{d/2}\left[ \frac{\partial_t R_k}{(P_k+\lambda_2)^3} \right]-
\lambda_6 Q_{d/2}\left[ \frac{\partial_t R_k}{(P_k+\lambda_2)^2} \right]
\right),
\\
\beta_6 &= \frac{1}{2(4\pi)^{d/2}} \left(
-90 \lambda_4^3 Q_{d/2}\left[ \frac{\partial_t R_k}{(P_k+\lambda_2)^4} \right]+
\right.
\nonumber\\
& \left.
+30 \lambda_4 \lambda_6 Q_{d/2}\left[ \frac{\partial_t R_k}{(P_k+\lambda_2)^3} \right] -\lambda_8 Q_{d/2}\left[ \frac{\partial_t R_k}{(P_k+\lambda_2)^2} \right]
\right).
\label{betaQ}
\end{align}
\end{subequations}
We note that these are  one-loop beta functions,
since no resummation is involved.
They coincide with the first three equations of the vertex
expansion, namely Eqs.~(\ref{eqs:vertexexp}),
when the $n$-point functions are evaluated at zero momentum.
In order to have more explicit formulae,
we can use the optimized regulator (\ref{opt}), that gives
\be
Q_n\left[ \frac{\partial_t R_k}{(P_k+\lambda_2)^\ell} \right] = \frac{2}{\Gamma(n+1)}\frac{k^{2(n+1)}}{(k^2+\lambda_2)^\ell}\ .
\label{Qopt}
\ee
Then, the first beta functions are
\begin{subequations}
\begin{align}
\beta_2 &=
-\frac{k^{d+2}}{(4\pi)^{d/2}\Gamma\left(\frac{d}{2}+1\right)}\frac{\lambda_4 }{(k^2+\lambda_2)^2} \ ,
\\
\beta_4 &= \frac{k^{d+2}}{(4\pi)^{d/2}\Gamma\left(\frac{d}{2}+1\right)} \left(
6 \frac{\lambda_4^2}{(k^2+\lambda_2)^3} -
 \frac{\lambda_6}{(k^2+\lambda_2)^2} 
\right)\ ,
\\
\beta_6 &= \frac{k^{d+2}}{(4\pi)^{d/2}\Gamma\left(\frac{d}{2}+1\right)} 
\left(-90\frac{\lambda_4^3}{(k^2+\lambda_2)^4}
+30 \frac{\lambda_4 \lambda_6 }{(k^2+\lambda_2)^3}  
- \frac{\lambda_8}{(k^2+\lambda_2)^2} 
\right)\ .
\label{betaopt}
\end{align}
\end{subequations}
One can also calculate the beta functions of this theory
at one loop using dimreg/\MSbar.
The corresponding expressions read
\begin{subequations}
\begin{align}
\beta_2 &=
\frac{(-)^{d/2}}{\Gamma\left(\frac{d}{2}\right)(4\pi)^{d/2}} \lambda_4 \lambda_2^{d/2-1}\ ,
\\
\beta_4 &= \frac{(-)^{d/2}}{(4\pi)^{d/2}}  \left(
3\lambda_4^2 \frac{\lambda_2^{d/2-2}}{\Gamma\left(\frac{d}{2}-1\right)}+
\lambda_6 \frac{\lambda_2^{d/2-1}}{\Gamma\left(\frac{d}{2}\right)}
\right)\ ,
\\
\beta_6 &= \frac{(-)^{d/2}}{(4\pi)^{d/2}} \left(
15 \lambda_4^3 \frac{\lambda_2^{d/2-3}}{\Gamma\left(\frac{d}{2}-2\right)} +
15 \lambda_4 \lambda_6 \frac{\lambda_2^{d/2-2} }{\Gamma\left(\frac{d}{2}-1\right)} 
+\lambda_8 \frac{\lambda_2^{d/2-1}}{\Gamma\left(\frac{d}{2}\right)} 
\right)\ .
\end{align}
\label{betadimreg}
\end{subequations}
In fact, one can even derive a functional perturbative
beta function for $V$, analogous to (\ref{betaV})
\cite{Codello:2017hhh}.
(We shall discuss this in Section \ref{sec:flpa}).

The beta functions obtained by the two procedures
are strikingly different.
In the beta functions derived from the FRG,
the dimension is carried by $k$,
and there are denominators that automatically produce
decoupling when one crosses the mass threshold $k^2=\lambda_2$.
In the beta functions of dimreg the dimension
is always carried by powers of $\lambda_2$,
and threshold effects are not accounted for.
In fact such beta functions are only valid
at energies much higher that $\lambda_2$.

Nevertheless, there is a relationship between these
two sets of beta functions.
To see this, note that for a generic regulator, 
the $Q$-functional with $\ell=n+1$ and $\lambda_2=0$ 
(which is dimensionless) is universal, {\it i.e.}
\be
Q_n\left[ \frac{\partial_t R_k}{P_k^{n+1}} \right] = \frac{2}{\Gamma(n+1)} \ ,
\label{universal}
\ee
independently of the shape of the regulator.
The reason for this is that in this case
the integrand is a total derivative:
\be
\int_0^\infty dz z^{n-1}\frac{\partial_t R_k}{P_k^{n+1}}=
\int_0^\infty dy y^{n-1}2\frac{r(y)-yr'(y)}{(y+r(y))^{n+1}}
=\int_0^\infty dy \frac{2}{n}\frac{d}{dy}\left(\frac{y}{y+r(y)}\right)^n\ .
\ee
The universal result will hold even if the regulator does not
satisfy all the requirements that are listed in the beginning
of Section \ref{sec:statement}: it is enough that $r(\infty)=0$ and $r(0)>0$.

In the presence of a mass $\lambda_2$,
we can expand the $Q$-functional for $k^2 > \lambda_2$:
\bea
Q_n\left[ \frac{\partial_t R_k}{(P_k+\lambda_2)^\ell} \right] &=& 
\sum_{j=0}^{\infty} \frac{(-)^j \Gamma(\ell+j)}{\Gamma(\ell)\Gamma(j+1)} \lambda_2^j 
Q_n\left[ \frac{\partial_t R_k}{P_k^{j+\ell}} \right]\ .
\eea
We see that the term $j=n-\ell+1$ in the sum is universal
and equal to
\be
\frac{2(-)^{n-\ell+1}}{\Gamma(\ell)\Gamma(n-\ell+2)} \lambda_2^{n-\ell+1}\ .
\ee
The beta functions of dimensional regularization consist
exactly of all these universal terms,
all the remaining ones being set simply to zero.

The main question we wish to address is this: 
is there a way to directly obtain the beta functions 
(\ref{betadimreg}) from the FRGE?
This will be the case provided $R_k$ is such that
\be
Q_n\left[ \frac{\partial_t R_k}{(P_k+\lambda_2)^\ell} \right] 
= \frac{2(-)^{n-\ell+1}}{\Gamma(\ell)\Gamma(n-\ell+2)} 
\lambda_2^{n-\ell+1}\ .
\label{desired}
\ee
Thus the question becomes one about the existence of a regulator
that gives (\ref{desired}).
It is immediately clear that any standard regulator,
satisfying the criteria given in the beginning of this section,
cannot do this.
To see this, it is sufficient to consider the case $\lambda_2=0$,
in which case the requirement (\ref{desired}) becomes
\be
Q_n\left[ \frac{\partial_t R_k}{P_k^\ell} \right] 
= \frac{2}{\Gamma(n+1)} \delta_{\ell, n+1}\ .
\label{desired2}
\ee
This implies that
\be
\frac{1}{\Gamma(n)} \int_0^\infty dy\, y^{n-1} \frac{r-yr'}{(y+r)^\ell} = \frac{1}{\Gamma(n+1)} \delta_{\ell , n+1}\ .
\ee
Using integration by parts and the standard properties of regulators,
we obtain 
\bea
\left( 1-\frac{\ell-1}{n} \right) Q_n\left[ \frac{R_k}{P_k^\ell} \right]
+ \left( \frac{\ell-1}{n } - \frac{\ell}{n+1} \right)\frac{\delta_{\ell, n+1}}{\Gamma(n+1)}=0
\eea
that, for $\ell \ne n+1$, gives
$Q_n\left[ \frac{R_k}{P_k^\ell} \right]=0$.
Since the integrand in this $Q$-functional is positive,
this implies $R_k=0$.
While $R_k$ cannot be identically
vanishing, it appears possible to reproduce \MSbar beta functions
by giving up some of the
requirements that are usually made of regulators and
taking the $R_k\to 0$ limit
in a suitable way, as we shall discuss in
the next section.

\section{The \texorpdfstring{\MSbar}{TEXT} pseudo-regulator}
\label{sec:LPA}

The desired ``pseudo-regulator'' depends, 
in addition to the scale $k$,
also on a dimensionless parameter $\epsilon$ and a mass $\mu$,
which play a similar role as the $\epsilon$ 
and $\mu$ parameters of dimreg:
\be
R_k(z) = \lim_{\epsilon \to 0} 
z\left[\left(\frac{z k^2}{\mu^4}\right)^\epsilon-1\right]  \ ,
\label{dimregcut}
\ee
or equivalently
\be
r(y)= \lim_{\epsilon \to 0} 
y\left[\left(\frac{y}{\tilde \mu^4}\right)^\epsilon-1\right]  \ ,
\ee
where $\tilde\mu=\mu/k$.
A derivation and an explaination of this
ansatz are given in Appendix~\ref{app:howtofind}.
Calculations have to be performed with a finite positive $\epsilon$
and the limit $\epsilon\to 0$ must be taken at the end of all calculations.
Note that expanding for small $\epsilon$
\be
R_k(z) = \epsilon z \log \left(\frac{k^2 z}{\mu^4}\right) 
+ O(\epsilon^2)\ .
\ee

The function (\ref{dimregcut}) grossly violates the
defining properties of a regulator, as spelled out 
in the beginning of Section \ref{sec:statement}.
Aside from the fact that it vanishes in the limit
$\epsilon\to0$, it is a growing function of $z$
and goes to zero for $z\to0$.
Nevertheless, it does what we asked for.
Calculating the $Q$-functional, 
we obtain
\bea
Q_n\left[\frac{\partial_t R_k}{(P_k+m^2)^\ell}\right]
&=&
k^{2(n-\ell+1)}\frac{2}{\Gamma(n)}
\int _0^\infty dy \,y^{n-1}\frac{r-y \partial_y r
-\tilde \mu/2\partial_{\tilde \mu}r}{(y+r+\tilde m^2)^\ell}
\nonumber
\\
&=&\lim_{\epsilon\to0}
\frac{2\epsilon }{1+\epsilon}m^{2(n-\ell+1)}
\left(\frac{\mu^2}{k m}\right)^{\frac{2n \epsilon}{1+\epsilon}}
\frac{\Gamma\left(1+\frac{n}{1+\epsilon}\right)
\Gamma\left(\ell-1-n+\frac{n\epsilon}{1+\epsilon}\right)}{\Gamma(n)\Gamma(\ell)}\ .
\nonumber
\eea
Here we introduced the more conventional
notation $m^2$ for the mass parameter $\lambda_2$,
and defined  $\tilde m=m/k$.
The integral in this $Q$-functional
is convergent for $\ell >(n+1+\epsilon)/(1+\epsilon)$
and is defined elsewhere by analytic continuation.
In the limit $\epsilon\to 0$ it goes to zero except
at the points where the second $\Gamma$-function in the numerator
has a pole, namely when $\ell-n-1$ is zero or a negative integer.
In this way we recover (\ref{desired}).
We note that for $n \le \ell-2$ the final result (\ref{desired}) is
identically zero because of the presence of the Gamma function on the denominator.
Since $\ell$ is an integer, $n$ must be integer in order to have a non zero result:
since, for the beta functions of the LPA, $n=d/2$,
this implies that only in even dimensions we get a non zero result.
This agrees with the standard lore that dimreg only works 
in even dimensions.

Sometimes one needs the $Q$-functionals for $n\le 0$.
One can obtain them by observing that
\be
Q_n\left[ \frac{\partial_t R_k}{(P_k+m^2)^\ell} \right] = \frac{(-1)^j}{\Gamma(n+j)}\int_0^\infty dz z^{n+j-1} \frac{d^j}{(dz)^j}\left(\frac{\partial_t R_k(z)}{(P_k(z)+m^2)^\ell}\right) \ ,
\ee
where $j$ is an integer such that $n+j>0$.
Evaluating this expression for the pseudo-regulator, we get
\be
Q_n\left[ \frac{\partial_t R_k}{(P_k+m^2)^\ell} \right] 
= 2 \,\delta_{-n,0}\,\delta_{\ell,1} \ .
\ee
This agrees with the analytic continuation of (\ref{desired}).

\subsection{The effective potential in the LPA}
\label{sec:flpa}

We complete the discussion of the LPA approximation
of a scalar theory by giving the functional equation
for the potential:
\be
\partial_t \tilde V_k = -d \tilde V_k + \left(\frac{d}{2}-1 \right) \tilde \phi \tilde V_k'+  c_d \left(- \tilde V_k'' \right)^{\frac{d}{2}}\ ,
\label{flowVdimreg}
\ee
where  $\tilde V_k = k^{-d} V_k $,
$\tilde \phi = k^{1-\frac{d}{2}} \phi$
and $c_d = \frac{1}{(4\pi)^{d/2}\Gamma \left[\frac{d}{2}+1 \right]}$. 
This agrees with the beta functional in $d=4$ discussed
in \cite{Codello:2017hhh}.
For comparison, the optimized regulator leads to the form
\be
\partial_t \tilde V_k = -d \tilde V_k + \left(\frac{d}{2}-1 \right) \tilde \phi \tilde V_k'+ c_d \frac{1}{1+ \tilde V_k'' }\ ,
\label{flowVopt}
\ee
where $c_d =  \frac{1}{(4\pi)^{d/2}\Gamma \left[\frac{d}{2}+1 \right]}$.
We observe that (\ref{flowVdimreg}) picks exactly the terms
of the expansion of (\ref{flowVopt})
with the right power of $\tilde V''$ to give a dimension-$d$ 
operator.
Eqs.~(\ref{flowVdimreg}) and
(\ref{flowVopt}) are one-loop results,
and in this sense can be said to be perturbative,
but they contain infinitely many terms of the vertex expansion
and thus are not perturbative in the standard sense.

Eq.~(\ref{flowVdimreg}) can be applied only to even dimensions, so it does not admit
the Wilson-Fisher fixed point as a solution in $d=3$.
This was to be expected, since dimreg only works in even dimensions.
We anticipate however that generalizations to continuous $d$ (including also odd integers) are possible, and
will be discussed in Section~\ref{sec:epsilonexp}.
Eq.~(\ref{flowVdimreg})
has been used in
\cite{Codello:2017epp,Safari:2017irw,Safari:2017tgs,Codello:2018nbe,Codello:2019vtg} to obtain several new results
on statistical models.
In $d=2$ the corresponding fixed-point equation
has the critical Sine-Gordon solution
\be
V_* = - \frac{m^2}{8 \pi} \cos \left(\sqrt{8 \pi} \phi\right)\ ,
\ee
where $m$ is an arbitrary mass.
This result holds independently of the shape of the regulator
\cite{Codello:2012sc}.

A related question is whether this pseudo-regulator
can reproduce some of the (multi)-critical theories in $d=2$.
It turns out that the answer is positive,
as we shall discuss in greater detail in Section~\ref{sec:multicritical},
where we consider a larger
truncation.

\subsection{An external field problem}
\label{sec:externalfield}

As a somewhat different application, let us consider
a free scalar field with mass $m^2$ in an external metric
$g_{\mu\nu}$. Let $\Delta=-\nabla_\mu\nabla^\mu$ 
be the covariant Laplacian.
In this case we can take over previous formulas for the pseudo-regulator,
simply reinterpreting $z=\Delta$.
We refer to \cite{baldazzi} for several examples of this type,
both in Lorentzian and Euclidean signature.
The Euclidean beta functional is
\bea
\partial_t \EAAe &=& 
\frac{1}{2(4\pi)^{d/2}}
\sum_{j=0}^\infty
Q_{\frac{d}{2}-j}\left[\frac{\partial_t R_k}{P_k+ \lambda_2}\right]B_{2j}(\Delta)\ 
\nonumber\\
&=& 
\frac{(-)^{d/2}\lambda_2^{d/2}}{(4\pi)^{d/2}}
\sum_{j=0}^{d/2}
\frac{(-)^j\lambda_2^{-j}}{\Gamma\left(\frac{d}{2}-j+1\right)} 
B_{2j}(\Delta) \ ,
\eea
where $B_{2j}(\Delta)$ are the heat kernel coefficients
of the operator $\Delta$ and consist of integrals of
powers of the curvature tensor and its covariant derivatives.
Note that the sum terminates at $j=d/2$,
because of the poles in the Gamma function in the denominator.
We get
\be
\partial_t \EAAe =
-\frac{1}{4\pi}
\left(\lambda_2 B_0(\Delta) -
B_2(\Delta) \right)
\nonumber
\ee
for $d=2$, and
\be
\partial_t \EAAe =
\frac{1}{(4\pi)^2}
\left(\frac{\lambda_2^2}{2} B_0(\Delta) -
\lambda_2 B_2(\Delta) + B_4(\Delta)\right) 
\nonumber
\ee
for $d=4$.
These formulae give the cutoff-dependence of the
effective action for the external metric,
generated by the scalar field.
By integrating these formulae from some ultraviolet 
scale $\Lambda$ down to $k=0$, one obtains
the effective action for the metric.
See \cite{Codello:2015oqa,Ohta:2020bsc} 
for some calculations of this type.

\subsection{A first generalization}
\label{sec:fistgen}

In the definition (\ref{dimregcut}) we have used an external,
arbitrary mass scale $\mu$.
One could use instead a dimensionful coupling of the theory.
In particular, in a massive theory, one could use $m$
instead of $\mu$.
In the discussion of the two-loop beta functions,
it will be convenient to actually use a mixture
of the two.
Therefore, let us generalize the pseudo-regulator to
\be
R_k(z) = \lim_{\epsilon \to 0} 
z\left[\left(\frac{z k^2}{m^{2b}\mu^{4-2b}}\right)^\epsilon-1\right]  \ .
\label{dimregcutb}
\ee
Note that $m$ is a running parameter, so when we
evaluate the $Q$-functional (\ref{desired})
it gives rise to an additional term
depending on the beta function of the
mass $\beta_{m^2}=\partial_t m^2$:
\be
Q_n\left[ \frac{\partial_t R_k}{(P_k+m^2)^\ell} \right] =  
\frac{2(-)^{n-\ell+1}}{\Gamma(\ell)\Gamma(n-\ell+2)} 
\left( 1-\frac{b}{2}\frac{\beta_{m^2}}{m^2} \right) m^{2(n-\ell+1)}\ .
\label{aliceb}
\ee
The term with the beta function of the mass is a higher-loop
effect, so at one loop this pseudo-regulator still reproduces
the result of dimreg.

We note that the above discussion could be generalized
replacing $m$ by any combination of couplings
with the dimension of mass.
This would give rise to additional beta functions 
in the r.h.s. of (\ref{aliceb})
and may be useful in higher-loop calculations.

In the massless case ($m=0$) one has to set $b=0$ and introduce by hand an IR regulator in the $Q$-functionals:
\be
Q_n\left[ \frac{\partial_t R_k}{P_k^\ell} \right]\mapsto
Q_n\left[ \frac{\partial_t R_k}{(P_k+\mu^2)^\ell} \right]\ .
\ee
The limit $\mu \to 0$ has to be taken in the very end.
Note that this IR regulator mass is not necessarily
 equal to
the dimreg parameter $\mu$,
but we will not need this degree of generality,
so the same mass will be used in both r\^oles.
Then we obtain
\be
Q_n\left[ \frac{\partial_t R_k}{(P_k+\mu^2)^\ell} \right] = \frac{2(-)^{n-\ell+1}}{\Gamma(\ell)\Gamma(n-\ell+2)} \mu^{2(n-\ell+1)} \ .
\ee
As we already said above, this formula gives zero for $n < l-1 $.
Taking the limit for $\mu \to 0$ we get zero for $n >l+1$.
So the result is
\be
Q_n\left[ \frac{\partial_t R_k}{P_k^\ell} \right] = \lim_{\mu \to 0}Q_n\left[ \frac{\partial_t R_k}{(P_k+\mu^2)^\ell} \right] = \frac{2}{\Gamma(n+1)} \delta_{\ell,n+1} \ .
\ee
We note that only one combination of $\ell$ and $n$ gives a non-vanishing result, which corresponds to 
the universal result of \Eqref{universal}.

\subsection{Interpolation with the optimized regulator}
\label{sec:interpolation}

The $Q$-functionals for the optimized regulator have been
given in (\ref{Qopt}).
Now let us consider the following one-parameter family of regulators:
\be
r_a(y)= a(1-y)\theta(1-y) \ .
\label{opta}
\ee
For $a\not=1$ they violate the normalization condition,
but otherwise they are acceptable regulators.
In fact, the parameter $a$ is used to optimize the results
\cite{Canet:2003qd,Balog:2019rrg,DePolsi:2020pjk}.
The corresponding $Q$-functionals are given by
\be
Q_n\left[ \frac{\partial_t R_k}{(P_k+m^2)^\ell} \right] = 
\frac{2a k^{2(n-l+1)}}{(a+\tilde m^2)^\ell} \frac{1}{\Gamma(n+1)}
\phantom{F}_2F_1\left(\ell,n,n+1,-\frac{1-a}{a+\tilde m^2}\right)
 \ .
\label{optregres2}
\ee
If $\ell<n+1$ and $\tilde m>0$, these are monotonically
increasing functions of $a$, which are equal to
(\ref{Qopt}) for $a=1$ and decrease monotonically to zero
when $a\to0$.
If $\ell>n+1$ and $\tilde m>0$, they grow as functions of $a$
and they go to zero when $a\to 0$.
For $\tilde m=0$ they are monotonic functions on the interval
$0<a<1$, with either a zero or a pole for $a\to 0$
depending whether $\ell<n+1$ or $\ell>n+1$.
Remarkably, the $Q$-functionals $\ell=n+1$, $\tilde m=0$
are independent of $a$ and equal to (\ref{universal}).
Thus, the universality of these $Q$-functionals is not spoiled
by the regulator not being normalized.

\begin{figure}
\begin{center}
\includegraphics[width=0.4\columnwidth]{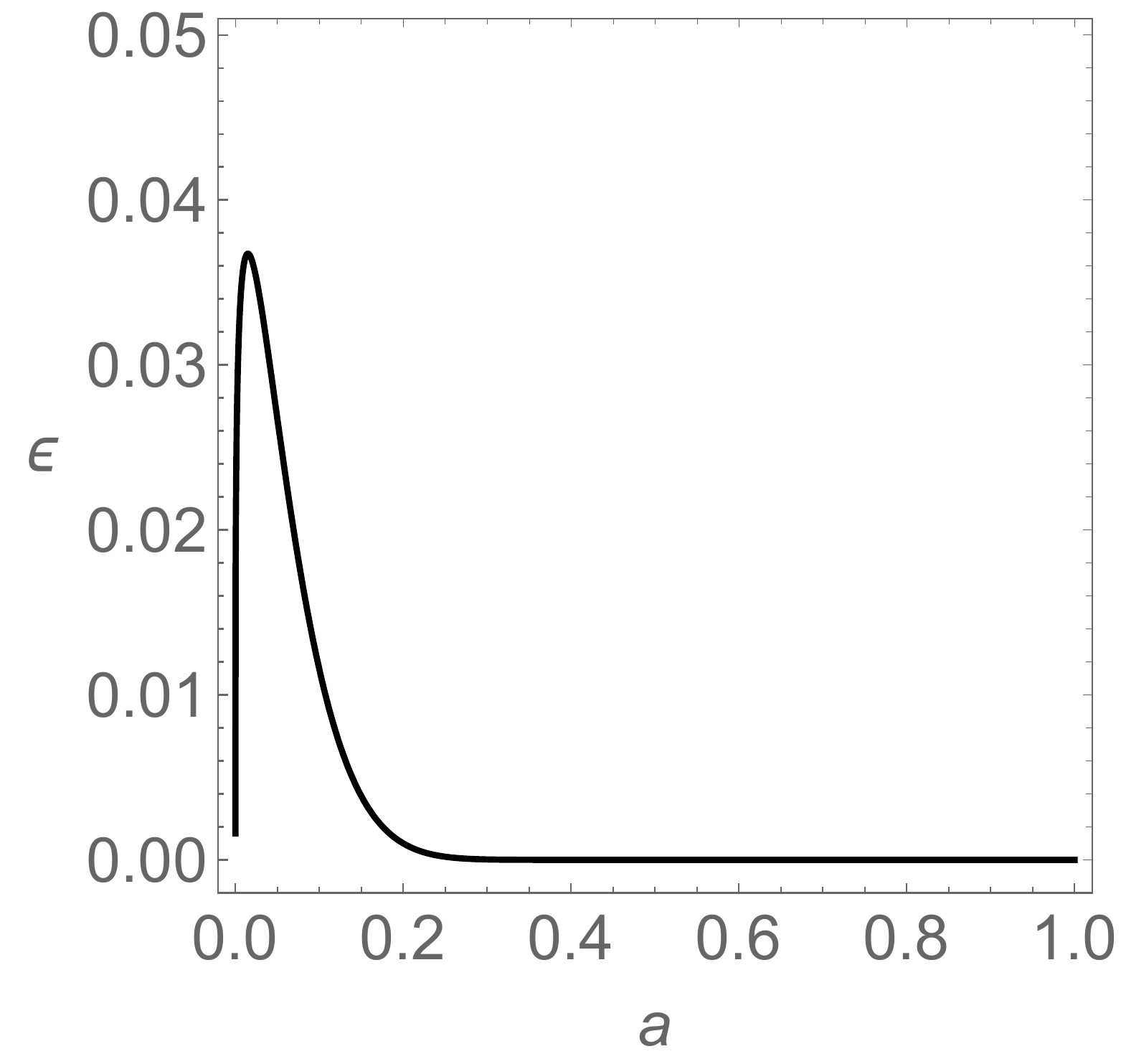}
\caption{A path in the $a$-$\epsilon$ plane
interpolating smoothly between the optimized cutoff
and the pseudo-cutoff of dimreg.}
\label{fig:example}
\end{center}
\end{figure}

On the other hand, if we set $a=0$ the regulator vanishes
identically and so do all the beta functions,
including the universal ones.
This means that the limit $a\to0$ is not continuous.
We would like to find a way to obtain at least the
universal beta functions also for $a=0$.
One can achieve this by introducing
an additional parameter $\epsilon$.
Consider the following interpolating regulator 
$R_k = k^2 r(y,\tilde m^2,\epsilon,a)$, with
\be
r(y,\tilde m^2,\epsilon,a) =  \left( a + (1-a) \tilde \mu^{-2(2-b)\epsilon} \tilde m^{-2b\epsilon} y ^{1+\epsilon}-y \right) \,\theta\left( 1 - \frac{a}{a+\epsilon}y \right) \ .
\label{rinterp}
\ee
For $\epsilon\to 0$ it reduces to (\ref{opta})
and for $a\to 0$ it reduces to (\ref{dimregcutb}).
Thus we can go continuously from the optimized regulator
to the pseudo-regulator reproducing dimreg
by following a curve of the form shown in Figure \ref{fig:example}.
In this way the limit $a\to0$ can  be made continuous.
The price one pays is that for $\epsilon\not=0$
one does not have a good regulator in the sense of
Section \ref{sec:statement}.
In any case we obtain the desired result
that all the non-universal beta functions go continuously to zero,
while the universal ones remain constant.

Let us see how the evaluation of the $Q$-functionals proceeds.
We have
\be
Q_n\left[\frac{\partial_t R_k}{(P_k+m^2)^\ell}\right]
=
\frac{k^{2(n-\ell+1)}}{\Gamma(n)}
\int_0^\infty dy y^{n-1}\frac{2r-2y \,\partial_y r-2 \tilde \mu^2 \partial_{\tilde \mu^2} r
- \tilde m^2\left(2-\frac{\beta_{m^2}}{m^2}\right)
\partial_{\tilde m^2}r}{(y+r+\tilde m^2)^\ell} \ .
\ee
Using (\ref{rinterp}), the fraction in the integral
can be written as the sum of three pieces:
\bea
&&
\frac{2a}{\left((1-a)\tilde m_b^{-4\epsilon}\, y^{1+\epsilon} 
+a+ \tilde m^2 \right)^\ell}\,
\theta\left( 1-\frac{ay}{a+\epsilon}\right)
\nonumber\\
&&
+\frac{2\epsilon(1-a)\left(1-\frac{b}{2}\frac{\beta_{m^2}}{m^2}\right)\tilde m_b^{-4\epsilon}\, 
y^{1+\epsilon}}{\left((1-a)\tilde m_b^{-4\epsilon}\, 
y^{1+\epsilon}+a+\tilde m^2\right)^\ell}\,
\theta\left(1-\frac{ay}{a+\epsilon}\right)
\nonumber\\
&& 
+\frac{ 2 \left( a + (1-a) \tilde m_b^{-4\epsilon}\, 
y^{1+\epsilon}-y\right)}{\left((1-a)\tilde m_b^{-4\epsilon}\, 
y^{1+\epsilon}+a+\tilde m^2\right)^\ell}\,\frac{a}{a+\epsilon}y\,
\delta\left(1-\frac{ay}{a+\epsilon}\right)
\ ,
\nonumber
\eea
where $\tilde m_b^2 = \tilde \mu^{2-b} \tilde m^{b}$ .\\
Performing the integral, the $Q$-functional is the sum
of three pieces 
\bea
1) \!\!\! &=& \!\!\!
2 a\frac{\left( \frac{a+\epsilon}{a} \right)^n}{\left(a+ \tilde m^2 \right)^\ell}
\frac{1}{\Gamma(n+1)}\,_2F_1 \left[\ell, \frac{n}{1+\epsilon}, 1+\frac{n}{1+\epsilon}; \frac{(a-1) \tilde m_b^{-4\epsilon}\,\left( \frac{a+\epsilon}{a} \right)^{1+\epsilon}}{a+\tilde m^2} \right] \ ,
\nonumber\\
2) \!\!\! &=& \!\!\!
\frac{2\epsilon (1-a)  \tilde m_b^{-4\epsilon}\, }{\Gamma(n)(n+1+\epsilon)} 
\left(\! 1-\frac{b}{2}\frac{\beta_{m^2}}{m^2}\!\right) 
\frac{\left( \frac{a+\epsilon}{a} \right)^{n\!+\!1\!+\!\epsilon}}{\left(a+ \tilde m^2 \right)^\ell}
\,_2F_1 \left[\ell,1\!+\!\frac{n}{1+\epsilon}, 2\!+\!\frac{n}{1+\epsilon}; \frac{(a-1)\tilde m_b^{-4\epsilon}\,\left(\frac{a+\epsilon}{a}\right)^{1+\epsilon}}{a+\tilde m^2}\right] \ ,
\nonumber\\
3) \!\!\! &=& \!\!\!
\frac{2}{\Gamma(n)}\frac{  \left( a + (1-a) \tilde m_b^{-4\epsilon}\, \left(\frac{a+\epsilon}{a} \right)^{1+\epsilon}- \frac{a+\epsilon}{a} \right) }{\left( (1-a) \tilde m_b^{-4\epsilon}\, \left( \frac{a+\epsilon}{a}  \right)^{1+\epsilon} +a+ \tilde m^2 \right)^\ell} \left( \frac{a+\epsilon}{a} \right)^{n-1}\ .
\nonumber
\eea
Taking the limit $\epsilon \to 0$
\bea
1) \!\!\! &\to& \!\!\!
2 \frac{a}{\left(a+ \tilde m^2 \right)^\ell}
\frac{1}{\Gamma(n+1)}\,_2F_1 \left[\ell, n, 1+n; \frac{a-1}{a+\tilde m^2} \right] \ ,
\nonumber\\
2) \!\!\! &\to& \!\!\!
0 \ ,
\nonumber\\
3) \!\!\! &\to& \!\!\!
0
\ .
\nonumber
\eea
Then sending $a \to 1$ we get (\ref{Qopt}) .\\
If $a\ne 0$ we can also take the two limits in the reverse order and obtain the same result.
For $a \to 0$ the previous expressions are not well defined, so in this case we have to take $\epsilon \to 0$ after the $a$ limit.
For $a \to 0$
\bea
1) &\to& 0 \ ,
\nonumber\\
2) &\to& 
2 \epsilon \left(1-\frac{b}{2}\frac{\beta_{m^2}}{m^2}\right)
\frac{\Gamma\left(1+\frac{n}{1+\epsilon} \right)\Gamma\left(\ell-1-\frac{n}{1+\epsilon}\right)}{(1+\epsilon)\Gamma(\ell)\Gamma(n)}
\left(\frac{\tilde m_b}{\tilde m}\right)^{\frac{4n\epsilon}{1+\epsilon}}
\tilde m^{2\left(n-\ell+1+\frac{\epsilon n}{1+\epsilon}\right)} \ ,
\nonumber\\
3) &\to& 0 \ .
\nonumber
\eea
Then taking the limit $\epsilon \to 0$ we get (\ref{aliceb}).

\section{Beyond the LPA}
\label{sec:beyondLPA}

In any quantum field theory application,
and in the FRG framework as well, the choice of a 
regularization scheme should  be
tailored to a specific model and computation.
In fact, 
although it is possible to devise 
regulators
which remove all possible divergences altogether,
much of the simplicity and power of the FRG
comes from the possibility to adopt
less drastic choices.
More minimalistic regularization schemes
allow for analytic, rather than numerical,
computations, thus rendering the optimization
process of such schemes simpler and more transparent.

In the process of relaxing the approximations
used to solve the exact FRG equations,
it is thus inevitable to reconsider 
the regulator choice.
In this section we discuss 
the adjustment of the \MSbar pseudo-regulator
to the transition from the
LPA to the inclusion of the
running wave function renormalization.
In the following, after the construction
of a more general family
of pseudo-regulators, we discuss its application
to scalar field theory.
We show how
these pseudo-regulators are appropriate for 
investigations within the LPA${}'$ approximation,
which differs from the LPA only
for the inclusion of a field- and momentum-independent
wave-function renormalization factor.
The next layer of complexity, namely
the O($\partial^2$) derivative
expansion including the field-dependence
of the wave-function renormalization,
will be addressed in Section \ref{sec:DE}.

From the point of view of standard perturbation theory,
the step from the LPA to the LPA${}'$ already involves 
the resummation of an infinite class of Feynman diagrams
 -- those self-energy-like one-particle-reducible corrections
to the internal propagator lines which 
are accounted for by a nontrivial field's anomalous dimension -- 
and therefore goes beyond 
finite-order perturbative calculations.

\subsection{R\^ole of the wave-function renormalization}
\label{sec:rolewavefun}

If the kinetic term in the action contains a nontrivial 
wave-function renormalization factor $Z_k\neq1$,
one usually includes this global factor
inside $R_k$ 
\begin{equation}
 R_k(z)\mapsto Z_k  {R}_k(z) \ .
 \label{eq:RproptoZ}
\end{equation}
There are several reasons in favor of this choice.
First of all, it allows to take over the
regulators already working in the LPA,
as the relevant regularized kinetic term 
is then in the functional form $z+R_k(z)$.
Furthermore, it is motivated by the desired invariance
under rigid rescalings of the fields,
also called reparameterizations.
In other words, it allows to remove $Z_k$ from the flow
equations by simply rescaling the fields
according to their quantum dimension
\begin{equation}
 d_\phi=\frac{d-2+\eta_k}{2} \ ,
\end{equation}
where 
\begin{equation}
\eta_k=-\partial_t\log Z_k
\end{equation}
is the field anomalous dimension.
While the former motivation is just a matter of convenience,
the latter is deeper and less arbitrary.
In fact, this choice is the one that minimizes the spurious
breaking of reparameterization invariance due to the
truncation of the exact FRG equation~\cite{Bagnuls:2000ae,Morris:1994ie,Comellas:1997ep,Osborn:2009vs}.

Following the choice of \Eqref{eq:RproptoZ}.
the flow equations receive further RG resummations
encoded in the appearance of $\eta_k$
on the r.h.s.~as
\begin{equation}
 \partial_t R_k(z)\mapsto Z_k  \left( \partial_t {R}_k(z)
 -\eta_k {R}_k(z) \right) \ .
 \label{eq:dRwithZeta}
\end{equation}
While the first term on the r.h.s.~gives rise 
to the $Q$-functionals already discussed in Section \ref{sec:statement},
the second term leads to the following new integrals
\be
Q_n\left[ \frac{ R_k}{(P_k+m^2)^\ell} \right] =
\sum_{j=0}^{\infty} \frac{(-)^j \Gamma(\ell+j)}{\Gamma(\ell)\Gamma(j+1)} m^{2j} 
Q_n\left[ \frac{ R_k}{P_k^{j+\ell}} \right] \ .
\ee
Also for these new $Q$-functionals we see that the term $j=n-\ell+1$ has no explicit $k$ dependence, but it is not universal.
For instance, the exponential regulator of Eq.~(\ref{cutoffs1})
would give
\begin{equation}
Q_n\left[ \frac{ R_k}{P_k^{n+1}} \right] =
\begin{cases} \log(2) & n=1 \\
\log \left( \frac{4}{3} \right) & n=2\\\frac{1}{2}\log \left( \frac{32}{27} \right) & n=3\end{cases} \ ,
\end{equation}
while the optimized regulator (\ref{opt})
leads to
\begin{equation}
 Q_n\left[ \frac{ R_k}{P_k^{n+1}} \right] = \frac{1}{\Gamma(n+2)} \ .
\end{equation}
This exemplifies the arbitrariness in the construction of an
\MSbar pseudo-regulator for calculations beyond the LPA.

If we straightforwardly apply the recipe 
\Eqref{eq:RproptoZ}, we obtain a divergent result:
\be
Q_n\left[ \frac{R_k}{(P_k+m^2)^\ell} \right] = 
\lim_{\epsilon\to0} 
\frac{(-m^2)^{n-\ell+1}}{\Gamma(\ell)\Gamma(n-\ell+2)} 
\left( 1-\frac{n}{n+1} \right) \frac{1}{\epsilon} \ .
\label{alicebb}
\ee
Therefore, including the wave function renormalization
in the pseudo-regulator requires some additional work. 
In the following we explore a family of pseudo-regulators which 
achieve the goal of reproducing one-loop \MSbar results,
plus RG resummations, in the $\epsilon\to0$ limit.

\subsection{An extended family of pseudo-regulators}

The first requirement on a new
pseudo-regulator which is appropriate for the LPA${}'$,
is that it reduces to the pseudo-regulator we have adopted for the LPA
in the $Z_k\to1$ limit.
\footnote{For simplicity we set $b=0$.}
Hence we consider a generalization 
of \Eqref{dimregcut} which amounts
to the introduction of two new parameters $Z_0>0$ and $\sigma$:
\be
R_k(z) = Z_0 Z_k^{\sigma \epsilon} \left[ \left( \frac{k^2}{\mu^4} \right)^\epsilon  z^{1+\epsilon} -z \right] \ .
\label{dimregblpa}
\ee
While the most common choice, as in \Eqref{eq:RproptoZ},
would be $Z_0=1$ and $\sigma=1/\epsilon$,
we prefer to keep the two variables arbitrary for the time
being.
We define
\bea
G_k(q^2) = \left(Z_k q^2 + V_k''+ R_k(q^2) \right)^{-1} 
\label{eq:genericG_k}
\eea
the regularized propagator.
With our pseudo-regulator this reads
\begin{equation}
G_k = \left( Z_0  \left( \frac{Z_k^\sigma k^2}{\mu^4} \right)^\epsilon  z^{1+\epsilon} -(Z_0 Z_k^{\sigma \epsilon}-Z_k)z + V_k'' \right)^{-1} ,
\end{equation}
from which it is manifest that having a vanishing
$(Z_0 Z_k^{\sigma \epsilon}-Z_k)$
would tremendously simplify the 
task of evaluating the loop integrals.
Though we restrain from this simplifying
assumption, we still assume that 
this difference is small.
We calculate all loop integrals
by means of their Taylor series in this
difference around zero.

Then the generic $Q$-functional becomes
\begin{align}
&Q_n\!\left[ G_k^\ell \partial_t R_k\right]\! =
\frac{\epsilon Z_0 Z_k^{\sigma \epsilon}}{\Gamma(n)\Gamma(\ell)} \sum_{p=0}^\infty (Z_0 Z_k^{\sigma \epsilon}-Z_k)^p \frac{\Gamma(\ell+p)}{\Gamma(p+1)}
\int_0^\infty \!\!\!dz \, z^{n+p} \frac{(2-\sigma\eta_k)\left( \frac{k^2}{\mu^4} \right)^\epsilon  z^{\epsilon} +\sigma\eta_k }{\left( Z_0  \left( \frac{Z_k^\sigma k^2}{\mu^4} \right)^\epsilon  z^{1+\epsilon}  + V_k'' \right)^{\ell+p}}
\nonumber
\end{align}
\begin{align}
&= \frac{\epsilon Z_0 Z_k^{\sigma \epsilon}}{(1+\epsilon)\Gamma(n)\Gamma(\ell)} \sum_{p=0}^\infty  \frac{(Z_0 Z_k^{\sigma \epsilon}-Z_k)^p}{\Gamma(p+1)}
\left(Z_0  \left( \frac{Z_k^\sigma k^2}{\mu^4} \right)^\epsilon\right)^{-\frac{n+p}{1+\epsilon}}
\nonumber\\
&\ \ \ \left[
(2-\sigma \eta_k)\Gamma\left( \frac{n+p}{1+\epsilon}+1 \right) \Gamma\left( \ell+p-\frac{n+p}{1+\epsilon}-1\right) \left(Z_0  \left( \frac{Z_k^\sigma k^2}{\mu^4} \right)^\epsilon\right)^{-1}\left(V_k''\right)^{\frac{n+p}{1+\epsilon}-\ell-p+1}
\right.
\nonumber\\
&\ \ \ \left. +\sigma \eta_k
\Gamma\left( \frac{n+p+1}{1+\epsilon} \right) \Gamma\left( \ell+p-\frac{n+p+1}{1+\epsilon}\right) \left(Z_0  \left( \frac{Z_k^\sigma k^2}{\mu^4} \right)^\epsilon\right)^{-\frac{1}{1+\epsilon}}\left(V_k''\right)^{\frac{n+p+1}{1+\epsilon}-\ell-p}
\right] \ .
\end{align}
From this expression it can be clearly seen that
$\sigma$ cannot diverge for vanishing $\epsilon$
(as a comparison of (\ref{eq:RproptoZ}) and (\ref{dimregblpa}) would suggest)
or both terms would also diverge.
On the other hand choosing a vanishing $\sigma$
in this limit would remove any
$Z_k$ and $\eta_k$ dependence,
thus reproducing the same results
of the LPA pseudo-regulator.
Finally, choosing $\sigma$ to stay constant
in the $\epsilon \to 0$ limit leads to
\begin{align}
&Q_n\!\left[ G_k^\ell \partial_t R_k\right]\! =
\frac{Z_0^{-n} \left(- V_k''\right)^{n-\ell+1}}{\Gamma(n)\Gamma(\ell)\Gamma(n-\ell+2)} \sum_{p=0}^\infty  \frac{\Gamma\left( n+p \right)}{\Gamma(p+1)} \left(1 -\frac{Z_k}{Z_0}\right)^p\!
\left[ (2-\sigma \eta_k) + \frac{\sigma \eta_k(n+p)}{n+p+1}\right]
\nonumber
\\
&=
\frac{Z_0^{-n} \left(- V_k''\right)^{n-\ell+1}}{\Gamma(\ell)\Gamma(n-\ell+2)} 
\left[ (2-\sigma \eta_k) \frac{Z_0^n}{Z_k^n} +\sigma \eta_k \frac{n}{n+1}{}_2F_1\!\left( 1+n,1+n,2+n; 1 -\frac{Z_k}{Z_0} \right)\right] \ .
\end{align}
Summarizing we have 
\begin{subequations}
\begin{align}
Q_n\!\left[ G_k^\ell \partial_t R_k\right] &=
\frac{Z_k^{-n} \left(-V_k'' \right)^{n-\ell+1}}{\Gamma(\ell)\Gamma(n-\ell+2)}\left( 2- \sigma \eta_k (1+H_0)\right) \ ,
\\
Q_n\!\left[ G_k^\ell G_k' \partial_t R_k\right] &=
-\frac{n}{n-1}\frac{ Z_k^{1-n} \left(-V_k''\right)^{n-\ell-1}}{\Gamma(\ell+2)\Gamma(n-\ell)}  
\left( 2- \sigma \eta_k (1+H_1) \right) \ ,
\\
Q_n\!\left[ G_k^\ell G_k'' \partial_t R_k\right] &=  
\frac{2n}{n-2}\frac{  Z_k^{2-n} \left(-V_k''\right)^{n-\ell-2}}{\Gamma(\ell+3)\Gamma(n-\ell-1)} 
\left( 2- \sigma \eta_k (1+H_2) \right) ,
\end{align}
\label{Qslpa'}
\end{subequations}
where primes denote differentiation with respect to $z$, and
we introduced the following notations:
\begin{subequations}
\begin{align}
H_0\!\left( n , Z_k , Z_0 \right) &=
-\frac{n }{n+1} \left( \frac{Z_k}{Z_0} \right)^{n} \,_2F_1\!\left( n+1,n+1,n+2;1- \frac{Z_k}{Z_0} \right),
\\
H_1\!\left( n, Z_k , Z_0 \right) &=
- \frac{n-1}{n} \frac{Z_0}{Z_k} +\frac{n-1}{n+1}\! \left( 1-\frac{Z_k}{Z_0} \right)\!\left( \frac{Z_k}{Z_0} \right)^{n-1}\!\,_2F_1\!\left( n+1,n+1,n+2;1-\frac{Z_k}{Z_0} \right),
\\
H_2\!\left( n, Z_k , Z_0 \right) &=
\frac{n-2}{n}\left( 1-\frac{2n-1}{n-1} \frac{Z_k}{Z_0} \right)\!\left(  \frac{Z_0}{Z_k} \right)^2\nonumber\\
&{}-\frac{n-2}{n+1} \left(1- \frac{Z_k}{Z_0}\right)^2\! \left( \frac{Z_k}{Z_0}\right)^{n-2}\,_2F_1\!\left( n+1,n+1,n+2;1- \frac{Z_k}{Z_0} \right).
\end{align}
\end{subequations}
The dependence of the $H_i$ functions on $Z_k/Z_0$
signals the expected breaking of reparameterization invariance,
which translates in non-autonomous flow equations 
for the dimensionless renormalized couplings.
An autonomous flow can be recovered in special
cases: besides the  $\sigma\to0$ limit,
other interesting choices are
\begin{align}
\lim_{Z_0 \to 0}H_i(n)&=0 \ ,\\
\lim_{Z_0 \to \infty}H_i(n)&=-1 \ . 
\end{align}
From \eqref{Qslpa'} we see that 
the second case suppresses the
``RG improvement'' terms, in the same way as setting $\sigma=0$,
and therefore gives back the LPA result.
Thus, in summary, the proper way to use the
pseudo-regulator (\ref{dimregblpa}) is to first evaluate the integrals,
then take the limit $\epsilon\to 0$
and finally the limit $Z_0\to 0$.

Finally, it is worth stressing that the previous identities are 
not restricted to the LPA${}'$ truncation.
For truncations where $Z_k$ depends on fields and/or momentum,
the relevant wave-function renormalization factor appearing 
inside the pseudo-regulator is to be identified with
$Z_k$ evaluated at preferred values of momentum and fields,
for instance minimizing the potential and the inverse propagator.
In the simplest cases the latter  are vanishing values.
Then the simple propagator $G_k$ of 
\Eqref{eq:genericG_k}, and the loop integrals given in the previous equations, would arise after a polynomial expansion
of $Z_k$ to obtain derivative vertices which are
local in field space and in spacetime.

\subsection{Scalar field with a generic potential and its anomalous dimension}
\label{sec:LPAprime}

As a first example of application of the above pseudo-regulator,
let us turn to a simple scalar field theory within the LPA${}'$ truncation. 
We consider a most generic effective potential,
which can be parameterized as follows
\begin{equation}
V_k = \sum_{n=0} Z_k^{n/2}\frac{\lambda_{n}}{n!}\phi^{n} \ .
\end{equation}
The anomalous dimension $\eta_k$ is computed
by extracting from the exact FRG equation the contributions 
to the quadratic part of the
two point function
\begin{align}
\eta_k & = -\frac{1}{Z_k} \lim_{p^2 \to 0}\frac{\partial}{\partial p^2}\partial_t\left. 
\frac{\delta^2 \Gamma_k}{\delta \phi_p \phi_{-p}}\right|_{\phi_0}
\nonumber\\
&= -\frac{\left(V_k^{(3)}(\phi_0)\right)^2}{Z_k}\left. \left(\lim_{p^2 \to 0}\frac{\partial}{\partial p^2}\int \frac{d^d q}{(2\pi)^d} G_k(q^2)^2 G_k((q+p)^2)\partial_t  R_k(q^2) \right)\right|_{\phi_0}\ .
\end{align}
Here $\phi_0$ is the minimum of the potential.
After taking the derivative and the limit, we obtain
\be
\eta_k = -\frac{\left(V_k^{(3)}(\phi_0)\right)^2}{Z_k(4\pi)^{d/2}}\left( 
Q_{\frac{d}{2}}\left[ G_k^2 G_k'\partial_t  R_k\right]
+Q_{\frac{d}{2}+1}\left[ G_k^2 G_k'' \partial_t  R_k\right]
\right) \ .
\ee
For $\phi_0=0$ and in the $Z_0\to0$ limit, this boils down to
\be
\eta_k = -\frac{(-)^{d/2}}{6\Gamma\left(\frac{d}{2}-2\right)} \frac{\lambda_3^2}{(4 \pi)^{d/2} } \lambda_2^{\frac{d}{2}-3} \left( 1-\sigma\frac{\eta_k}{2} \right) \ .
\ee
Within the LPA$'$ truncation for this pseudo-regulator we obtain
what is essentially a one-loop equation with RG improvement.
Therefore in $d=2,4$ 
we find $\eta=0$.
\footnote{Note that we drop the $k$-subscript in $\eta_k$ from this point on.}
To reproduce a non-vanishing $\eta$ with this pseudo-regulator we need
to consider larger truncations, as detailed in Section~\ref{sec:twoloops}.\\

Now we consider the equation for the potential and we calculate the beta functions
\begin{align}
\partial_t V_k &= \frac{1}{2(4\pi)^d} Q_{\frac{d}{2}}\left[
G_k \partial_t\! \left( Z_k R_k \right)\right] 
 \ ,\\
\beta_n &= \frac{n}{2}\eta \lambda_n + Z_k^{-n/2}\frac{\partial^n \partial_t V_k}{(\partial \phi)^n} \Big|_{\phi=0} \ .
\end{align}
We list the initial beta functions
in the $Z_0\to0$ limit:
\begin{subequations}
\begin{align}
\beta_1 &= \frac{1}{2}\eta \lambda_1+ 
\frac{(-)^{d/2}}{\Gamma\left(\frac{d}{2}\right)(4\pi)^{d/2}}
\left(1-\sigma\frac{\eta}{2}\right)
\lambda_3 
\lambda_2^{d/2-1} \ ,
\\
\beta_2 &= \eta \lambda_2 +
\frac{(-)^{d/2}}{(4\pi)^{d/2}} \left(1-\sigma\frac{\eta}{2}\right)
\left(
\lambda_3^2\frac{ \lambda_2^{d/2-2}}{\Gamma\left(\frac{d}{2}-1\right)} +
\lambda_4 \frac{\lambda_2^{d/2-1}}{\Gamma\left(\frac{d}{2}\right)} 
\right) \ ,
\\
\beta_3 &= \frac{3}{2}\eta \lambda_3 +
\frac{(-)^{d/2}}{(4\pi)^{d/2}} \left(1-\sigma\frac{\eta}{2}\right)
\left(
\lambda_3^3 \frac{\lambda_2^{d/2-3}}{\Gamma\left(\frac{d}{2}-2\right)} 
+3\lambda_3 \lambda_4 \frac{
\lambda_2^{d/2-2}}{\Gamma\left(\frac{d}{2}-1\right)} 
+\lambda_5 \frac{\lambda_2^{d/2-1}}{\Gamma\left(\frac{d}{2}\right)}
\right) \ ,
\\
\beta_4 &= 2\eta \lambda_4 +  
\frac{(-)^{d/2}}{(4\pi)^{d/2}} \left(1-\sigma\frac{\eta}{2}\right)
\times\nonumber\\
&\ \ \
\left(
\lambda_3^4 \frac{\lambda_2^{d/2-4}}{\Gamma\left(\frac{d}{2}-3\right)} 
+6\lambda_3^2 \lambda_4 \frac{\lambda_2^{d/2-3}}{\Gamma\left(\frac{d}{2}-2\right)} 
+\left( 4\lambda_3 \lambda_5 +  3 \lambda_4^2 \right) \frac{\lambda_2^{d/2-2}}{\Gamma\left(\frac{d}{2}-1\right)} 
+\lambda_6 \frac{\lambda_2^{d/2-1}}{\Gamma\left(\frac{d}{2}\right)}
\right) \ .
\end{align}
\end{subequations}
In $d=6$,  
 if $\lambda_2 \to 0$ and $\lambda_{n \ge 4}=0$,  we get the universal one-loop result, plus RG resummations
\begin{align}
\eta &=   \frac{\lambda_3^2}{6(4\pi)^3} \left(1-\sigma\frac{\eta}{2}\right) \ ,
\\
\beta_3 &=  -  \frac{3\lambda_3 ^3 }{4(4\pi)^3} 
\left(1-\sigma\frac{\eta}{2}\right) \ .
\end{align}
Solving these equations and expanding in $\lambda_3$
\begin{align}
\eta &=   \frac{\lambda_3^2}{6(4\pi)^3+\frac{\sigma}{2}\lambda_3^2} 
=\frac{\lambda_3^2}{6(4\pi)^3}-\frac{\sigma}{2}\frac{\lambda_3^4}{36(4\pi)^6}+O(\lambda_3^6) \ ,
\end{align}
it can be checked that it is possible to adjust $\sigma$
to reproduce the correct two-loop result~\cite{Gracey:2015tta}
for either
$\eta$ ($\sigma=-26/3$) 
or $\beta_3$ ($\sigma=-250/9$),
but not both simultaneously
\footnote{The situation does not change if we insert also the RG improvement proportional to $\beta_2/\lambda_2$ , which is finite in the limit $\lambda_2 \to 0$ .}.
In fact, recovering the full
two-loop RG equations requires larger
truncations, as we discuss in Section~\ref{sec:twoloops} .

\section{The \texorpdfstring{$\boldsymbol{O(\partial^2)}$}{TEXT} derivative expansion}
\label{sec:DE}

The pseudo-regulators introduced in the previous section are also
apt for application to a larger class of truncations
which  accounts for a possible field dependence of the
wave function renormalization, the $O(\partial^2)$ of
the derivative expansion.
While this kind of more elaborate approximation is often 
an optional for many models, it is in some cases a necessity
already as a zeroth order approach, such as for instance
in the applications to nonlinear sigma models
or for conformal field theories in two dimensions.
For this reason, in this section we address these
two examples.
They allow us to account for a trivial generalization of the LPA${}^\prime$
formulas given in the previous section, and also to 
discuss more subtle points about the 
scope of an \MSbar pseudo-regulator, such as
its applicability  to
strongly interacting field theories
and to models 
with nonlinear symmetries.

\subsection{Multicritical models}
\label{sec:multicritical}
\label{sec:DEmulti}

We consider the following truncation of $\Gamma_k$
\be
\Gamma_k[\phi] =  \int dx \left( V_k(\phi) 
+ \frac{1}{2}Z_k(\phi) \partial_\mu \phi \partial^\mu \phi \right) \ .
\label{eq:DEsinglefield}
\ee
This kind of ansatz is general enough
to capture the emergence of a tower of
multicritical $\phi^{2p}$ scalar field theories below the fractional
upper critical dimensions $d_p=2p/(p-1)$,
and to provide good estimates of their properties
in $d=2$~\cite{Morris:1994jc,Defenu:2017dec,Defenu:2020cji}.
As these conclusions apply to conventional 
FRG regulator choices, it is interesting to check
whether these nice results can be obtained
even with an \MSbar pseudo-regulator.

The flow equations of the functions $V_k$ and $Z_k$
for the pseudo-regulator (\ref{dimregblpa})
can be obtained from those presented in Appendix~\ref{app:floweqDE}
for the more general case of $O(N)$ models.
More specifically, they correspond to 
Eqs.~(\ref{eq:DEV}) and (\ref{eq:DEZ}),
for $d=2$ and $N=1$
\footnote{Here and in the following sections 
we identify the wave function 
renormalization of the single-field theory
of \Eqref{eq:DEsinglefield} with the one of the radial mode
in the linear $O(N)$ model, which is $\tilde{Z}_k(\phi)$
in \Eqref{eq:DEZ}.
Notice that this choice is not the conventional
one for FRG studies, which usually 
associate the single-field $Z_k$
to the $N\to1$ limit
of the Goldstone-modes wave function renormalization.}.
Rescaling the field 
\begin{equation}
\phi = Z_k(0)^{-1/2} \tilde \phi\ ,\quad
\eta=-\partial_t\log Z_k(0)
\end{equation}
and introducing
the dimensionless renormalized
functions
\begin{equation}
v_k(\tilde\phi) = k^{-2} V_k(\phi) \ ,\quad
\zeta_k(\tilde\phi)=Z_k(0)^{-1}Z_k(\phi) \ ,
\label{eq:dimlessvzeta}
\end{equation}
these flow equations read
\begin{subequations}
\begin{align}
\partial_t v_k =& -2 v_k +\frac{\eta}{2} \tilde \phi v'_k
-\frac{1}{4\pi}\left( 1-\sigma \frac{\eta}{2} \right) \zeta_k^{-1} v''_k \ ,
\\
\partial_t  \zeta_k  =& \eta \zeta_k +\frac{\eta}{2} \tilde \phi \zeta'_k
+\frac{1}{8\pi}\left( 1-\sigma \frac{\eta}{2} \right) 
\left[-2\frac{ \zeta_k''}{ \zeta_k}
+3  \left(\frac{  \zeta_k'}{ \zeta_k}\right)^2 \right] \ .
\end{align}
\label{eq:RGflow2D}
\end{subequations}

\begin{table}
\centering
\begin{tabular}{ |c|c|c|c|c| } 
 \hline
  & this work & optimized \cite{Defenu:2017dec,Defenu:2020cji} 
  & Morris \cite{Morris:1994jc} & exact \cite{Zamolodchikov:1986db}\\ 
 \hline 
$\eta_2$  & $0.25$ & $0.2132$ & $0.309$ & $0.25$  \\ 
$\nu_2$  & $0.666667$ & $\dots$ & $0.863$ & $1$  \\ 
\hline
$\eta_3$  & $0.111111$ & $0.1310$ & $0.200$ & $0.15$ \\ 
$\nu_3$  & $0.5625$ & $\dots$ & $0.566$ & $0.556$  \\ 
\hline
$\eta_4$  & $0.0625$ & $0.0910$ & $0.131$ & $0.1$ \\ 
$\nu_4$  & $0.533333$ & $\dots$ & $0.545$ & $0.536$  \\ 
\hline
$\eta_5$  & $0.04$ & $0.0679$ & $0.0920$ & $0.0714$ \\ 
$\nu_5$  & $0.520833$ & $\dots$ & $0.531$ & $0.525$  \\ 
\hline
$\eta_6$  & $0.0277778$ & $0.0522$ & $0.0679$ & $0.0535714$ \\ 
$\nu_6$  & $0.514286$ & $\dots$ & $0.523$ & $0.519$  \\ 
\hline
$\eta_7$  & $0.0204082$ & $\dots$ & $0.0521$ & $0.0416667$ \\ 
$\nu_7$  & $0.510417$ & $\dots$ & $0.517$ & $0.514$  \\ 
\hline
$\eta_8$  & $0.015625$ & $\dots$ & $0.0412$ & $0.0333333$ \\ 
$\nu_8$  & $0.507937$ & $\dots$ & $0.514$ & $0.511$  \\ 
\hline
$\eta_9$  & $0.0123457$ & $\dots$ & $0.0334$ & $0.0272727$ \\ 
$\nu_9$  & $0.50625$ & $\dots$ & $0.511$ & $0.509$  \\ 
\hline
$\eta_{10}$  & $0.01$ & $\dots$ & $0.0277$ & $0.0227273$ \\ 
$\nu_{10}$  & $0.505051$ & $\dots$ & $0.509$ & $0.508$  \\ 
\hline
$\eta_{11}$  & $0.00826446$ & $\dots$ & $0.0233$ & $0.0192308$ \\ 
$\nu_{11}$  & $0.504167$ & $\dots$ & $0.508$ & $0.506$  \\ 
\hline
\end{tabular}
\caption{Estimates of the critical exponents $\eta_p$ and $\nu_p$
for the two dimensional $\phi^{2p}$
multicritical scalar models.
The first three columns present FRG estimates:
the first obtained
with the \MSbar pseudo-regulator, the second with the
optimized regulator of \Eqref{opt}, the third with
an homogeneous regulator. Finally, the last
column shows the exact results, from CFT methods.
\label{tab:multicritical}}
\end{table}

Now we search for the scaling solutions for this system of equations.\\
Setting the $\mathbb{Z}_2$-parity and normalizations conditions
\begin{subequations}
\begin{align}
v'(0)&=0 \ ,&& &&v''(0)=\zeta_0\tilde m^2 \ ,&&  \\
\zeta(0)&=\zeta_0 \ ,&& &&\zeta'(0)=0 \ ,&& 
\end{align}
\end{subequations}
the previous system of equations has the following family of 
fixed points
\begin{subequations}
\begin{align}
v_* &= - \frac{2-\sigma \eta}{16\pi} \tilde m^2
\cos \left[ \frac{2}{\sqrt{\eta}} 
\arctan \left( \sqrt{\frac{  \Phi^2 }{1- \Phi^2 } }\right) \right] \ ,
\\
\zeta_*  &= \zeta_0\left( 1- \Phi^2\right)^{-1} \ ,\\
\Phi&=\sqrt{\frac{4\pi\eta\zeta_0}{2-\sigma\eta}}\,\tilde\phi \ .
\end{align}
\end{subequations}
It is remarkable that with the \MSbar pseudo-regulator
the scaling solutions can be written in closed form:
usually in the FRGE they are only known numerically.
These solutions manifestly preserve reparameterization invariance:
the normalization factor $\zeta_0$ cannot
influence observable quantities as it can be
eliminated from the action (\ref{eq:DEsinglefield})
by rescaling $\tilde \phi$.
Depending on the sign of $m^2$, the fixed-point potential
can have a maximum or a minimum for zero field.\\
Note that $\zeta_*$ diverges for $\Phi^2 = 1 $.
In order to have a potential $v_*$ which is smooth at this 
point and is bounded from below we impose
\begin{subequations}
\begin{align}
\left. v^{(n)}_*\right|_{\Phi^2\to1} &= \mbox{finite} \ , \quad \forall n
\label{eq:vsmooth}
\\
\lim_{\tilde \phi \to \infty} v_*  & = +\infty \ .
\label{eq:vbounded}
\end{align}
\end{subequations}
From (\ref{eq:vsmooth}) we get the quantization rule: 
\be
\sin \frac{\pi}{\sqrt{\eta}}=0 \ ,
\ee
such that
\be
\eta= \frac{1}{p^2} \ , \quad p=1,2,3 \dots \ ,
\label{eq:etamulticrit}
\ee
while (\ref{eq:vbounded}) can be fulfilled by adjusting 
the sign of $\tilde m^2$ (while the modulus remains free):
\be
\tilde m^2\lessgtr0 \quad  \mbox{if}\quad  (-1)^{1/\sqrt{\eta}}=\pm1 \ .
\ee
In this way $v_*$ acquires the typical shape of
a $(p-1)$-critical potential.

To compute the critical exponents associated to
these fixed points, we linearize the RG
flow around them and look for eigenperturbations.
In other words, we insert  $v_k \to v_* + e^{\theta t} \,\delta v$, $\zeta_k \to \zeta_* + e^{\theta t}\, \delta \zeta$ 
and $\eta=\eta_*+\delta \eta$ 
into  \Eqref{eq:RGflow2D}, and expand them to first order
in the perturbations $\delta v$, $\delta \zeta$ and $\delta \eta$.
For $\delta \eta\neq0$ the corresponding $\delta\zeta$ is complex
and furthermore singular at $\Phi^2=1$. We therefore
impose $\delta\eta=0$.
In this simplified case the linearized equations read
\begin{subequations}
\begin{align}
\theta \, \delta v =& -2\, \delta v +\frac{\eta}{2} \tilde \phi\,\delta v'
-\frac{1}{4\pi}\left( 1-\sigma \frac{\eta}{2} \right) \zeta_*^{-1}
\left[ \delta v'' -v_*''\, \frac{ \delta \zeta}{\zeta_*}  \right] \ ,
\\
\theta \, \delta \zeta  =& \eta \, \delta \zeta +\frac{\eta}{2} \tilde \phi \, \delta \zeta'
+\frac{1}{4\pi}\left( 1-\sigma \frac{\eta}{2} \right) 
\left[ -\frac{\delta \zeta''}{\zeta_*}+3\frac{\zeta_*'\, \delta \zeta'}{\zeta_*^2}+
\left(\frac{\zeta_*'' }{\zeta_*}- 3\left(\frac{\zeta_*'}{\zeta_*}\right)^2 \right)
\frac{\delta \zeta}{\zeta_*}
   \right] \ .
\end{align}
\label{eq:eigenmulticrit2d}
\end{subequations}
The condition of fixed $\eta$
results in LPA-like perturbations
with vanishing $\delta\zeta$.\\
Besides the trivial solutions
\begin{subequations}
\begin{align}
 \theta &= -2 \ , && &&\delta v = 1 \ , && \hspace{1cm} 
\\
 \theta &= -2 + \frac{\eta}{2} \ ,&& \hspace{1cm} 
 &&\delta v = \tilde \phi \ ,  &&
\\
\theta &= 0 \ ,&& \hspace{1cm} 
&&\delta v = v_* \ ,&&
\end{align}
\end{subequations}
we find the even and odd eigenperturbations respectively
\begin{subequations}
\begin{align}
\theta &=-2 +2\eta\,n^2 \ , 
&&\delta v  =  \cos \left[ \sqrt{\frac{4+2 \theta}{\eta}} 
\arctan \left( \sqrt{\frac{  \Phi^2 }{1- \Phi^2 } } \right) \right] \ ,&&
\\
\theta &=-2+2\eta \left(\! n+\frac{1}{2} \right)^2\!\!, 
&&\delta v  =\sqrt{\frac{2-\sigma \eta}{8\pi (2+\theta)}} \sin \left[ \sqrt{\frac{4+2 \theta}{\eta}} 
\arctan \left( \sqrt{\frac{  \Phi^2 }{1- \Phi^2 } }\right) \right] \ ,&&
\end{align}
\end{subequations}
where $\eta$ assumes its fixed-point value (\ref{eq:etamulticrit}).
Enforcing regularity of $\delta v$ at the pole of $\zeta$
requires $n=1,2,3 \dots$. From the largest even parity eigenvalue ($n=1$), excluding the unit operator, we get the critical exponent $\nu$
\be
\nu = \frac{1}{2-2 \eta} \ .
\ee
In Table \ref{tab:multicritical} 
we summarize these results
for the critical exponents $\eta$ and $\nu$
and compare them to FRG estimates obtained
by means of the optimized regulator
and with the homogeneous regulator,
as well as with the exact values.
Comparing \Eqref{eq:etamulticrit} with the
exact result
\be
\eta=\frac{3}{(p+1)(p+2)}
\ee
we see that for large $p$ our result is off by a factor $3$,
whereas $\nu$ correctly tends to $1/2$.

\subsection{The nonlinear \texorpdfstring{$\sigma$}{TEXT} model}
\label{sec:nonlineaON}

Addressing the nonlinear $\sigma$ model with the
\MSbar pseudo-regulator requires only a simple generalization
of the truncation we just studied, to account
for a multiplet of fields, rather than a single one.
We therefore start from the following truncation of $\Gamma_k$
for a $O(N)$-invariant multiplet of scalars:
\be
\Gamma_k[\phi] =  \int dx \left( U_k(\rho) 
+ \frac{1}{2}Z_k(\rho) \partial_\mu \phi_a \partial^\mu \phi^a + 
\frac{1}{4}Y_k(\rho) \partial_\mu \rho \partial^\mu \rho \right) \ ,
\label{eq:GammaDE}
\ee
where the $N$ fields $\phi^a$ are in
the fundamental representation of $O(N)$,
and $\rho=\phi^a\phi^a/2$ is the corresponding local invariant.
We further define the radial wave function renormalization
\be
\tilde Z_k \left( \rho \right) = Z_k\left( \rho \right)+\rho Y_k\left( \rho \right) \ .
\label{eq:tildeZ}
\ee
In Appendix~\ref{app:floweqDE} we show the flow equations
of this model in the present truncation,
for general $d$ and adopting the pseudo-regulator
of \Eqref{dimregblpa}
in the $Z_0\to0$ limit.
For the especially interesting case $d=2$,
we obtain
\begin{subequations}
\begin{align}
\partial_t U_k =&
-\frac{1}{4\pi} \left( \tilde Z_k^{-1}\left(U'_k+2 \rho U''_k\right)
+ (N-1)Z_k^{-1} U'_k \right) \ ,
\\
\partial_t  \tilde Z_k  =&
- \frac{ \left(\tilde Z_k'+2\rho \tilde Z_k''\right)}{4\pi \tilde Z_k}
-(N-1) \frac{\left(  Z_k'+\rho Y_k'\right)}{4\pi Z_k} 
\nonumber\\
&
+\frac{ 3\rho  \left( \tilde Z_k'\right)^2}{4\pi \tilde Z_k^2} 
+ (N-1)\frac{ \rho  Z'_k \left( Y_k- Z_k'\right)}{2\pi Z_k^{2}} \ .
\end{align}
\label{eq:1loopDE}
\end{subequations}
Here we suppressed the RG improvement by
setting $\sigma=0$; the effect of a non-vanishing $\sigma$
will be addressed in a moment.


As it stands, this action could still describe a linear model.
If we make the assumptions
\begin{subequations}
\begin{align}
Z_k(\rho) &= \frac{Z_k}{g_k^2} \ ,
\\
\tilde Z_k (\rho) &=\frac{1}{g_k^2} \left( \frac{1}{Z_k}-2 \rho \right)^{-1},
\\
U_k &= - h_k \sqrt{\frac{1}{Z_k}-2 \rho } \ ,
\end{align}
\label{eq:nlsmZV}
\end{subequations}
the EAA becomes 
\be
\Gamma_k[\phi] =  \int dx 
\left[
\frac{Z_k}{2g_k^2}\left(\delta_{ab}
+\frac{\phi^a\phi^b}{\frac{1}{Z_k}-2\rho}
\right)
\partial_\mu \phi_a \partial^\mu \phi^b
- h_k \sqrt{\frac{1}{Z_k}-2 \rho }\right]
\label{eq:GammaNLSM}
\ee
which describes a nonlinear $\sigma$ model
with values in a sphere $S^N$ of radius $Z_k^{-1/2}$
and coupled to an external source $h_k$ \cite{zinnjustin}.
In this case the symmetry group is extended to $O(N+1)$.
Inserting this ansatz in the flow equations (\ref{eq:1loopDE})
one deduces the correct one-loop beta functions
\begin{subequations}
 \begin{align}
\partial_t g_k &= -\frac{N-1}{4\pi} g_k^3 \ ,
\\
\eta &= -\partial_t \log Z_k = \frac{N}{2\pi} g_k^2 \ ,
\\
\partial_t h_k &= 0 \ .
\end{align}
\end{subequations}
Thus, the flow equations (\ref{eq:1loopDE})
mantain the form of the ansatz in \Eqref{eq:nlsmZV},
that is to say, they preserve the nonlinearly realized
$O(N+1)/O(N)$ symmetry.
This might appear trivial as we are applying one-loop
RG equations, but it is not so for two reasons.
First, this compatibility extends beyond one-loop order
as we observe  in the following by the inclusion of the RG improvement.
Second, because this conclusion does not hold for finite $\epsilon$,
i.e.~within the realm of ordinary FRG computations.
In fact, it is  well known that the FRG regulator,
being a deformation of the two point function of the $N$ fields,
explicitly breaks the nonlinear part of the $O(N+1)$ symmetry.
For this reason, most FRG applications to nonlinear sigma models
have adopted different formulations based on the
background field method~\cite{Codello:2008qq,Percacci:2009fh,Fabbrichesi:2010xy,Bazzocchi:2011vr,Flore:2012ma,Flore:2012wh,Percacci:2013jpa,Safari:2014hqa,Codello:2015oqa}.

Let's then turn to the RG improvement,
which leads us beyond the one-loop approximation.
To this end we should note that the pseudo-regulator in \Eqref{dimregblpa}
has a factor $Z_k^{\sigma \epsilon}$, 
but here $Z_k$ should be replaced by
$Z_k g_k^{-2 }$ to be compatible with the ansatz
of \Eqref{eq:nlsmZV}.
Then, with this little adjustment of the pseudo-regulator,
for a generic $\sigma$ we get
\begin{subequations}
\begin{align}
\partial_t g_k &= -\frac{(N-1)g_k^3}{4\pi+\sigma g_k^2} \ ,
\\
\eta &= \frac{2N g_k^2}{4\pi+\sigma g_k^2} \ ,
\\
\partial_t h_k &= 0 \ .
\end{align}
\end{subequations}
Even though the previous flow equations hold in $d=2$,
it is possible to apply them in $d=2+\varepsilon$  by simply augmenting them with their $\varepsilon$-dependent canonical dimensional part.
In so doing, one can recover 
the $\varepsilon$-expansion description 
of the nontrivial
fixed point which exists for $\varepsilon>0$.
We defer this discussion to the end of Section \ref{sec:epsilonexp}.

\section{The two-loop beta functions}
\label{sec:twoloops}

In the previous sections we have
shown that the $\epsilon\to 0$ limit 
of the FRG beta functions for the \MSbar pseudo-regulator 
reduces them to well-known \MSbar one-loop RG equations,
possibly up to a resummation.
We have shown this in the LPA, in the LPA${}^\prime$
and in the $O(\partial^2)$ derivative expansion.
In this section we show how to reproduce the two-loop result
in four dimensions,
by considering larger truncations and by
taking the $\epsilon\to 0$ limit in a suitable way.

Although the computation  of the beta function of the quartic coupling
was discussed by several authors
already, see Refs.~\cite{Papenbrock:1994kf,Bonini:1996bk,Bonanno:1997dj,Pernici:1998tp,Kopietz:2000bh,Zappala:2002nx,Arnone:2003pa,Codello:2013bra}, part of the arguments
adopted in those works do not apply to the
 \MSbar pseudo-regulator,
which is not an IR regulator.
Furthermore, we crucially
rely on analytic continuation of divergent integrals,
such that parametric limits are allowed to not commute,
whereas standard FRG regulators 
render all integrals convergent.
In addition, we also compute the two-loop
running of the mass.

We closely follow the notations and the arguments of the first FRG work addressing this task, namely Ref.~\cite{Papenbrock:1994kf}.
We therefore focus on the linear $O(N)$ models with bare action
\begin{subequations}
\begin{align}
S[\phi]&=\!\int\!\dd^d x\left\{ \frac{1}{2}\partial_\mu\phi^a\partial^\mu\phi^a
+U_\Lambda(\rho)\right\} \ ,\\
U_\Lambda(\rho)&=\bar{m}^2\rho+\frac{\bar{\lambda}}{2}\rho^2 \ .
\end{align}
\label{eq:bareaction}
\end{subequations}
Note that compared to \Eqref{potexp}, we have changed the
notation to $\lambda_2=\bar m^2$ and $\lambda_4=3\bar\lambda$,
and the bars denote bare couplings.
In a massless scheme such as \MSbar,
the two-loop beta function of the quartic coupling is
universal and mass independent, such that it is usually
possible to assume $\bar{m}^2=0$ right from the start.
We instead focus on a massive theory in the symmetric regime
for technical reasons. In fact,  we are going to 
adopt an FRG pseudo-regulator which does not regulate IR
divergences.
This does not prevent us from analysing the massless theory though,
as we are allowed to take the $\bar{m}^2\to0$ limit
of any IR safe quantity after the loop integrals are computed.

In our regularization scheme, 
it is furthermore essential to account for the $k$-dependence of
the renormalized mass parameter ${m}^2$,
or else the correct two-loop beta function would not be reproduced.
In fact, as the latter contributes to the running of 
$\lambda$ in any mass-dependent scheme, it does so
also in our computations at non-vanishing $\epsilon$.
Interestingly, this contribution will survive
the $\epsilon\to 0$ limit, 
if the latter is taken carefully enough.
This computation will thus serve as an example
of a more general mechanism,
according to which the super-renormalizable
and the non-renormalizable sectors of a theory,
which show a nontrivial running in any mass-dependent
scheme, do feed back into the running
of renormalizable operators even in a 
massless scheme such as \MSbar,
provided the mass-thresholds effects
are correctly accounted for~\footnote{This 
mechanism has been observed also in
Ref.~\cite{Gies:2019nij}}.

As discussed in the introduction,
a two-loop result involves arbitrarily-high orders of
the derivative expansion.
We therefore cannot use an ansatz such as (\ref{derexp}),
or its multi-field generalization.
Instead, we must make the ansatz
\be
\Gamma_k[\phi] =
\int \dd^dx \left\{ U_k(\rho) +
\frac{1}{2} \partial_\mu \phi^a \,Z_k\left( \rho, -\partial^2 \right) \partial^\mu \phi_a
+\frac{1}{4} \partial_\mu \rho\, Y_k\left( \rho, -\partial^2 \right) \partial^\mu \rho \right\} \ .
\ee
Equivalently, expressing the field in the angular (Goldstone) modes,
and in the radial (massive), mode $\sqrt\rho$,
one finds that their wave function renormalizations
are given by the function $Z_k$ and
\be
\tilde Z_k \left( \rho, q^2 \right) = Z_k\left( \rho, q^2 \right)+\rho Y_k\left( \rho, q^2 \right) \ ,
\ee
respectively.
In order to compute the beta function of the
quartic coupling it may seem sufficient to stop at the fourth
order of the vertex expansion, and hence
assume that $U_k$ is quadratic in $\rho$ and $Z_k$ is linear.
However, in the vertex expansion,
the beta function of $\Gamma^{(n)}$ involves also
$\Gamma^{(n+1)}$ and $\Gamma^{(n+2)}$,
so we need $U_k$ up to order $\rho^3$
and $Z_k$ up to order $\rho^2$.
In general we shall use the following terminology for the
expansion of these functions:
\begin{subequations}
\begin{align}
U_k(\rho)&=\sum_{n=1}^\infty\frac{u_n}{n!}Z_k^nk^{d-n(d-2)}(\rho-\rho_0)^n\ ,\\
Z_k(\rho,q^2)&=\sum_{n=0}^\infty\frac{z_n(q^2/k^2)}{n!}Z_k^{n+1}k^{-n(d-2)}(\rho-\rho_0)^n\ ,\\
\tilde Z_k(\rho,q^2)&=\sum_{n=0}^\infty\frac{\tilde z_n(q^2/k^2)}{n!}Z_k^{n+1}k^{-n(d-2)}(\rho-\rho_0)^n\ ,
\end{align}
\end{subequations}
where $Z_k = Z_k(\rho_0,0)$ is the wave function renormalization
and $\rho_0$ is the minimum of the potential.
In any scheme, the effective potential is already renormalized at one loop. 
The functions $Z_k(\rho,q^2)$ and $\tilde Z_k(\rho,q^2)$
also receive one-loop radiative corrections
in mass-dependent schemes.
The one-loop contributions to these functions
are however field dependent corrections,
and therefore correspond to radiatively generated
momentum-dependent vertices.
In any scheme the field-independent part of the wave function
renormalizations, i.e.~the fields anomalous dimensions,
receive corrections from the two-loop order on.  
These well known perturbative facts are
recovered from the FRG equations, straightforwardly
in mass-dependent schemes, and with a little care
also for mass-independent schemes, as we show in this section.

Although the FRGE looks like a one-loop equation,
this is only true as long as one uses the full propagators
and vertices (double lines and black dots in Eqs.~(\ref{eqs:vertexexp})).
The full propagators and vertices can be expanded in loops,
as briefly explained in the introduction,
giving rise to infinite series
that can be represented in terms of standard Feynman diagrams. 
This introduces resummations of perturbative diagrams
of two kinds.
The first is the so called ``spectral adjustment''
of the regulator, i.e.~the possible dependence of the regulator
on the couplings of the theory, 
most commonly the wave function renormalization $Z_k$
(as already discussed in Section \ref{sec:rolewavefun}).
This produces terms depending on the field's 
anomalous dimensions.
The second source of resummations is provided by
the mass thresholds, which in a functional setup
may also depend on 
the point of expansion
in the space of field amplitudes,
\be
\kappa = k^{2-d} Z_k \rho_0 \ ,
\ee
at which we define local couplings.
These include the perturbatively
renormalizable ones
\begin{subequations}
\begin{align}
  m^2\equiv k^2 u_1 = Z_k^{-1} U'(\rho_0) \ ,\quad
 \lambda\equiv u_2 = k^{d-4} Z_k^{-2} U''_k(\rho_0) \ ,
\end{align}
as well as the non-renormalizable ones
\begin{align}
u_n &= k^{n(d-2)-d} Z_k^{-n} U^{(n)}_k(\rho_0) \ ,
\quad \quad \quad \,\, n\geq3 \ ,\\
z_n\!\left( y \right) &= k^{n(d-2)} Z_k^{-(n+1)}  Z^{(n)}_k(\rho_0,k^2 y)\ ,\quad n\geq1 \ ,\\
\tilde  z_n \!\left( y \right) 
&= k^{n(d-2)} Z_k^{-(n+1)}  \tilde Z^{(n)}_k(\rho_0,k^2 y)\ ,\quad n\geq 1 \ .
\end{align}
\label{eqs:defsu_n&z_n}
\end{subequations}
If one were to suppress both these portals towards higher order
corrections, the FRGE would boil down to
a pure one-loop result.
However, thanks to these two contributions,
higher loops are generated by solving the RG
equations and constructing the RG trajectory,
i.e.~in the process of renormalizing the theory.

Let us now address the task of integrating the $d=4$ flow
equations from the UV initial condition $\Gamma_{k=\Lambda}=S$
down to $k<\Lambda$~
\footnote{ Here the limit $\Lambda\to\infty$ is allowed
as part of the regularization choice, and should not be
confused with the possibility to remove a UV cutoff, thus
defining a UV complete theory. The latter question is instead
emerging when trying to take such a limit at for a fixed IR 
action $\Gamma_{k=0}$.}.
We recall that for the bare theory of
\Eqref{eq:bareaction} the loop expansion
corresponds to the expansion in the coupling $\lambda$.
From now on it is more convenient to think in this way.
To reproduce perturbation theory, we need to 
compute the RG vector field in vicinity
of the Gaussian fixed point up to next-to-leading order
in $\lambda$. This is tantamount to integrating
the flow order by order in a Taylor expansion
for small $\lambda$.
We first input the initial condition $\Gamma_{\Lambda}$
on the right hand sides of the RG equations.
This produces a one-loop beta function for 
the renormalizable couplings $m^2$ and $\lambda$.
On the other hand, the RG equations radiatively generate
further couplings, namely those  whose $t$-derivative
at this initial point is non-vanishing.
By considering the Feynman diagrams mentioned above, we deduce a
power counting for the radiatively generated couplings
in terms of $\lambda$:
\begin{subequations}
\begin{align}
\eta &= O\left( \lambda^2 \right),&&
&&  z_1  = O\left( \lambda^2 \right),&&
&&  \tilde z_1 = O\left( \lambda^2 \right),&&\\
u_3 &= O\left( \lambda^3 \right),&&
&& z_2  = O\left( \lambda^3 \right),&&
&& \tilde z_2  = O\left( \lambda^3 \right),&&\\
u_4 &= O\left( \lambda^4 \right),&&
&& z_1^2 = O\left( \lambda^4 \right),&&
&& \tilde z_1^2  = O\left( \lambda^4 \right),&&   
\end{align}
\label{eq:powercounting}
\end{subequations}
and similar relations for $u_n$ and $z_n$,
with progressively higher powers of $\lambda$ for higher $n$.
Thus after an infinitesimal
RG step from $k=\Lambda$ to $k=\Lambda-\delta k$, the 
effective average action changes and the
perturbative expansion of the FRG vector field correspondingly  adjusts.
To compute the most general form of $\beta_\lambda$ along such a flow,
which is
exact at the order $\lambda^3$ (still within a local expansion
around vanishing fields),
we can use the power counting of \Eqref{eq:powercounting}
to eliminate the higher order terms.
This results in
\bea
\beta_{m^2}
&=&-\frac{k^2}{16 \pi^2}\left[ 
(N-1) \lambda\, l_{1,0}^4(0) +3 \lambda\, l_{0,1}^4(2 \lambda \kappa ) 
+ (N-1) \langle z_1 \rangle _{1,0}^6 (0) + \,\langle \tilde z_1 \rangle _{0,1}^6(2 \lambda \kappa )  \right]\nonumber\\
&&+\eta m^2 + k^2 \left(2 \kappa +\partial_t \kappa \right) \lambda 
+ O(\lambda^3) \ ,
\label{eq:FRGbetamtorderlambda3}
\\
\beta_\lambda &=&  \frac{N-1}{16\pi^2}l^4_{2,0}(0) \lambda^2
+ \frac{9}{16\pi^2}l^4_{0,2} (2\lambda \kappa) \lambda^2 \nonumber\\
&& -\frac{N-1}{16\pi^2}l^4_{1,0}(0)  u_3-\frac{5}{16\pi^2}l^4_{0,1} (2\lambda \kappa) u_3+ \left(2 \kappa +\partial_t \kappa \right)  u_3
\nonumber\\
&& +\frac{N-1}{8\pi^2} \lambda \langle z_1 \rangle^{6}_{2,0} (0) +\frac{3}{8\pi^2} \lambda \langle \tilde z_1 \rangle^{6}_{0,2}\left( 2 \lambda \kappa \right)
- \frac{N-1}{16\pi^2} \langle z_2 \rangle^{6}_{1,0}(0) -\frac{1}{16\pi^2} \langle \tilde z_2 \rangle^{6}_{0,1} \left( 2 \lambda \kappa \right)
\nonumber\\
&& +2 \eta \lambda + O(\lambda^4) \ .
\label{eq:FRGbetaatorderlambda3}
\eea
In these equations the threshold functions $l_{n,m}^4$ and 
the averages $\langle z_n\rangle^{6}_{n,m}$ denote
one-loop integrals over virtual momenta, with 
momentum-independent and -dependent vertices respectively.
The precise definitions  can be found in \Eqref{eq:thresholds&averages}.
The $\lambda$ dependence of mass thresholds 
should also be expanded, for instance:
\be
l_{0,n}^d (2 \lambda \kappa ) = l_{n,0}^d(0) -2 n \lambda \kappa \, l_{n+1,0}^d(0) + O \left( \lambda^2 \right) \ .
\ee
However this would bring corrections only for
non-vanishing $\kappa$, which is not generated
at the two-loop order.
We can thus set $\kappa=0$ in Eqs. (\ref{eq:FRGbetamtorderlambda3}) and (\ref{eq:FRGbetaatorderlambda3}).
While the contribution of the non-renormalizable couplings
$u_3$, $z_{1,2}$ and $\tilde{z}_{1,2}$ is obvious
in any mass-dependent scheme, one might expect that
it would not be present in \MSbar,
since all dimensionful integrals, in absence
of mass thresholds, need to vanish in the
$\epsilon\to0$ limit.
This expectation is however incorrect,
because when the computation is performed 
at non-vanishing mass, and the $m\to0$ limit
is taken after the $\epsilon\to0$ limit,
the contribution of the beta functions of the mass and of the
non-renormalizable couplings inside $\beta_\lambda$
attains a finite non-vanishing value.

To illustrate the details of this mechanism, 
we should first choose a specific
form of the $\epsilon$-dependent pseudo-regulator which is suitable
for the present computation.
We adopt the following function
\footnote{ As 
$\eta$ is vanishing at one loop in this model,
we discard the precise form of the
$Z_0$ and $Z_k$ dependence of the pseudo-regulator:
in fact, the $\eta$ dependence appearing 
on the r.h.s.~of the flow equations through the regularization
only contributes to the r.h.s.~of the flow from three loops on,
and so we can safely replace $Z_k\to1$.
\label{nonuni1}}
\be
R_k(z) =
\left[ \left(\frac{ k^2}{\mu^{4-2b} M^{2b}}\right)^\epsilon \left( z+ M^2 \right)^{1+\epsilon}-(z+m^2) \right] \ .
\label{eq:massdepReg}
\ee
Here $\mu$ is a $k$-independent momentum scale and $M$ plays the role
of a regularized mass, which is assumed to be 
$k$-dependent. $M$ should be an analytic function of $m$,
such that the $m\to0$ limit  smoothly removes also 
$M$.~\footnote{If $M$ depended also on other couplings 
aside from $m$, their contributions would only appear
from three loops.
\label{nonuni2}}
The precise form of $M$ making all
relevant integrals finite and ensuring the
$k$ independence of the beta functions
in the $\epsilon\to0$ limit
is derived in Appendix~\ref{app:thresholdsmass}.
Imposing the analiticity requirement on $M$  fixes $b=1$.

We now turn to the beta function of \Eqref{eq:FRGbetaatorderlambda3},
where on the r.h.s.~we organized different kinds of contributions
on different lines.
The first line provides the one-loop expression, as well as 
also a first type of higher order contribution, due to the
RG improvement of the pseudo-regulator.
More specifically, the threshold functions
are responsible for the appearance of $\beta_{m^2}$ on 
the r.h.s.~of $\beta_\lambda$, as
\be
l^4_{0,2} (2\lambda \kappa )  
= l_{2,0}^4(0) + O\left( \lambda^2 \right)
= 1-\frac{\beta_{m^2}}{2m^2} + O\left( \lambda^2 \right) 
= 1- \frac{N+2}{32\pi^2} \lambda 
+ O\left( \lambda^2 \right) \ .
\label{eq:lsatorderlambda}
\ee
As already anticipated, the ratio $\beta_{m^2}/m^2$ 
attains a finite mass-independent value. 
In fact,  the one-loop \MSbar result
for $\beta_{m^2}$ is recovered also with the present pseudo-regulator.
Even the two-loop \MSbar coefficient for $\beta_{m^2}$
can be correctly reproduced,
although this requires a careful choice of the function
$M^2$, which is described in Appendix~\ref{app:2loopsbetam}.

The second line of \Eqref{eq:FRGbetaatorderlambda3}
encodes the effect of $u_3$, which is generated
by the flow equation itself, as detailed 
in Appendix~\ref{app:2loops}.
It is a general feature of the FRG equations that
solving the flow equation for $u_3$ as
a function of $\lambda$,  at leading order in $\lambda$, is equivalent to setting $u_3$
at its $\lambda$-dependent fixed point value.
With the present pseudo-regulator this value of
the sextic coupling reads
\be
 u_3^{(1)} = \frac{N+26}{32\pi^2} \lambda^3 \frac{k^2}{m^2} \ .
\label{eq:u3FP}
\ee
This illustrates a second mechanism that generates two-loop terms,
even with the \MSbar pseudo-regulator.
In fact, despite all momentum integrals appearing 
in the beta function of $u_3$,
and any other non-renormalizable couplings, 
being dimensionful and
thus vanishing in the $\epsilon\to0$ limit,
some of the integrals appearing in the solution of the
flow and fixed-point equations for these couplings
are dimensionless and therefore survive
in \MSbar.
In other words, the flow equations should be 
solved before the $\epsilon\to0$ limit is taken.
Then, replacing \Eqref{eq:u3FP} in the second line
of \Eqref{eq:FRGbetaatorderlambda3} produces
further $\lambda^3$ terms in the beta function.

A similar fate applies to the third line of \Eqref{eq:FRGbetaatorderlambda3},
although the computational details this time are
somewhat more intricate. This is due to the momentum dependence
of the non-renormalizable couplings appearing inside $Z$ and
$\tilde{Z}$. In the process of solving
the flow equations for these couplings at leading
order in $\lambda$, and plugging the solution in
\Eqref{eq:FRGbetaatorderlambda3},
the following momentum averages are
generated
\begin{subequations}
\begin{align}
\langle z_1\rangle^{6}_{2,0}(0) &=
-8 (16\pi^2) \lambda^2 A \ ,&&
&&\langle \tilde z_1\rangle^6_{0,2}(0) = - 4 (16\pi^2) (N+8) \lambda^2 A \ ,&&\\
\langle z_2\rangle^6_{1,0}(0) &=  32 (16\pi^2) \lambda^3 B \ ,&&
&&\langle \tilde z_2\rangle^6_{0,1}(0) =  8 (16\pi^2)(N+26) \lambda^3 B \ .&&
\end{align}
\label{eq:z12intermsofAB}
\end{subequations}
Here $A$ and $B$ are dimensionless double momentum integrals whose precise
form is given in Appendix~\ref{sec:Zcontrib2loops}.
Although these are two-loop integrals, they involve only one copy
of $\partial_t R_k$, because one of them
disappears in the process of solving the flow equations for the
non-renormalizable couplings.
As a consequence, the $1/\epsilon^2$ pole of the integrals
is not balanced by the $\epsilon$ factor coming from the
single $\partial_t R_k$. Thus both $A$ and $B$ 
exhibit a $1/\epsilon$ pole~\footnote{
Incidentally, neither $A$ nor $B$ would be divergent
within a strict derivative expansion where the
RG-generated momentum dependence of $Z$ and $\tilde{Z}$
is truncated to its power series expansion around
$p^2=0$, because in the latter case
the two-loop integrals would exhibit only a
$1/\epsilon$ pole.
However this truncation would not reproduce the full
two-loop beta function, but just part of the $O(\lambda^3)$
contributions.}.
Despite this divergence, the flow
equation itself is finite, at least at order $\lambda^3$,
as in fact the only appearance of $A$ and $B$ on 
the r.h.s.~of \Eqref{eq:FRGbetaatorderlambda3}
is through the combination $A+B$, in which
the $1/\epsilon$ poles cancel.
The final result of this process is therefore
\be
A+B
= \frac{1}{2\left(16\pi^2\right)^2} \ .
\label{eq:AplusB}
\ee
Also for these terms, taking the $\epsilon\to0$ limit
too early, i.e.~before the flow for $Z$ and $\tilde{Z}$ is
solved and fed back inside $\beta_\lambda$, would fail
to unveil higher order corrections.

Putting all these contributions together,
the truncated beta function of \Eqref{eq:FRGbetaatorderlambda3}
in the $\epsilon\to 0$ limit reduces to
\be
\beta_\lambda
=  \frac{N+8}{16\pi^2} \lambda^2 - \frac{2(5N+22)}{(16\pi^2)^2} \lambda^3  +2 \eta \lambda \ .
\label{eq:twoloopswouteta}
\ee
We next turn to the computation of the anomalous dimension.
Following Ref.~\cite{Papenbrock:1994kf}, we split
$\eta$ in the sum
\begin{equation}
 \eta= \eta^{(1)}+\eta^{(2)} \ ,
\end{equation}
the two terms on the r.h.s.~being the contributions
of the momentum-independent and -dependent parts of 
the wave function renormalizations, respectively.
Notice that both contributions would
vanish in a truncation neglecting the field
dependence of the wave function renormalizations,
as the vacuum expectation value $\kappa$ vanishes
at this order.
Thus both terms are entirely due to the four-point
function $\Gamma_k^{(4)}$.
In the first part, $\eta^{(1)}$ is proportional
to the derivative couplings at zero momenta
\be
\eta^{(1)} =
 \frac{1}{16\pi^2} l_{1,0}^4(0) \left[ (N-1) z_1(0)+ \tilde z_1(0) \right] \ .
 \label{eq:preeta1}
\ee
The $O(\lambda^2)$ solution of the flow equation
gives
\begin{subequations}
\begin{align}
z_1(0) & = \frac{1}{3(16\pi^2)} \frac{k^2}{m^2} \lambda^2 \ ,\\
\tilde z_1(0) &=
 \frac{(N+8)}{6(16\pi^2)} \frac{k^2}{m^2} \lambda^2 \ ,
\end{align}
\label{eq:z1sat1loop}
\end{subequations}
such that the first contribution to the anomalous
dimension reads
\be
\eta^{(1)} = 
- \frac{(N+2)}{2(16\pi^2)} \lambda^2 \ .
\label{eq:theeta1}
\ee

The second part of the anomalous dimension is instead
proportional to the nontrivial momentum dependence
of $\Gamma_k^{(4)}$.
Again taking the limit for $\epsilon \to 0$ at the end 
of the nesting process, we find
\bea
\eta^{(2)}=  \frac{(N+2)}{(16\pi^2)^2}\lambda^2 \ .
\label{eq:theeta2}
\eea
Thus the whole  two-loop  anomalous dimension is recovered
\be
\eta = \eta^{(1)} + \eta^{(2)}=  \frac{(N+2)}{2(16\pi^2)^2}\lambda^2 \ .
\ee

Inserting in \Eqref{eq:twoloopswouteta} we finally arrive at
\be
\beta_\lambda
=\frac{N+8}{16\pi^2} \lambda^2 
-\frac{9N+42}{(16\pi^2)^2} \lambda^3 \ .
\label{eq:twoloopswitheta}
\ee
This is the universal part of the beta function at two loops.
With different mass-dependent regulators one would obtain
additional non-universal terms depending on the mass.
The contributions to the beta function
from three loops up is known not to be universal.
In our approach this regulator-dependence arises
at least from two sources:
the freedom of inserting other couplings in the pseudo-regulator,
as discussed in Section \ref{sec:fistgen} and footnote \ref{nonuni2},
and the contributions coming from $Z_k$,
as mentioned in Section \ref{sec:rolewavefun}
and footnote \ref{nonuni1}.

A similar treatment of \Eqref{eq:FRGbetamtorderlambda3} leads to
\be
\beta_{m^2} =
m^2\left[
\frac{(N+2)}{16\pi^2} \lambda -\frac{(N+2)}{4(16\pi^2)^2}  \left((1+2 f_1) (N+2)- 8 \sqrt{3} \pi +70 \right)\lambda^2\right]\ .
\label{eq:betam2twoloops}
\ee
Some more details are reported in Appendix \ref{app:2loopsbetam}.
We observe that the two-loop term is not universal,
and that the \MSbar result can be reproduced by 
suitably fixing the parameter $f_1$,
which enters the pseudo-regulator (\ref{eq:massdepReg})
through the choice
of $M$.

\section{Generalization to continuous dimensions}
\label{sec:epsilonexp}

Despite the fact that \MSbar is limited to applications in an 
even number of dimensions, the pseudo-regulator we discussed
lends itself to generalizations to any continuous
$d$, thanks to the intimate relation that exists
between dispersion relations and the dimensionality
of spacetime.
Consider the following pseudo-regulator
\be 
R_k(z) = Z_0 Z_k^{\sigma \epsilon} \left[\mu^{2(1-\alpha)} \left( \frac{k^2}{\mu^4} \right)^{\epsilon} z^{\alpha+\epsilon} -z \right] \ ,
\label{eq:regulatoralpha}
\ee
which generalizes (\ref{dimregblpa}) in that the
regularized propagator is now an homogeneous
function of momentum with power $\alpha+\epsilon$,
rather than  $1+\epsilon$.
This allows to correspondingly generalize
the formulae (\ref{Qslpa'}) for
the $Q$-functionals, whenever 
the dimension of the momentum integrals,
after having factored out all $\mu$ dependence,
is a nonnegative integer.
In fact, in this case the $\epsilon \to 0$ and the 
$Z_0\to 0$ limits give
\begin{subequations}
\begin{align}
Q_n\left( G_k^\ell \partial_t R_k\right) &=
2 \mu^{2 n\left( 1-\frac{1}{\alpha}\right)} \frac{\Gamma \left( \frac{n}{\alpha} \right) Z_k^{-\frac{n}{\alpha}}\left( -V_k''\right)^{\frac{n}{\alpha} -\ell+1}}{\Gamma(n)\Gamma(\ell)}
\frac{\left(1-\frac{\sigma\eta}{2}\right)}{\Gamma \left( \frac{n}{\alpha} -\ell+2 \right)}\ ,
\label{eq:firstQwithalpha}
\\
Q_n\left( G_k^\ell  G_k' \partial_t R_k\right) &=  -
\frac{\alpha ^2 \Gamma \left(\frac{n-1}{\alpha }+2\right)\mu^{2 (n-1)\left( 1-\frac{1}{\alpha}\right)}}{(n-1) \Gamma (n)}\frac{ 2 Z_k^{1-\frac{n}{\alpha}} \left(-V_k''\right)^{\frac{n-1}{\alpha}-\ell}}{\Gamma(\ell+2)}
\frac{\left(1-\frac{\sigma\eta}{2}\right)}{\Gamma\left( \frac{n-1}{\alpha} -\ell +1\right)}
\ ,
\\
Q_n\left( G_k^\ell  G_k'' \partial_t R_k\right) &=
\frac{\alpha ^2 \left(2 \alpha -\frac{(\alpha -1) (\ell+2)}{\frac{n-2}{\alpha }+2}\right) \Gamma \left(\frac{n-2}{\alpha }+3\right)\mu^{2 (n-2)\left( 1-\frac{1}{\alpha}\right)}}{(n-2) \Gamma (n)}
\nonumber\\
&\qquad\qquad
\times\frac{ 2 Z_k^{2-\frac{n}{\alpha}} \left(-V_k''\right)^{\frac{n-2}{\alpha}-\ell}}{\Gamma(\ell+3)} 
\frac{\left(1-\frac{\sigma\eta}{2}\right)}{\Gamma\left( \frac{n-2}{\alpha} -\ell+1 \right)}\ ,
\end{align} 
\label{eqs:Qsforalpha}
\end{subequations}
where the $\alpha$-dependent arguments of the
Gamma functions in the denominators are positive integers.
Recall that for a scalar field theory
and in the derivative expansion 
the 
index $n$ takes the values $d/2+l$ with
$l=0,1,2,\dots$.
Hence, if $\alpha$ is a continuous power, these formulae
are applicable to continuous $d$.

We illustrate the use of this generalized pseudo-regulator,
by addressing the description of the Wilson-Fisher
fixed point for $2<d<4$, for the linear $O(N)$ models.
We focus on the flow equations we presented in Appendix~\ref{app:floweqDE}
within the derivative expansion,
namely Eqs.~(\ref{eq:DEbetaUQ}) and (\ref{eq:DEbetaZQ}).
As we expect the 
effective potential to play a dominant role
 in the description of the Wilson-Fisher fixed point,
we demand that the corresponding quantum contributions 
be non-vanishing in our regularization scheme.
Specifically, the first kind of
 $Q$-functional, given in \Eqref{eq:firstQwithalpha}, 
is nontrivial in $2<d<4$ only if $d/2\alpha$ is a positive
integer.
Under this assumption the flow equations of the
derivative expansion become
\begin{subequations}
\begin{align}
\partial_t u &= -d u + (d-2+\eta) \tilde\rho u'\nonumber\\
&+  
\frac{\alpha \tilde \mu^{d\left( 1-\frac{1}{\alpha}\right)} }{(4\pi)^{d/2}\Gamma\left( \frac{d}{2}+1 \right)}
\left[ \tilde\zeta^{-\frac{d}{2\alpha}}\left( -u'-2\tilde\rho u'' \right)^{\frac{d}{2\alpha}} +(N-1)\zeta^{-\frac{d}{2\alpha}}\left( -u' \right)^{\frac{d}{2\alpha}}
\right] \ ,
\\
\partial_t \tilde\zeta &= \eta \tilde\zeta + (d-2+\eta) \tilde\rho \tilde\zeta' 
\nonumber\\
&-\frac{\tilde \mu^{d\left( 1-\frac{1}{\alpha}\right)} }{(4\pi)^{d/2}\Gamma \left(\frac{d}{2}\right)}
\left[ (\tilde\zeta'+2\tilde\rho \tilde\zeta'')\tilde\zeta^{-\frac{d}{2\alpha}}\left( -u'-2\tilde\rho u'' \right)^{\frac{d}{2\alpha}-1}
+(N-1)\tilde\zeta' \zeta^{-\frac{d}{2\alpha}}\left( -u' \right)^{\frac{d}{2\alpha}-1}\right]
\nonumber\\
& - \frac{(d+2 (\alpha -6)) (d-2 \alpha )\tilde \mu^{d\left( 1-\frac{1}{\alpha}\right)} }{6 \alpha  (4\pi)^{d/2}\Gamma \left(\frac{d}{2}\right)} \tilde\rho (3u''+2\tilde\rho u''') \tilde\zeta'\tilde\zeta^{-\frac{d}{2\alpha}}\left( -u'-2\tilde\rho u'' \right)^{\frac{d}{2\alpha}-2}
\nonumber\\
& - (N-1) \frac{d(d+2)(d^2-4 \alpha ^2 )\tilde \mu^{d\left( 1-\frac{1}{\alpha}\right)}}{24 \alpha(4\pi)^{d/2} \Gamma \left(\frac{d}{2}+2\right)}\tilde\rho u'' \zeta'\zeta^{-\frac{d}{2\alpha}}\left( -u' \right)^{\frac{d}{2\alpha}-2}
\nonumber\\
& + (N-1)\frac{(d-2 \alpha)\tilde \mu^{d\left( 1-\frac{1}{\alpha}\right)}}{\alpha(4\pi)^{d/2}   \Gamma \left(\frac{d}{2}\right)}u''
(\tilde\zeta-\zeta)\zeta^{-\frac{d}{2\alpha}}\left( -u' \right)^{\frac{d}{2\alpha}-2} \ ,
\end{align}
\label{eqs:DEforalpha}
\end{subequations}
where $u$, $\zeta$ and $\tilde\zeta$
are the  dimensionless
renormalized counterparts of the
$U_k$, $Z_k$ and $\tilde Z_k$ of Eqs.~(\ref{eq:GammaDE})
and (\ref{eq:tildeZ}),
defined in analogy to Eq. (\ref{eq:dimlessvzeta}).
Here for notational simplicity we dropped the RG improvement,
by setting $\sigma=0$; the full equations contain
the factor $(1-\sigma\eta/2)$ in front of every quantum
contribution.
The $\tilde{\mu}$ dependence of these flow equations
can be cancelled by a further rescaling 
of all dimensionful quantities with
respect to $\tilde{\mu}$, which casts
the RG equations in a genuine \MSbar form.

We look for fixed points of the previous flow equations
by means of a small-fields polynomial expansion
\be
u = \sum_{i=0} \frac{\lambda_{2i}}{i!}\tilde\rho^i \ ,
\hspace{1cm}
\tilde\zeta = 1+\sum_{i=1} \frac{\tilde z_{2i}}{i!}\tilde\rho^i \ ,
\hspace{1cm}
\zeta = 1+\sum_{k=i} \frac{z_{2i}}{i!}\tilde\rho^i \ .\\
\ee
We find the Gaussian fixed point for every value of $\alpha$ and a
nontrivial fixed point only for $\alpha=d/4$, located at
\be
\lambda_4^* = \frac{(4\pi)^{d/2}\Gamma\left( \frac{d}{2} \right)(4-d)}{8+N}  \tilde{\mu}^{4-d}\ ,
\label{eq:WFFPcontinuousd}
\ee
with $\eta=0$, and all others couplings being vanishing.
We note that with standard regulators the
dimensionless potential of the WF fixed point 
has a nontrivial minimum, but the dimensionful mass 
(deduced from the limit $k\to0$) is zero,
in accordance with the fact that the theory is scale invariant
at quantum level.
With the \MSbar pseudo-regulator the ``dimensionless mass''
$\lambda_2$ is zero even for finite $k$.
The same phenomenon happens also in
the functional perturbative approach \cite{Codello:2017hhh}.

For $\alpha=d/4$ the power of $\tilde{\mu}$ appearing in the Eqs.~(\ref{eqs:DEforalpha})  is $(d-4)$. Therefore, the rescaling which
maps these equations into those of \MSbar, is 
 effectively declaring the dimensionality of the couplings to be the one expected in four dimensions, the rescaling factors differing only by powers of 
$(k/\tilde{\mu})^{4-d}$.
Furthermore, the value $\alpha=d/4$ is precisely the one that
makes the quartic interaction marginal for
continuous $d$. In fact, the effective kinetic
term of the regularized theory
has a dispersion relation $z^\alpha$, which changes
the dimensionality
of the scalar field from $(d-2)/2$ to $(d-2 \alpha)/2$.
In other words, within the present truncation, 
our one-loop-like equations for
$\phi^4$ theory are able to detect the Wilson-Fisher
fixed point in continuous $d$ only when the pseudo-regulator
turns $d$ into an ``effective upper critical dimension''.
This interpretation is also consistent with the apparent
absence of the multicritical models with $\rho^{p}$ interactions, for $p>2$. In fact, the effective upper 
critical dimensions for these models would 
be at $d=2p\alpha/(p-1)$, which is not
compatible with our simplifying assumption
of an integer $d/2\alpha$.

The stability matrix at the fixed point of
\Eqref{eq:WFFPcontinuousd}
is triangular and the eigenvalues are
\begin{subequations}
\begin{align}
\theta_i &= -d + i(d-2) + i \left[ 1+ \frac{6(i-2)}{N+8}\right](4-d) \ ,
\label{eq:thetas}
\\
\omega_i &= i(d-2) + \left[ i+1+ \frac{3(2i^2-2i-3)}{N+8}\right](4-d)
+ \frac{3i}{2(N+8)}(4-d)^2 \ .
\label{eq:omegas}
\end{align}
\label{eq:exponentsepsilon}
\end{subequations}
By setting $d= 4-\varepsilon$
we recognize that this prediction agrees with the
usual first order of the $\varepsilon$ expansion.
For instance, for $N=1$ we get
\begin{subequations}
\begin{align}
\theta &=
\left( -2+\frac{\varepsilon}{3} , \varepsilon, 2+3\varepsilon , 4+\frac{19}{3}\varepsilon, 6+11 \varepsilon, 8+17\varepsilon \dots \!\right) \ ,
\\
\omega &=\left(  2 , 4+\frac{4}{3}\varepsilon, 6+4 \varepsilon, 8+8\varepsilon \dots \!\right) \ .
\end{align}
\end{subequations}
Notice that $\varepsilon$ here should
not be confused with the $\epsilon$ of Eq. (\ref{eq:regulatoralpha}),
the latter having been removed by the limit $\epsilon\to 0$.
Also, the one-loop 
predictions of \Eqref{eq:exponentsepsilon}
become exact in the $N\to\infty$ limit.

Order $\varepsilon^2$ corrections affect the
estimate of $\omega_i$ in \Eqref{eq:omegas}
but are missing in \Eqref{eq:thetas}.
This is related to the fact that
the fixed point value of $\eta=0$
is vanishing in this truncation, such that the
RG resummations triggered by the 
dependence of $R_k$ on $Z_k$
are ineffective at the fixed point. 
In fact,
improvements on the estimate of $\theta$'s
can be obtained by
allowing for the feedback of other couplings in 
the pseudo-regulator. For instance,
it is natural to allow for
the replacement of the mass parameter $\mu^2$
with the running $\lambda_2$ through a tunable
parameter $b$~\footnote{
We cannot replace every occurence of $\mu^2$ with $\lambda_2$,
otherwise singularities of the form $\beta_2/\epsilon$ would
arise in the beta functions. The insertion of $\lambda_2$
in the pseudo-regulator must preserve the cancellation
of such poles.},
and write
\begin{equation}
 R_k(z) = Z_0 Z_k^{\sigma \epsilon} \left[\mu^{2(1-\alpha)} \left( \frac{k^2}{\mu^{2(2-b)}\lambda_2^{b}} \right)^{\epsilon} z^{\alpha+\epsilon} -z \right] \ .
\label{eq:regulatoralphab}
\end{equation}
This would result in a different RG improvement of 
Eqs. (\ref{eqs:DEforalpha}), where
each quantum term is now multiplied
by the factor $(1-b\beta_2/(2\lambda_2)-\sigma\eta/2)$,
which leads to the following $b$-dependent
quartic coupling evaluated at the fixed point and critical exponents
\begin{subequations}
\begin{align}
\lambda_4^* &= \frac{(4\pi)^{d/2}\Gamma\left( \frac{d}{2} \right)(4-d)}{N+8-\frac{b}{2}(N+2)(4-d)}  \tilde{\mu}^{4-d}\ ,
\label{eq:WFFPcontinuousdb}
\\
\theta_i &= -d + i(d-2) + i \left[1+\frac{6(i-2)}{N+8}\right](4-d) 
 +\delta_{i,4} \frac{b}{2}\left(\frac{N+2}{N+8}\right)(4-d)^2 \ ,
\label{eq:thetasb}
\\
\omega_i &= i(d-2) + \left[ i+1+ \frac{3(2i^2-2i-3)}{N+8}\right](4-d)
+ \frac{3i}{2(N+8)}(4-d)^2 \ .
\label{eq:omegasb}
\end{align}
\label{eq:exponentsepsilonb}
\end{subequations}
Getting better estimates of the critical $\eta$
and the correlation-length exponent $\nu$
does instead require a larger truncation,
accounting for at least part of the two-loop
contributions, as discussed in Section~\ref{sec:twoloops}.
This kind of more elaborate analysis of the Wilson-Fisher
fixed point by means of the \MSbar pseudo-regulator
is left for future studies.

We reiterate that the flow
equations (\ref{eqs:DEforalpha})
have been obtained under the
assumptions that $2<d<4$ and
that $d/2\alpha$ is a positive integer.
In even dimensions the equations
have additional terms.
In fact, taking the $d\to 2$ limit
in (\ref{eqs:DEforalpha}) would miss relevant
contributions which are 
present in the flow equations studied in
Section \ref{sec:DEmulti}.
The latter can indeed be reproduced
by the $\alpha$-generalized \MSbar pseudo-regulator,
by applying \Eqref{eqs:Qsforalpha} directly
in $d=2$ and taking the $\alpha\to1$ limit.

 Finally, let us comment on the extension of the
nonlinear sigma model of Section \ref{sec:nonlineaON}
to dimension $d>2$.
Instead of just using the $\varepsilon$-expansion,
it is possible to use directly the generalized
\MSbar pseudo-regulator (\ref{eq:regulatoralpha}).
As a result, in $d=2+\varepsilon$ we recover
the well-known nontrivial fixed point 
\begin{subequations}
\begin{align}
g_{*}^{2} &= \frac{2 \pi  \varepsilon }{N-1}+\frac{\pi  \sigma  \varepsilon ^2}{(N-1)^2}+O\left(\varepsilon ^3\right) \ ,
\\
\nu^{-1} &= \varepsilon -\frac{\sigma  \varepsilon^2 }{2 (N-1)}+O\left(\varepsilon ^3\right) \ ,
\\
\eta &= \frac{\varepsilon }{N-1}+O\left(\varepsilon ^3\right) \ .
\end{align}
\end{subequations}
Here it is possible to adjust $\sigma$ ($\sigma = -2$) to get the full two-loop result for $\nu$, 
but not for $\eta$.
\footnote{This situation is similar to
the one observed Section \ref{sec:LPAprime}
for $\phi^3$ theory in six dimensions,
although there we could tune $\sigma$
to get the right two-loop result for either 
$\beta_3$ or $\eta$.}.
The latter correction would arise by considering a truncation where $Z_k$ depends also on the momenta, as is discussed in Section~\ref{sec:twoloops}.

\section{Conclusions}
\label{sec:conclusions}

Mass-dependent Wilsonian RG schemes,
such as for instance momentum subtraction with a sharp UV cutoff,
simultaneously achieve the two goals of
regularizing a field theory and of 
defining the heavy modes to be integrated
out while constructing an effective 
description of the system.
In these schemes, information about mass thresholds is essential and
built in the effective theory at all scales.
Mass-independent schemes instead,
e.g.~dimreg/\MSbar,
remove the latter piece of information
by taking the limit of
infinite separation between the
physical scales of applicability
of the effective field theory
and the heavy masses of the
underlying microscopic description.
It is therefore natural that,
in the construction of a mass-independent
scheme out of a Wilsonian one,
the infinite-separation-of-scales limit
also becomes a regularization-removal
process.

This is precisely what has been observed
in the present work. More specifically,
we have focused on the
functional renormalization group (FRG) equations,
a prototypical Wilsonian representation of
field theory based on shell-by-shell
integration of  modes according 
to a coarse-graining-defining function
$R_k(q^2)$, which acts as a smooth infrared
cutoff on modes with  momentum $q^2\ll k^2$.
As a matter of fact, we have found that
it is indeed possible to achieve a
continuous transition from this exact
mass-dependent scheme
to functional RG equations within \MSbar,
at the price of taking a parametric limit
$\epsilon\to0$ that 
in even dimensions
also results in
the removal of the cutoff: $R_k\to0$~\footnote{ 
The generalization to continuous dimensions
discussed in Section~\ref{sec:epsilonexp}
is an exception, as in this case the deformation of
the dispersion relation
operated by the pseudo-regulator survives
the $\epsilon\to0$ limit.}.

The dependence of $R_k$ on continuous parameters,
such as our $\epsilon$, is allowed and welcome
in the FRG setup. In fact, it is often used
in FRG applications as a diagnostic tool 
(weak dependence on such parameters is taken as a sign of
a good truncation) or even as
a selection criterion for the ``best'' regulator
(e.g.~through the principle of minimum sensitivity \cite{Balog:2019rrg,DePolsi:2020pjk}).
However we find that, while taking 
the $\epsilon\to0$ limit,
the regulator $R_k$ at some point must
leave the domain of acceptable IR
Wilsonian cutoffs and violate some
of the conditions that define physical
coarse grainings.
This is quite to be expected, as
dimreg is by no means a physical
IR cutoff. As such, also the pseudo-cutoff
form which should be attained by $R_k$
for asymptotically small $\epsilon$,
our \Eqref{dimregcut},
defies every interpretation as a
conventional regulator, and is well
suited for its goal only when augmented
by analytic continuation of the 
momentum integrals in $\epsilon$.

Quite interestingly,
there is a certain degree of freedom in
how to attain the \MSbar limit,
and it is even possible to 
achieve it starting from
the most popular functional forms
of $R_k$, as is explained in Section \ref{sec:interpolation}.
This exercise renews interest on several
nontrivial aspects of vanishing-$R_k$ limits
of FRG equations, to which we will devote 
a subsequent paper \cite{next}.
There, we will also show 
that this limit can help
to weed out certain unphysical features that are
introduced in the FRGE by some choices of regulator,
and that are intimately related to the preservation
of nonlinear symmetries.

Indeed, one of the most tantalizing aspects
of this research direction, is the possibility
to look at the \MSbar limit of FRG equations
as a novel way to approach the challenging
problem of gauge and nonlinear symmetries.
Here, we have limited ourselves to
explore these aspects in Section \ref{sec:nonlineaON},
where we have observed that the $\epsilon\to0$
limit of the RG equations of a linear $O(N)$ model
have the pleasant property of
preserving also a nonlinearly realized 
$O(N+1)$ symmetry. Further systematic
studies of this problem are in order,
to assess whether taking the \MSbar limit
might ease the task of
fulfilling Ward-Takahashi identities
and master equations
(actually $R_k$-deformed versions of the latter).

Although it might be possible to 
study the problem of reproducing \MSbar
at the level of the
exact FRG equation (\ref{FRGE}),
we have only addressed this goal within 
specific approximation schemes,
mainly the first orders of the derivative
expansion and up to two loops.
One major conceptual problem which might be
raised against our efforts, is whether it makes 
any sense to try to join the FRG framework,
whose strength is in the exact and nonperturbative
nature of the formalism,
with dimreg/\MSbar, which is widely believed
to be applicable only within perturbation theory.
Another way of formulating this question,
would be to ask for a way of performing
numerical FRG computations and still
make sense of an \MSbar limit.

Keeping this aspect in mind,
after the construction of an FRG pseudo-regulator which successfully
reproduces the one-loop \MSbar{}
beta functions for vanishing $\epsilon$, see
Section \ref{sec:LPA},
we have addressed the question as to whether
this pseudo-regulator choice and the $\epsilon\to0$ limit
spoil the nonperturbative nature of the exact FRG equation.
We have provided reasons to argue for a negative answer.
In Sections \ref{sec:beyondLPA} and \ref{sec:DE}
we have first illustrated the physical content
of the RG resummations contained in
the RG improvement of one-loop beta functions,
showing that they account for higher-order
perturbative contributions and can
even fairly describe some nonperturbative
critical phenomena in two dimensions,
see for instance Table \ref{tab:multicritical}.

Furthermore, we have found strong hints suggesting that
the FRG equations  remain nonperturbative, as
long as the limit of vanishing $\epsilon$ is taken
at the end of the relevant computations 
(indeed, analytic continuation of integrals and
parametric limits are processes
which can be performed also numerically).
To substantiate this conclusion, and to illustrate
how the limit of vanishing pseudo-regulator should be dealt with,
in Section \ref{sec:twoloops} we have derived the 
two-loop beta functions for massive $\phi^4$ theory in four dimensions.
The details of the computation, presented in 
Appendix \ref{app:2loops}, show how the 
typical higher-orders FRG contributions,
which are tied to dimensionful momentum integrals and
stay rightly non-vanishing for $\epsilon\neq0$,
survive even for $\epsilon\to0$
as long as the RG equations are solved before
taking the latter limit.


Other interesting aspects we have just touched upon,
and that certainly deserve further attention,
are the use of the \MSbar pseudo-regulator in 
 theories featuring background fields (Section \ref{sec:externalfield})
 and in continuous dimensions (Section \ref{sec:epsilonexp}).
Finally, as in this work we have confined our attention
to scalar field theories, future works will need to explore
the generalization to fermions, gauge theories, and gravity.

\section*{Acknowledgments}
We are grateful to O.~Zanusso for seminal
discussions on this topic, and to J.~M.~ Pawlowski for 
his challenging questions. L.Z. acknowledges
support by the DFG under Grants Nos. GRK1523/2,
 Gi328/9-1, 
 and  406116891 within the Research Training Group RTG 2522/1,
 for some of the duration of this project.

\newpage
\begin{appendix}

\section{Derivation of the pseudo-regulator}
\label{app:howtofind}

In this section we present a derivation of
the functional form of the \MSbar pseudo-regulator
in \Eqref{dimregcutb}.
We want to obtain the result (\ref{desired}) from the FRGE.
Inspired by the deformation of integrands
which takes place in dimensional regularization, we
consider the following family of regulators
\be
R_k(z) =  \mu^{-2\epsilon} F(k, \mu,  m, \epsilon) z^{1+\epsilon} - z  \ .
\label{eq:howtofirstR}
\ee
Here the first term containing the momentum power
$(1+\epsilon)$ is needed to reproduce the $1/\epsilon$ pole
of the analytically continued dimensionless integral,
and can be obtained for instance by
starting with a $(d+\epsilon)$ dimensional
integral, and by integrating out the $(+\epsilon)$ dimensions
\cite{Georgi:1994qn}.
In it, $F$ is an arbitrary dimensionless
function and $\mu$ is a classical arbitrary mass parameter.
The second term in \Eqref{eq:howtofirstR}
is there to cancel the original inverse propagator 
of the bare theory
(we recall that the regulator $R_k$ acts additively,
and that the regulated inverse propagator is the combination
$z+R_k(z)$).

Note that $\partial_t R_k(z)$ has two contributions: one coming from the explicit
dependence of $F$ on $k$ and another one proportional to $\beta_{m^2} = \partial_t m^2$.
Assuming that $\partial_t F \propto \epsilon$ and using the following identity
\be
\Gamma \left[ -n + \epsilon \right] = \frac{(-)^{n}}{\Gamma[n+1]} \frac{1}{\epsilon} + O(\epsilon^0) \ ,
\ee
the Mellin transform of the first term inside $\partial_tR_k$
is
\begin{align}
&\frac{\partial_t F}{F^{1+\frac{n}{1+\epsilon}}}
\left(\frac{\mu}{m}\right)^{\frac{2n \epsilon}{1+\epsilon}}
\frac{\Gamma\left(1+\frac{n}{1+\epsilon}\right)\Gamma(l-n-1+\frac{n \epsilon}{1+\epsilon})}{(1+\epsilon)\Gamma(n)\Gamma(l)} \left(m^2\right)^{n-l+1}
=\nonumber\\
&\frac{\partial_t F}{ \epsilon\,F^{1+n}} \frac{(-)^{n-l+1}}{\Gamma(l)\Gamma(n-l+2)} \left(m^2\right) ^{n-l+1}  
+ O(\epsilon) \ .
\end{align}
So taking the limit for $\epsilon \to 0$, we find (\ref{desired}) if
\be\partial_t F(k , \mu, m, \epsilon) = 2 \epsilon F(k, \mu, m, \epsilon)^{1+n} \ ,
\ee
that is
\be
F = 1+ \epsilon  \log \left(\frac{ k^2}{ \mu^{2-2b} m^{2b}} \right) +  O(\epsilon^2) 
\approx \left( \frac{k^2}{ \mu^{2-2b} m^{2b}} \right)^\epsilon  .
\ee
With this $F$ , the second piece of 
$\partial_t R_k$
proportional to $\beta_{m^2}$ can be calculated in the same way.
This agrees with \Eqref{dimregcutb}, of which 
\Eqref{dimregcut} is a special case, corresponding
to $b=0$.
As described in the main text,
different choices of $b$ only affect higher-order corrections.

\newpage

\section{Flow equations in the \texorpdfstring{$\boldsymbol{O(\partial^2)}$}{TEXT}
derivative expansion}
\label{app:floweqDE}
In this section we present part of the
flow equations of the linear $O(N)$ model
in the $O(\partial^2)$ of the derivative expansion,
which can be found for instance in Ref.~\cite{Tetradis:1993ts}.
These descend from the exact FRG equation
upon specifying the truncation of \Eqref{eq:GammaDE}.
We introduce the following notations
\begin{subequations}
\bea
G_0 &=& \left(Z_k q^2 + R_k(q^2) + U_k' \right)^{-1} ,
\\
G_1 &=& \left(\tilde Z_k q^2 + R_k(q^2) + U_k'+2\rho U_k''\right)^{-1} ,
\eea
for the Goldstone-bosons and radial-mode propagators.
\end{subequations}
The flow equations for $U_k$ and $\tilde Z_k$,
which is defined in \Eqref{eq:tildeZ}, are
\begin{align}
\partial_t U_k &=
\frac{1}{2(4\pi)^{d/2}} \left( Q_{\frac{d}{2}}\left[ G_1 \partial_t  R_k \right]
+ (N-1)Q_{\frac{d}{2}}\left[ G_0 \partial_t  R_k \right] \right) \ ,
\label{eq:DEbetaUQ}
\\
\partial_t  \tilde Z_k  &=
- \frac{\left(\tilde Z_k'+2\rho \tilde Z_k''\right)}{2(4\pi)^{d/2}} Q_{\frac{d}{2}}\left[ G_1^2 \partial_t  R_k \right]
-(N-1) \frac{\left(  Z_k'+\rho Y_k'\right)}{2(4\pi)^{d/2}} Q_{\frac{d}{2}}\left[ G_0^2 \partial_t  R_k \right]
\nonumber\\
&
+\frac{2 \rho \left( \tilde Z_k'\right)^2}{(4\pi)^{d/2}} \left[
 \frac{2d+1}{2}  Q_{\frac{d}{2}+1}\left[ G_1^3 \partial_t  R_k \right]
+ \frac{(d+2)(d+4)}{4}\left( Q_{\frac{d}{2}+2}\left[ G_1^2 G_1' \partial_t R_k\right]  \right.\right.
\nonumber\\
&
\left.\left.+
 Q_{\frac{d}{2}+3}\left[ G_1^2 G_1'' \partial_t  R_k\right]
\right)\right] 
+ \frac{2 \rho\left( 3 U_k''+2 \rho U_k''' \right)^2}{(4\pi)^{d/2}} \left( 
Q_{\frac{d}{2}}\left[ G_1^2 G_1'\partial_t  R_k \right]
+Q_{\frac{d}{2}+1}\left[ G_1^2 G_1'' \partial_t R_k \right]
\right)
\nonumber\\
&
+\frac{2 \rho \tilde  Z_k' \left( 3 U_k''+2 \rho U_k'''\right)}{(4\pi)^{d/2}} \left[
\left( d+2 \right) \left( Q_{\frac{d}{2}+1}\left[ G_1^2 G_1' \partial_t  R_k\right] +
 Q_{\frac{d}{2}+2}\left[ G_1^2 G_1'' \partial_t  R_k\right] \right)\right.
\nonumber\\
&
\left.+2 Q_{\frac{d}{2}}\left[ G_1^3 \partial_t  R_k\right]\right] 
+ (N-1)\frac{ \rho Y_k }{(4\pi)^{d/2}} \left( 
2U''_k Q_{\frac{d}{2}}\left[ G_0^3\partial_t  R_k \right]
+d Z'_k Q_{\frac{d}{2}+1}\left[ G_0^3 \partial_t R_k \right]
\right)\nonumber\\
&
+(N-1)\frac{2 \rho \left(  Z_k'\right)^2}{(4\pi)^{d/2}} \left[
\frac{(d+2)(d+4)}{4} \left( Q_{\frac{d}{2}+2}\left[ G_0^2 G_0' \partial_t R_k\right] \!+\!
 Q_{\frac{d}{2}+3}\left[ G_0^2 G_0'' \partial_t  R_k\right]
\right) \right.
\nonumber\\
&
\left.+\frac{1}{2} Q_{\frac{d}{2}+1}\left[ G_0^3 \partial_t  R_k \right]\right]
+ (N-1)\frac{2 \rho\left( U_k'' \right)^2}{(4\pi)^{d/2}} \left( 
Q_{\frac{d}{2}}\left[ G_0^2 G_0'\partial_t  R_k \right]
+Q_{\frac{d}{2}+1}\left[ G_0^2 G_0'' \partial_t R_k \right]
\right)
\nonumber\\
&
+(N-1)\frac{2 \rho  Z_k' U_k''}{(4\pi)^{d/2}}
\left( d+2 \right) \left( Q_{\frac{d}{2}+1}\left[ G_0^2 G_0' \partial_t  R_k\right] +
 Q_{\frac{d}{2}+2}\left[ G_0^2 G_0'' \partial_t  R_k\right] \right) \ .
 \label{eq:DEbetaZQ}
\end{align}
The beta functional for $Z_k$ is instead presented
with slightly different notations in \Eqref{eq:dtZrhoDE}.

Using the pseudo-regulator (\ref{dimregblpa}), in the $Z_0\to 0$ limit and suppressing the RG improvement by setting $\sigma=0$, we find
\bea
\partial_t U_k \!\!\!\! &=& \!\!\!\!
\frac{1}{(4\pi)^{d/2}\Gamma\left( \frac{d}{2}+1 \right)} \left( \tilde Z_k^{-\frac{d}{2}}\left(-U'_k-2 \rho U''_k\right)^{\frac{d}{2}}
+ (N-1)Z_k^{-\frac{d}{2}}\left(-U'_k\right)^{\frac{d}{2}} \right) \ ,
\label{eq:DEV}\\
\partial_t  \tilde Z_k  \!\!\!\!&=&\!\!\!\!
- \frac{\left(\tilde Z_k'+2\rho \tilde Z_k''\right)}{(4\pi)^{d/2}\Gamma\left( \frac{d}{2}\right)} \tilde Z_k^{-\frac{d}{2}}\left(-U'_k-2 \rho U''_k\right)^{\frac{d}{2}-1}
-(N-1) \frac{\left(  Z_k'+\rho Y_k'\right)}{(4\pi)^{d/2}\Gamma\left( \frac{d}{2}\right)} Z_k^{-\frac{d}{2}}\left(-U'_k\right)^{\frac{d}{2}-1}
\nonumber\\
&&\!\!\!\!
+\frac{\left( 4+18d-d^2\right) }{12(4\pi)^{d/2}\Gamma\left( \frac{d}{2} \right)} 
\rho \left( \tilde Z_k'\right)^2\tilde Z_k^{-\frac{d}{2}-1}\left(-U'_k-2 \rho U''_k\right)^{\frac{d}{2}-1}
\nonumber\\
&&\!\!\!\!
+\frac{(10-d) }{3(4\pi)^{d/2}\Gamma\left( \frac{d}{2}-1 \right)}
\rho \tilde  Z_k' \left( 3 U_k''+2 \rho U_k'''\right)
\tilde Z_k^{-\frac{d}{2}}\left(-U'_k-2 \rho U''_k\right)^{\frac{d}{2}-2}
\nonumber\\
&&\!\!\!\!
- \frac{ \rho\left( 3 U_k''+2 \rho U_k''' \right)^2}{3(4\pi)^{d/2}\Gamma \left( \frac{d}{2}-2 \right)} 
\tilde Z_k^{-\frac{d}{2}+1}\left(-U'_k-2 \rho U''_k\right)^{\frac{d}{2}-3}
\nonumber\\
&&\!\!\!\!
+(N-1)\frac{(4-6d-d^2) }{6(4\pi)^{d/2}\Gamma\left( \frac{d}{2} \right)} 
\rho \left(  Z_k'\right)^2 Z_k^{-\frac{d}{2}-1}\left(-U'_k\right)^{\frac{d}{2}-1} 
\nonumber\\
&&\!\!\!\!
-(N-1)\frac{\left( d+2 \right) }{3(4\pi)^{d/2}\Gamma\left( \frac{d}{2}-1 \right)}
\rho  Z_k' U_k'' Z_k^{-\frac{d}{2}}\left(-U'_k\right)^{\frac{d}{2}-2} 
\nonumber\\
&&\!\!\!\!
- (N-1)\frac{ \rho\left( U_k'' \right)^2}{3(4\pi)^{d/2}\Gamma\left( \frac{d}{2}-2 \right)} 
Z_k^{-\frac{d}{2}+1}\left(-U_k'\right)^{\frac{d}{2}-3} 
\nonumber\\
&&\!\!\!\!
+ (N-1)\frac{2 \rho Y_k }{(4\pi)^{d/2}\Gamma\left( \frac{d}{2}-1 \right)} \left( 
U''_k -\frac{d}{d-2} Z'_k Z_k^{-1} U'_k
\right)
Z_k^{-\frac{d}{2}}\left(-U'_k\right)^{\frac{d}{2}-2} \ .
\label{eq:DEZ}
\eea
For general $\sigma$, the r.h.s. have to be simply multiplied by $(1-\sigma\eta/2)$.

\newpage

\section{Threshold functions for a mass-dependent pseudo-regulator}
\label{app:thresholdsmass}

In this appendix we detail the computation of the
following threshold functions
\be
l_{n,0}^d(0) = \frac{n Z_k^{n}}{2 } k^{2n -d} \int_0^\infty\!\!\! dz \, z^{\frac{d}{2}-1} \frac{\partial_t R_k(z)}{P_k(z)^{n+1}} \ ,
\ee
where
\be
P_k(z)=Z_k z+R_k(z) \ ,
\ee
by means of the mass-dependent pseudo-regulator of 
\Eqref{eq:massdepReg}.
As we need the result for the computation of the two-loop
beta function in four dimensional $\lambda\phi^4$ theory,
we content ourselves of the first orders
in a perturbative expansion in $\lambda$.
In particular, we neglect the $\eta$ dependence appearing 
on the r.h.s.~of the flow equations through the regularization,
as it would lead to higher orders in $\lambda$.
Our pseudo-regulator choice results in simple propagators
but a somewhat more convoluted contribution
of the differentiated pseudo-regulator:
\begin{subequations}
\bea
P_k \left( z \right) &=& Z_k \left(\frac{ k^2}{\mu^{4-2b} M^{2b}}\right)^\epsilon \left( z+ M^2 \right)^{1+\epsilon} \ ,
\\
\partial_t R_k(z) &=& 2 \epsilon \left( 1-\frac{b \,\partial_t M^2 }{2 M^2} \right) P_k \left( z \right) +  (1+\epsilon)  \partial_t M^2 \, P_k \left( z \right)-  \beta_{m^2} \ .
\eea
\label{eq:massdepPanddtR}
\end{subequations}
The loop integral can then be split into three different kinds of contributions, corresponding to the three pieces of $\partial_tR_k$
\begin{align}
\frac{2\, l^{d}_{n,0}}{nk^{2n-d}}
&=
\frac{2 \epsilon}{\Gamma \left( \frac{d}{2} \right)} \left( 1-\frac{b \,\partial_t M^2 }{2 M^2} \right) \left(\frac{k^2}{\mu^{4-2b} M^{2b}}\right)^{-n\epsilon} 
\int_0^\infty \!\!\!dz  \frac{z^{\frac{d}{2}-1} }{\left( z+ M^2 \right)^{n(1+\epsilon)}}
\nonumber\\
& +  \frac{(1+\epsilon)}{\Gamma \left( \frac{d}{2} \right)} \partial_t M^2 \left(\frac{k^2}{\mu^{4-2b} M^{2b}}\right)^{-n\epsilon}
\int_0^\infty \!\!\! dz  \frac{z^{\frac{d}{2}-1} }{\left( z + M^2 \right)^{1+n(1+\epsilon)}}
\nonumber\\
& -  \frac{\beta_{m^2} }{\Gamma \left( \frac{d}{2} \right)}
\left(\frac{k^2}{\mu^{4-2b} M^{2b}}\right)^{-(n+1)\epsilon}
\int_0^\infty \!\!\!dz  \frac{z^{\frac{d}{2}-1} }{\left( z + M^2 \right)^{(1+\epsilon)(n+1)}}
\nonumber\\
&=
2 \epsilon \left( 1-\frac{b \,\partial_t M^2 }{2 M^2} \right) \left(\frac{ k^2}{\mu^{4-2b} M^{2b}}\right)^{-n\epsilon} 
\frac{\Gamma \left( n+n \epsilon -\frac{d}{2} \right)}{\Gamma \left( n+n\epsilon \right)} \left( M^2 \right)^{\frac{d}{2}-n-n\epsilon}
\nonumber\\
& +  (1+\epsilon) \partial_t M^2 \left(\frac{k^2}{\mu^{4-2b} M^{2b}}\right)^{-n\epsilon}
\frac{\Gamma \left( n+1+n \epsilon -\frac{d}{2} \right)}{\Gamma \left( n+1+n\epsilon \right)} \left( M^2 \right)^{\frac{d}{2}-n-1-n\epsilon}
\nonumber\\
& -  \beta_{m^2} 
\left(\frac{ k^2}{\mu^{4-2b} M^{2b}}\right)^{-(n+1)\epsilon}
\frac{\Gamma \left( n+1+(n+1) \epsilon -\frac{d}{2} \right)}{\Gamma \left( n+1+(n+1)\epsilon \right)} \left( M^2 \right)^{\frac{d}{2}-(1+\epsilon)(n+1)} \ .
\end{align}

To extract the $\epsilon\to 0$ asymptotics
we make use of the standard expansion
\be
\Gamma \left( -n +\epsilon \right) =
\frac{(-)^n}{\Gamma (n+1)} \left[ \frac{1}{\epsilon} -\gamma + h(n) \right] +O(\epsilon) \ ,
\ee
where $h(n) = \sum_{i=1}^n \frac{1}{i} $.
Furthermore, we need to parameterize the 
possible dependence of $M^2$ on $\epsilon$.
Recalling that for vanishing $\epsilon$
also $R_k$ needs to vanish, i.e.~$M^2$
should reduce to $m^2$,
we can write
\bea
M^2 &=& m^2 + \epsilon\, m_1^2(k,m,\mu) + O\left(\epsilon^2\right) \ ,
\\
\partial_t M^2 &=& \beta_{m^2} \left(f_0 + \epsilon F_1(k,m,\mu) \right)+ O\left(\epsilon^2\right) \ .
\eea
Here $m_1^2$ and $F_1$ are two independent functions
and $f_0$ is a proportionality factor. 
Thus, we  allow for the possibility that $\lim_{\epsilon \to 0} \partial_t M^2 \ne \partial_t \lim_{\epsilon \to 0} M^2$,
which can be achieved e.g.~by means of the choice
\be
M^2 = \left( 1+(f_0-1) \int_{\epsilon^2}^{\epsilon^2 \frac{m^2}{\mu^2}} ds \Gamma(s) \right) \,m^2
+\epsilon \,m_1^2 \ .
\ee

The need for this behavior of $M^2$ can be appreciated 
by inspecting the dimensionful integrals $l^d_{n,0}$ with
$n\leq d/2-1$:
\begin{align}
&\left.\frac{2\, l^{d}_{n,0}}{nk^{2n-d}}\right|_{n\leq\frac{d}{2}-1}
=
\frac{\beta_{m^2}\left( (n+1) f_0-n\right) \left( -m^2\right)^{\frac{d}{2}-n-1}}{ n \Gamma (n+2) \Gamma \left(\frac{d}{2}-n\right) \epsilon}
+ \frac{2 \left(1-\frac{b\, f_0\, \beta_{m^2} }{2 m^2}\right) \left(-m^2  \right)^{\frac{d}{2}-n}}{ \Gamma (n+1) \Gamma \left(\frac{d}{2}-n+1\right)}
\nonumber\\
&+
\frac{\beta_{m^2} \left( - m^2\right)^{\frac{d}{2}-n-1} \left\{ n(1-f_0) \left[ \log \left( \frac{k^2}{\mu ^{4-2 b} m^{2 b-2}}\right)+h(n)-
   h\left(\frac{d}{2}-n-1 \right)\right]+f_0+ F_1(k)\right \}}{n \Gamma (n+1) \Gamma \left(\frac{d}{2}-n\right)}
\nonumber\\
& - \frac{\beta_{m^2} m_1^2 \left(- m^2\right)^{\frac{d}{2}-n-2} ( (n+1)f_0 -n) }{ n \Gamma (n+2) \Gamma \left(\frac{d}{2}-n-1\right)}
+O(\epsilon) \ .
\end{align}
These exhibit a $1/\epsilon$ pole which can be eliminated by
tuning $f_0\neq1$. To fulfill this, as well as
the condition of removing the renormalization scale $k$
from the beta functions, we set
\bea
f_0 &=& \frac{n}{n+1} \ ,
\label{eq:thef0}
\\
F_1 &=&
f_1+\frac{n \left( h \left(\frac{d}{2}-n-1 \right)- h(n)-1 -\log \left(\frac{ k^2 }{ \mu ^{4-2 b} m^{2 b-2}}\right)\right)}{n+1} \ ,
\\
m_1^2 &=&
\left[ f_1+\frac{n \left( h \left(\frac{d}{2}-n-1 \right)- h(n)-1 -\log \left(\frac{ k }{\mu}\right) \right)}{n+1} \right] \beta_{m^2} \log \frac{k}{\mu}
\nonumber\\
&& \hspace{2cm}+(b-1) \frac{n}{2(n+1)} m^2 \left( \log \frac{m^2}{\mu^2} \right)^2 +O(\lambda^2) \ .
\label{eq:them1}
\eea
As a result we have
\begin{align}
&\left.\frac{2\, l^{d}_{n,0}}{nk^{2n-d}}\right|_{n\leq\frac{d}{2}-1}
=
\frac{2  \left(1-\frac{b\,n\, \beta_{m^2} }{2 (n+1)m^2}\right) \left(-m^2\right)^{\frac{d}{2}-n}}{\Gamma (n+1) \Gamma \left(\frac{d}{2}-n+1\right)}
+ f_1
\frac{\beta_{m^2} \left( -  m^2\right)^{\frac{d}{2}-n-1} }{n \Gamma (n+1) \Gamma \left(\frac{d}{2}-n\right)} +O(\epsilon)
\nonumber\\
&=
\frac{2 \left( -m^2 \right)^{\frac{d}{2}-n}}{\Gamma (n+1) \Gamma \left(\frac{d}{2}-n+1\right)} -
\frac{\beta_{m^2}}{m^2} \frac{ \left(- m^2 \right)^{\frac{d}{2}-n}}{\Gamma (n+2) \Gamma \left(\frac{d}{2}-n\right)}
\left[ \frac{b\,n}{\frac{d}{2}-n}+ f_1\frac{n+1}{n}
\right]
+O(\epsilon) \ .
\end{align}
Recall that $M^2$ must be analytic around $m^2=0$.
From \Eqref{eq:them1} we see that this can be achieved only if $b=1$.
   
On the other hand, the remaining loop integrals 
are harmless, as they read
\bea
\left.\frac{2\, l^{d}_{n,0}}{nk^{2n-d}}\right|_{n\ge \frac{d}{2}-1}&=&
(f_0-1)\frac{\beta_{m^2}}{m^2}
\frac{ \Gamma \left(n-\frac{d}{2}+1\right) }{ \Gamma (n+1)}\left( m^2\right)^{\frac{d}{2}- n} + O(\epsilon) \ , \\
\left.\frac{2\, l^{d}_{n,0}}{nk^{2n-d}}\right|_{n=\frac{d}{2}}&=&
\frac{2}{\Gamma \left(\frac{d}{2}+1\right)} 
\left[ 1- ((b-1) f_0 +1)\frac{ \beta_{m^2}}{2m^2}
\right] + O(\epsilon) \ .
\eea

For completeness we list some of these integrals in the
lowest even numbers of dimensions.
If $d=2$ there is no divergent $l$-function, and in particular
\begin{subequations}
\bea
l_{1,0}^2(0) &=&
1- \frac{1}{2 } \frac{\beta_{m^2}}{m^2} \ ,
\\
l_{n>1,0}^2(0) &=&
\frac{\beta_{m^2}}{m^2}
\frac{ ( f_0-1)  }{2 }\left( \frac{k^2}{m^2}\right)^{n-1} \ .
\eea
\end{subequations}

If $d=4$, the function $l_{1,0}^4$ which enters
in the determination of $\beta_{m^2}$
has a pole unless we choose $f_0 = 1/2$ according
to \Eqref{eq:thef0}. This results in:
\begin{subequations}
\bea
l_{1,0}^4(0) &=& - \frac{m^2}{k^2} +
\left( 1 + 2f_1\right) \frac{\beta_{m^2}}{4 k^2} \ ,
\\
l_{2,0}^4(0) &=&
1- \frac{\beta_{m^2}}{2m^2} \ ,
\\
l_{n>2,0}^4(0) &=&
-\frac{ 1 }{ 4 (n-1)}
\left(\frac{k^2}{ m^2}\right)^{n-2} \frac{\beta_{m^2}}{m^2} \ .
\eea
\end{subequations}
These equations are easily interpreted by 
applying them e.g.~to a $\lambda\phi^4$ theory
within the LPA.
To zeroth order in $\lambda$, i.e. neglecting
$\beta_{m^2}$ on the r.h.s., 
we recover the standard result that
integrals with negative mass dimension
do not contribute to the 
one-loop beta functions.
Moreover the positive dimensional integral
leads to the usual one-loop RG equation 
for the mass:
\be
\beta_{m^2} = \frac{N+2}{ 16\pi^2 } \lambda m^2 + O \left( \lambda^2 \right) \ .
\ee
Further details of the pseudo-regulator choice,
such as the coefficient $f_1$,
would affect higher perturbative
orders. In fact, in Appendix~\ref{app:2loopsbetam}
we show that the latter coefficient
is fixed by requiring that $\beta_{m^2}$
agrees with the \MSbar result also at two loops.

\newpage

\section{Two-loop computation}
\label{app:2loops}

In this appendix we detail the computation of the universal part of the two-loop
beta function in $\phi^4$ theory in four dimensions.

According to our priority, i.e.~the computation of $\beta_\lambda$ 
at order $\lambda^3$,
we first focus on the flow equation for the effective potential,
\begin{subequations}
\be
\partial_t U_k(\rho) =
\frac{1}{2} \int \frac{\dd^d q}{(2\pi)^d} \partial_t R_k(q) \left( \frac{N-1}{M_0 \left( \rho, q^2 \right) } +\frac{1}{M_1 \left( \rho, q^2 \right) }  \right) \ ,
\label{eq:betaU}
\ee
where
\bea
M_0 \left( \rho, q^2 \right) &=&
Z_k \left( \rho, q^2 \right)q^2+R_k(q)+U'_k(\rho) \ ,\\
M_1 \left( \rho, q^2 \right) &=&
\tilde Z_k \left( \rho, q^2 \right)q^2+R_k(q)+U'_k(\rho)+2 \rho U''_k(\rho) \ .
\eea
\end{subequations}
From this functional equation, the beta functions
of the mass and of the quartic coupling can be
derived by differentiation with respect to\ $\rho$.
Defining $\rho_0$ as the field expansion point and 
\begin{subequations}
\begin{align}
w_0 &= 2 \rho_0 \, U''_k(\rho_0) \ ,\\
P &= Z_k \left( \rho_0, z \right) z +R_k(z) + U'_k(\rho_0 ) \ ,\\
\tilde P &= \tilde Z_k \left( \rho_0, z \right) z +R_k(z) + U'_k(\rho_0) \ ,
\end{align}
\end{subequations}
the flow equations for the two renormalizable couplings read
\begin{align}
&\frac{d}{dt} U_k'(\rho_0) =
\partial_t U'_k(\rho_0) + U''_k(\rho_0) \frac{d}{dt} \rho_0
\nonumber\\
&= -2 v_d (N-1) k^{d-2} Z_k^{-1} U''_k(\rho_0) l_{1,0}^d(0) -2 v_d (N-1) k^d Z_k^{-1}
\langle Z_k'(\rho_0) \rangle_{1,0}^{d+2}(0) ,
\nonumber\\
& -2 v_d k^{d-2} Z_k^{-1} \left( 3 U''_k(\rho_0) + 2 \rho_0  U'''_k(\rho_0) \right) l_{0,1}^d( w_0)
-2 v_d k^d Z_k^{-1} \langle \tilde Z_k'(\rho_0) \rangle_{0,1}^{d+2}( w_0) 
\nonumber\\
& + U''_k(\rho_0) \frac{d}{dt} \rho_0 \ ,
\label{eq:dtU'}
\\
&\frac{d}{dt} U''_k(\rho_0) =
\partial_t U''_k(\rho_0) + U'''_k (\rho_0) \frac{d}{dt} \rho_0
\nonumber\\
&= 2v_d (N-1) k^{d-4} Z_k^{-2} \left( U''_k(\rho_0) \right)^2 l_{2,0}^d(0)
+2 v_d k^{d-4} Z_k^{-2} \left( 3 U''_k(\rho_0) + 2\rho_0 U'''_k(\rho_0) \right)^2 l_{0,2}^d( w_0)
\nonumber\\
& +4 v_d (N-1) k^{d-2} Z_k^{-2} U''_k(\rho_0) \langle Z_k'(\rho_0)\rangle_{2,0}^{d+2}(0)
\nonumber\\
& +4 v_d k^{d-2} Z_k^{-2} \left( 3 U''_k(\rho_0) + 2\rho_0 U'''_k(\rho_0) \right) \langle \tilde Z_k'(\rho_0)\rangle_{0,2}^{d+2}( w_0)
\nonumber\\
& +2 v_d(N-1) k^d Z_k^{-2} \langle  Z'_k(\rho_0) ^2 \rangle_{2,0}^{d+4}(0)
+2v_d  k^d Z_k^{-2} \langle \tilde Z'_k(\rho_0)^2 \rangle_{0,2}^{d+4}( w_0)
\nonumber\\
& -2 v_d (N-1) k^{d-2} Z_k^{-1} U'''_k(\rho_0) l_{1,0}^d(0)
-2 v_d k^{d-2} Z_k^{-1} \left( 5 U'''_k (\rho_0)+2\rho_0 U_k^{(4)}(\rho_0) \right) l_{0,1}^d( w_0)
\nonumber\\
&-2 v_d (N-1) k^d Z_k^{-1} \langle Z''_k(\rho_0) \rangle_{1,0}^{d+2}(0)
-2 v_d k^d Z_k^{-1} \langle \tilde Z_k''(\rho_0) \rangle_{0,1}^{d+2}( w_0)
\nonumber\\
& + U'''_k(\rho_0) \frac{d}{dt} \rho_0 \ .
\label{eq:dtU''}
\end{align}
Here we adopted standard notations for the loop integrals
\begin{subequations}
\begin{align}
l_{n_1,n_2}^d (w) &= - \frac{Z_k^{n_1+n_2}}{2} k^{2(n_1+n_2)-d}
\int_0^\infty dz \, z^{\frac{d}{2}-1}\partial_t \left\{ P(z) ^{-n_1}( \tilde P(z)+w)^{-n_2} \right\} \ ,\\
\langle D_k(\rho_0)  \rangle_{n_1,n_2}^{d}(w) &= - \frac{Z_k^{n_1+n_2}}{2} k^{2(n_1+n_2)-d}
\int_0^\infty dz \, z^{\frac{d}{2}-1} D_k(\rho_0, z) \partial_t \left\{ P(z)^{-n_1}( \tilde P(z)+w)^{-n_2} \right\} \ ,
\end{align}
\label{eq:thresholds&averages}
\end{subequations}
and $v_d^{-1}= 2(4\pi)^{d/2}\Gamma(d/2)$.
Notice however that our convention for the inverse propagators $P_k$
and $\tilde{P}_k$ slightly departs from the most common choice~\cite{Papenbrock:1994kf}, in that we include the
mass parameter $\bar{m}^2$ therein.
Furthermore, while $\rho_0$ is usually chosen 
as the running minimum of the potential,
such that $\rho_0>0$ corresponds to
a regime of spontaneous symmetry breaking,
we instead assume that $U_k'(\rho_0)>0$.
We can safely choose $\rho_0=0$ for our goals,
as no dynamical symmetry breaking is within reach
of a two-loop computation in the present model.

Eqs.~(\ref{eq:dtU'}) and (\ref{eq:dtU''}) can be rewritten as
\bea
\left( \beta_{m^2} -\eta m^2 \right) k^{-2} &=& 
-2 v_d (N-1) \left( \lambda\, l_{1,0}^d(0) +\langle z_1 \rangle_{1,0}^{d+2}(0)  \right)
\nonumber\\
&& 
-2 v_d  \left( 3 \lambda + 2 \kappa  u_3  \right) l_{0,1}^d \!\left( 2 \lambda \kappa \right)
-2v_d \,\langle \tilde z_1 \rangle_{0,1}^{d+2}   \left( 2 \lambda \kappa \right)
\nonumber\\
&& +\lambda \left( \left( d-2+\eta \right) \kappa +  \partial_t \kappa \right) \ ,
\label{eq:betamcomplete}
\\
\beta_\lambda &=& \left( d-4+2 \eta \right) \lambda
+2v_d (N-1) \lambda^2 l_{2,0}^d(0)
+2 v_d  \left( 3 \lambda +2\kappa u_3 \right)^2 l_{0,2}^d \! \left( 2 \lambda \kappa \right)
\nonumber\\
&& -2 v_d (N-1)  u_3 l_{1,0}^d(0) -2 v_d \left( 5 u_3 +2 \kappa u_4 \right) l_{0,1}^d \! \left( 2 \lambda \kappa \right)
\nonumber\\
&& +4 v_d (N-1) \lambda \langle z_1\rangle^{d+2}_{2,0}(0) +4 v_d \left( 3 \lambda + 2 \kappa u_3 \right)
\langle \tilde z_1\rangle^{d+2}_{0,2}\left( 2 \lambda \kappa \right)
\nonumber\\
&& -2 v_d (N-1) \langle z_2\rangle^{d+2}_{1,0}(0) -2 v_d \langle \tilde z_2 \rangle^{d+2}_{0,1}\left( 2 \lambda \kappa \right)
\nonumber\\
&& +2 v_d (N-1) \langle z_1^2\rangle^{d+4}_{2,0}(0) 
+2 v_d \langle \tilde z_1^2\rangle^{d+4}_{0,2}\left( 2 \lambda \kappa \right)
\nonumber\\
&& + u_3 \left( (d-2+\eta)\kappa +\partial_t \kappa \right) \ ,
\label{eq:betalambdacomplete}
\eea
where $\eta = -\partial_t \log Z_k$ is the field anomalous
dimension.
As described in the main text,
introducing the powercounting of \Eqref{eq:powercounting},
which is generated by the flow equation itself,
into Eqs.~(\ref{eq:betamcomplete}) and (\ref{eq:betalambdacomplete}),
and truncating them to order $\lambda^3$,
result in the simplified perturbative Eqs.~
(\ref{eq:FRGbetamtorderlambda3}) and (\ref{eq:FRGbetaatorderlambda3}) for $d=4$.
In the following we address  the $O(\lambda^3)$ contributions
arising on the r.h.s.~of \Eqref{eq:FRGbetaatorderlambda3},
organizing them line by line, as these also correspond to
different kinds of corrections.

\subsection{\texorpdfstring{$\mathbf{\boldsymbol{\beta}_{m^2}}$}{TEXT} contribution}

Using the previous pseudo-regulator and the one-loop beta function for $m^2$
the threshold functions can be expanded at leading order in $\lambda$,
as in \Eqref{eq:lsatorderlambda}.
Inserting this into the beta function (\ref{eq:FRGbetaatorderlambda3})
we get
\bea
\beta_\lambda &=&  \frac{N+8}{16\pi^2} \lambda^2- \frac{(N+8)(N+2)}{2(16\pi^2)^2} \lambda^3
\nonumber\\
&& -\frac{N-1}{16\pi^2}l^4_{1,0}(0)  u_3-\frac{5}{16\pi^2}l^4_{0,1} (2\lambda \kappa) u_3+2 \kappa  u_3
\nonumber\\
&& +\frac{N-1}{8\pi^2} \lambda \langle z_1 \rangle^6_{2,0} (0) +\frac{3}{8\pi^2} \lambda \langle \tilde z_1 \rangle^6_{0,2}
\left( 2 \lambda \kappa \right)
 - \frac{N-1}{16\pi^2} \langle z_2 \rangle^6_{1,0}(0)
 -\frac{1}{16\pi^2} \langle \tilde z_2 \rangle^6_{0,1} \left( 2 \lambda \kappa \right)
\nonumber\\
&& +2 \eta \lambda \ .
\label{eq:betalambdastep1}
\eea

\subsection{\texorpdfstring{$\mathbf{u_3}$}{TEXT} contribution}

To evaluate the contribution of the sextic coupling
generated by the flow equation, it is enough to
consider a uniform and field-independent wave function renormalization 
for all modes, as in the LPA${}^\prime$; that is, it is safe to
set $Z_k \left( \rho, q^2 \right) = \tilde Z_k \left( \rho, q^2 \right) = Z_k $ at order $O(\lambda^3)$.
The  flow of the sextic coupling can be deduced by
taking the third derivative of \Eqref{eq:betaU}
\begin{align}
\partial_t U'''_k(\rho) &=
+4v_d (N-1) k^{d-4} Z_k^{-2}  U''_k(\rho) U'''_k(\rho)  l_{2,0}^d(0)
-4v_d (N-1) k^{d-6} Z_k^{-3} \left( U''_k(\rho) \right)^3 l_{3,0}^d(0)
\nonumber\\
& +4 v_d k^{d-4} Z_k^{-2} \left( 3 U''_k(\rho) +2\rho U'''_k(\rho) \right) \left( 5 U'''_k(\rho) +2\rho U^{(4)}_k(\rho) \right) l_{0,2}^d(w)
\nonumber\\
&-4 v_d k^{d-6} Z_k^{-3} \left( 3 U''_k(\rho) +2\rho U'''_k(\rho) \right)^3 l_{0,3}^d(w)
\nonumber\\
& -2 v_d (N-1) k^{d-2} Z_k^{-1} U^{(4)}_k(\rho) l_{1,0}^d(0)
+2 v_d (N-1) k^{d-4} Z_k^{-2} U'''_k(\rho) U''_k(\rho) l_{2,0}^d(0)
\nonumber\\
& -2 v_d k^{d-2} Z_k^{-1} \left( 7 U^{(4)}_k (\rho)+2\rho U_k^{(5)}(\rho) \right) l_{0,1}^d(w)
\nonumber\\
& +2 v_d k^{d-4} Z_k^{-2} \left( 5 U'''_k (\rho)+2\rho U_k^{(4)}(\rho) \right)
\left( 3 U''_k (\rho)+2\rho U_k'''(\rho) \right) l_{0,2}^d(w)
\end{align}
and evaluating it at $\rho=\rho_0$,
such that $w\to w_0$.
Using the fact that 
\be
\partial_t u_3 = (2d-6) u_3 +
Z_k^{-3} \left( \partial_t U'''(\rho_0)+ U^{(4)}(\rho_0)\frac{d}{dt}\rho_0 \right) \ ,
\ee
one deduces
\bea
\partial_t u_3 &=& (2d-6) u_3
-4v_d (N+26) \lambda^3 l_{3,0}^d(0)
+O\left( \lambda^4 \right) \ .
\eea

At one loop and for $d=4$, $u_3$ is given
by the fixed-point solution of the previous equation
\be
u_3^{(1)} = \frac{N+26}{16\pi^2} \lambda^3 k^2 Z_k^3\int_0^\infty dz\, \frac{z}{P^3} \ .
\ee
By evaluating the momentum integral with the previous pseudo-regulator we find an expression
which is finite in the $\epsilon\to0$ limit, namely 
\Eqref{eq:u3FP} 
in the main text.~
Now let's compute $\kappa$ at one loop, by looking for a scaling solution for it, i.e.~by solving $\partial_t \kappa =0$, which gives
\be
\beta_{m^2}  k^{-2} =
-2 v_d (N-1) l_{1,0}^d(0) \lambda 
-2 v_d  \left( 3 \lambda + 2 \kappa  u_3  \right) l_{0,1}^d \left( 2 \lambda \kappa \right)
 +\lambda \left( d-2 \right) \kappa \ .
\ee
Specifying $d=4$ and using the previous identities we get
\be
\frac{N+2}{16\pi^2} \lambda m^2  k^{-2} =
\frac{(N-1)}{16\pi^2} m^2 k^{-2} \lambda 
+\frac{1}{16 \pi^2} 3 \lambda m^2 k^{-2} +2 \lambda  \kappa 
+O(\lambda^2) \ .
\ee
So as anticipated in the main text, $\kappa = O(\lambda)$ 
and as such would not affect the $O(\lambda^3)$ of $\beta_\lambda$.
Inserting this result for $u_3$ into the beta function 
(\ref{eq:betalambdastep1}) we obtain
\bea
\beta_\lambda &=& \frac{N+8}{16\pi^2} \lambda^2 + \frac{2(5N+22)}{(16\pi^2)^2} \lambda^3
\nonumber\\
&& +\frac{N-1}{8\pi^2} \lambda \langle z_1 \rangle^6_{2,0} (0) +\frac{3}{8\pi^2} \lambda \langle \tilde z_1 \rangle^6_{0,2}\left( 2 \lambda \kappa \right)
 - \frac{N-1}{16\pi^2} \langle z_2 \rangle^6_{1,0}(0) 
 -\frac{1}{16\pi^2} \langle \tilde z_2 \rangle^6_{0,1} \left( 2 \lambda \kappa \right)\nonumber\\
&& +2 \eta \lambda \ . 
\label{eq:betalambdastep2}
\eea

\subsection{Wave-function renormalization contribution}
\label{sec:Zcontrib2loops}

Recalling that $\kappa$ and $Z_k$ can be neglected in the third
line of \Eqref{eq:FRGbetaatorderlambda3}, as 
they would give 
higher order corrections, the wave-function renormalization contribution is encoded
in the following averages 
\begin{subequations}
\bea
\langle z_1\rangle^6_{2,0}(0) &=& 
16\pi^2  \int \frac{\dd^4p}{(2\pi)^4} p^2 Z_k'(0,p^2)  \frac{\partial_t R_k(p^2)}{P(p^2)^3} \ ,
\\
\langle z_2\rangle^6_{1,0}(0)  &=& 
8\pi^2  \int \frac{\dd^4p}{(2\pi)^4} p^2 Z_k''(0,p^2)  \frac{\partial_t R_k(p^2)}{P(p^2)^2} \ ,
\eea
 \end{subequations}
and similar relations for $\tilde z_1$ and $\tilde z_2$.
Here we should input the momentum dependence of
the wave function renormalization as generated
at one loop, that is
\begin{subequations}
\bea
Z_k'(\rho_0,p^2) &=&
-4 \lambda^2 \, 
\frac{I_k (p^2)}{p^2} \ ,
\label{eq:oneloopZ'}\\
\tilde Z_k'(\rho_0,p^2) &=&
-2(N+8) \lambda^2 \, 
\frac{I_k (p^2)}{p^2} \ ,
\label{eq:onelooptildeZ'}\\
Z_k''(\rho_0,p^2) &=& 
32 \lambda^3 \, 
\frac{J_k (p^2)}{p^2} \ ,\\
\tilde Z_k''(\rho_0,p^2) &=&
8(N+26)\lambda^3 \, 
\frac{J_k (p^2)}{p^2} \ ,
\eea
\end{subequations}
where $I_k$ and $J_k$ are the
following  one-loop integrals
\begin{subequations}
\bea
I_k (p^2) &=&
\frac{1}{2} \int \frac{\dd^4q}{(2\pi)^4} \frac{1}{P_k(q)} \left( \frac{1}{P_k(q+p)} - \frac{1}{P_k(q)} \right) \ ,\\
J_k (p^2) &=&
\frac{1}{2} \int \frac{\dd^4q}{(2\pi)^4} \frac{1}{P_k(q)^2} \left( \frac{1}{P_k(q+p)} - \frac{1}{P_k(q)} \right) \ .
\eea
\end{subequations}
Nesting these expressions leads to
\Eqref{eq:z12intermsofAB},
where the averages of $z_1$, $\tilde z_1$ and 
$z_2$, $\tilde z_2$
are respectively proportional to
the dimensionless two-loop integrals
\begin{subequations}
\bea
A &=& 
\frac{1}{2}
\int \frac{\dd^4 p}{(2\pi)^4} I_k(p^2)\frac{\partial_t R_k(p)}{P_k(p)^3} \ ,
\label{eq:theA}\\
B &=& 
 \frac{1}{2}
\int \frac{\dd^4 p}{(2\pi)^4} J_k(p^2)\frac{\partial_t R_k(p)}{P_k(p)^2} \ .
\label{eq:theB}
\eea
\end{subequations}

We first compute $I_k (p^2)$ with the pseudo-regulator
(\ref{eq:massdepReg}):
\bea
&& I_k (p^2,\epsilon) =
\frac{1}{2} \int \frac{\dd^4q}{(2\pi)^4} \frac{1}{P_k(q)} \left( \frac{1}{P_k(q+p)} - \frac{1}{P_k(q)} \right)
\nonumber\\
&& = \frac{1}{2} 
\left(\frac{\mu^2 m^2}{k^2}\right)^{2\epsilon}
\int \frac{\dd^4q}{(2\pi)^4} \frac{1}{\left( q^2 +m^2 \right)^{1+\epsilon}\left( (q+p)^2 +m^2 \right)^{1+\epsilon}}-(p \to 0 ) 
\nonumber\\
&& = \frac{1}{2} 
\left(\frac{\mu^2 m^2}{k^2}\right)^{2\epsilon}
\frac{\Gamma(2+2\epsilon)}{\Gamma(1+\epsilon)^2} \int \frac{\dd^4q}{(2\pi)^4} \int_0^1 dx 
\frac{x^{\epsilon}(1-x)^{\epsilon}}{\left( x\, q^2 +(1-x)(q+p)^2 +m^2 \right)^{2+2\epsilon}}-(p \to 0) 
\nonumber\\
&& = \frac{1}{2} 
\left(\frac{\mu^2 m^2}{k^2}\right)^{2\epsilon}
\frac{\Gamma(2+2\epsilon)}{\Gamma(1+\epsilon)^2} \int_0^1 dx\int \frac{\dd^4q}{(2\pi)^4}  
\frac{x^{\epsilon}(1-x)^{\epsilon}}{\left(  q^2 +x(1-x)p^2 + m^2 \right)^{2+2\epsilon}}-(p \to 0) 
\nonumber\\
&& = \frac{1}{32\pi^2}  
\left(\frac{\mu^2}{ k^2}\right)^{2 \epsilon}
\frac{\Gamma(2\epsilon)}{\Gamma(1+\epsilon)^2} \int_0^1 \!\!dx \,
x^{\epsilon}(1-x)^{\epsilon}\left( 1+ x(1-x)\frac{p^2}{m^2} \right)^{-2\epsilon}- (p \to 0) \ .
\eea
Taking the limit for $\epsilon \to 0$ results in
the following finite expression
\bea
I_k (p^2) &=& 
-\frac{1}{32\pi^2}
\int_0^1 dx 
\log \left( 1+ x(1-x)\frac{p^2}{m^2} \right)
\nonumber\\
&=& 
\frac{1}{16\pi^2} 
\left[ 1- \sqrt{\frac{4 m^2+p^2}{p^2}}\atanh \left(\sqrt{\frac{p^2}{4 m^2+p^2}}\right) \right] \ .
\eea
We can then insert this result in the expression
(\ref{eq:theA}) for the  $A$ coefficient
\begin{align}
A &= \epsilon \left(\frac{\mu^2 m^2}{k^2}\right)^{2\epsilon} 
\int \frac{\dd^4 p}{(2\pi)^4}
\frac{ I_k(p^2,\epsilon) }{\left( p^2 +m^2 \right)^{2+2\epsilon}}
\nonumber\\
&= \frac{\epsilon}{(16 \pi^2 )^2} \left(\frac{\mu^2 m^2}{k^2}\right)^{4\epsilon}
\frac{\Gamma(2\epsilon)}{\Gamma(1+\epsilon)^2} \int_0^1 \!\!dx \,
x^{\epsilon}(1-x)^{\epsilon}
\int_0^{\infty} \!\! dp\,p^3 \frac{\left( m^2 +x(1-x) p^2 \right)^{-2\epsilon}}{\left( p^2 + m^2 \right)^{2+2\epsilon}}
\nonumber\\
&- \frac{\epsilon}{(16 \pi^2 )^2} \left(\frac{\mu^2 m^2}{k^2}\right)^{4\epsilon} 
\frac{\Gamma(2\epsilon)}{\Gamma(1+\epsilon)^2} \int_0^1 \!\!dx \,
x^{\epsilon}(1-x)^{\epsilon}
\int_0^{\infty} \!\! dp\,p^3 \frac{ m^{-4\epsilon} }{\left( p^2 +m^2 \right)^{2+2\epsilon}}
\nonumber\\
&= \frac{\epsilon}{(16 \pi^2 )^2} \left(\frac{ \mu^2 }{k^2}\right)^{4 \epsilon}
\frac{\Gamma(2\epsilon)}{\Gamma(1+\epsilon)^2} \int_0^1 \!\!dx \,
x^{\epsilon}(1-x)^{\epsilon}\frac{ 1}{4(1-x(1-x))}
\nonumber\\
& 
\hspace{1cm}\times  \left[
\frac{\sqrt{\pi } 16^{\epsilon } \Gamma \left(2 \epsilon +\frac{1}{2}\right) ((1-x) x)^{2 \epsilon } (1+2 \epsilon -(1-x) x (1-2 \epsilon )) }{\sin (2 \pi  \epsilon )\Gamma (2 \epsilon +2) ((1-x) x-1)^{4 \epsilon }}
   \right.
\nonumber\\
& \left.\hspace{1cm} -
\frac{2 (1-\epsilon )x(1-x) +(1+2 \epsilon -(1-x) x (1-2 \epsilon )) \, _2F_1\left(1,2+2 \epsilon ;3-2 \epsilon ;\frac{1}{x-x^2}\right)}{(1-x)^2 x^2 (1-2 \epsilon ) (1-\epsilon )}
\right]
\nonumber\\
& - \frac{\epsilon}{(16 \pi^2 )^2} \left(\frac{\mu^2 }{k^2}\right)^{4\epsilon}
\frac{\Gamma(2\epsilon)}{\Gamma(1+\epsilon)^2} \int_0^1 \!\!dx \,
x^{\epsilon}(1-x)^{\epsilon}
\frac{ 1}{4 \epsilon  (1+2 \epsilon )} \ .
\end{align}
If we first expand the integrand
around $\epsilon =0$ and then perform the integral over $x$
we find
\bea
A 
&=& \frac{1}{\left(16\pi^2\right)^2}
\left[ - \frac{1}{16 \epsilon}
+ \frac{3-2 \log \left( \frac{\mu^2}{k^2} \right)}{8} 
+O(\epsilon)
\right] \ .
\label{eq:Aresult}
\eea
Notice that the coefficient of the pole is equal to one fourth of the coefficient in front of $\log \left( \frac{\mu^2}{k^2 } \right)$.\\

To demonstrate that the $\epsilon\to0$ limit and the
$x$ integration do commute, let's compute the two also
in the opposite order. Thus, we first perform the integral over $x$ and then take $\epsilon \to 0$.
For notational convenience we split $A$ in four different terms
\begin{subequations}
\begin{align}
A =&\ a_1+a_2+a_3+a_4 \ ,\\
a_1 =&\ \frac{1}{(16 \pi^2 )^2} \left(\frac{\mu^2}{k^2}\right)^{4 \epsilon}
\frac{\epsilon\Gamma (-1+2 \epsilon )}{2\Gamma (\epsilon +1)^2}\int_0^1 dx 
\frac{ ((1-x) x)^{\epsilon -1}  }{ (1+x(1-x) ) } \ ,\\
a_2 =&\, - \frac{1}{(16 \pi^2 )^2} \left(\frac{\mu^2}{k^2}\right)^{4 \epsilon} 
\frac{\sqrt{\pi } 2^{4 \epsilon -3} \Gamma \left(\frac{1}{2}+2 \epsilon\right) }{\sin (2 \pi  \epsilon )(2 \epsilon +1) \Gamma (\epsilon+1)^2}\nonumber\\
&\times\int_0^1 dx
\frac{((1-x) x)^{3 \epsilon }  (1+2 \epsilon -x(1-x) (1-2 \epsilon )) }{\left(-1+x(1-x)\right)^{1+4 \epsilon }} \ ,\\
a_3 =& -\frac{1}{(16 \pi^2 )^2} \left(\frac{\mu^2}{k^2}\right)^{4 \epsilon} 
\frac{\pi \epsilon}{ 2 \sin (2 \pi  \epsilon ) \Gamma (3-2 \epsilon ) \Gamma (\epsilon +1)^2}\nonumber\\
& \times 
\int_0^1 dx
\frac{  ((1-x) x)^{\epsilon } (1+2 \epsilon -(1-x) x (1-2 \epsilon ))  \, _2F_1\left(1,2 (\epsilon +1);3-2 \epsilon ;\frac{1}{x-x^2}\right)}{ x^2 (1-x)^2 (1-x(1-x) )} \ ,\\
a_4 =& - \frac{1}{(16 \pi^2 )^2} \left(\frac{\mu^2}{k^2}\right)^{4\epsilon}
\frac{1}{8\epsilon  (1+2 \epsilon )^2 } \ .
\end{align}
\end{subequations}
By performing the  integrals over $x$
and then expanding them around $\epsilon =0$ 
they become
\begin{subequations}
\bea
a_1 &=& \frac{1}{(16\pi^2)^2 } \left[ -\frac{1}{2\epsilon} -1-\frac{\pi \sqrt{3}}{18 }-2 \log \left(\frac{\mu^2}{k^2}\right) +O(\epsilon) \right] \ ,
\\
a_2 &=& \frac{1}{(16\pi^2)^2 } \left[ \frac{1}{16\epsilon} - \frac{1}{8} -\frac{\pi}{36} \left(\sqrt{3}+9 i\right)  +\frac{1}{4} \log \left(\frac{\mu ^2}{k^2}\right) +O(\epsilon)  \right] \ ,
\\
a_3 &=& \frac{1}{(16\pi^2)^2 } \left[ \frac{1}{2\epsilon} +1+\frac{\pi}{12}\left(\sqrt{3}+3 i\right) + 2 \log \left(\frac{\mu ^2}{k^2}\right) +O(\epsilon) \right] \ ,
\\
a_4 &=& \frac{1}{(16\pi^2)^2 }  \left[ -\frac{1}{8 \epsilon}+\frac{1}{2}-\frac{1}{2} \log \left(\frac{\mu^2}{k^2}\right) +O(\epsilon)\right] \ .
\eea
\end{subequations}
Combining these results we recover \Eqref{eq:Aresult}.

Then we turn to the computation of $ J_k (p^2)$
\begin{align}
& J_k (p^2,\epsilon) =
\frac{1}{2} \int \frac{\dd^4q}{(2\pi)^4} \frac{1}{P_k(q)^2} \left( \frac{1}{P_k(q+p)} - \frac{1}{P_k(q)} \right)
\nonumber\\
&= \frac{1}{2} 
\left(\frac{\mu^2 m^2 }{ k^2}\right)^{3\epsilon}
\int \frac{\dd^4q}{(2\pi)^4} \frac{1}{\left( q^2 + m^2 \right)^{2+2\epsilon}\left( (q+p)^2 + m^2 \right)^{1+\epsilon}}-(p \to 0) 
\nonumber\\
& = \frac{1}{2} 
\!\left(\frac{\mu^2 m^2}{k^2}\right)^{2\epsilon} \!\!\!\!
\frac{\Gamma(3+3\epsilon)}{\Gamma(1+\epsilon)\Gamma(2+2\epsilon)} \! \int \!\! \frac{\dd^4q}{(2\pi)^4} \! \int_0^1 \!\!dx 
\frac{x^{2\epsilon+1}(1-x)^{\epsilon}}{\left( x\, q^2 +(1-x)(q+p)^2+ m^2 \right)^{3+3\epsilon}}-(p \to 0) \quad
\nonumber\\
& = \frac{1}{2} 
\! \left(\frac{\mu^2 m^2}{k^2}\right)^{3\epsilon} \!\!\!
\frac{\Gamma(3+3\epsilon)}{\Gamma(1+\epsilon) \Gamma(2+2\epsilon)} \int_0^1\! dx\int \frac{\dd^4q}{(2\pi)^4}  
\frac{x^{2\epsilon+1}(1-x)^{\epsilon}}{\left(  q^2 +x(1-x)p^2+ m^2 \right)^{3+3\epsilon}}-(p \to 0) 
\nonumber\\
& =\frac{1}{32\pi^2 m^2} 
\left(\frac{\mu^2}{ k^2}\right)^{3 \epsilon}\!\!\!
\frac{\Gamma(1+3\epsilon)}{\Gamma(1+\epsilon)\Gamma(1+2\epsilon)} \int_0^1 \!\!dx \,
\frac{x^{2\epsilon+1}(1-x)^{\epsilon}}{\left( 1+x(1-x)\frac{p^2}{m^2} \right)^{1+3\epsilon}}- (p \to 0) \ .
\end{align}
Taking the limit $\epsilon \to 0$ we again find a finite 
one-loop result
\begin{align}
J_k(p^2) &=  \frac{1}{32\pi^2 } 
 \int_0^1 dx 
\left[\frac{x}{ x(1-x)p^2+m^2 }- \frac{x}{m^2 } \right]
\nonumber\\
&= \frac{1}{64\pi^2 m^2} 
\left[
\frac{2m^2}{\sqrt{p^2(4 m^2+p^2)}}\log \left(1+\frac{ \left(\sqrt{p^2(4 m^2+p^2)}+p^2\right)}{ 2m^2}\right)
-1\right] \ ,
\end{align}
which enters the computation of the $B$ coefficient
through \Eqref{eq:theB}.
The latter proceeds along the same lines as for
$A$. Namely, we exchange again the $p$ and the $x$
integrals
\begin{align}
&B =  \epsilon \left(\frac{\mu^2 m^2}{k^2}\right)^{\epsilon} 
\int \frac{\dd^4 p}{(2\pi)^4}
\frac{ J_k (p^2,\epsilon) }{\left( p^2 +  m^2 \right)^{1+\epsilon}}
\nonumber\\
&= \frac{\epsilon}{(16\pi^2)^2 }\! \left(\frac{\mu^2 m^2}{ k^2}\right)^{4\epsilon}\!\!\!
\frac{\Gamma(1+3\epsilon)}{\Gamma(1+\epsilon)\Gamma(1+2\epsilon)} \int_0^1 \!\!dx \,
x^{2\epsilon+1}(1-x)^{\epsilon} \!\!\!
\int_0^{\infty} \!\!\!\!
dp\,p^3 \frac{\left(  m^2 +x(1-x) p^2 \right)^{-1-3\epsilon}}{\left( p^2 +  m^2 \right)^{1+\epsilon}} 
\nonumber\\
&
\ \ \ \ -\frac{\epsilon}{(16\pi^2)^2 } \left(\frac{\mu^2 m^2}{ k^2 }\right)^{4\epsilon} \!\!\!
\frac{\Gamma(1+3\epsilon)}{\Gamma(1+\epsilon)\Gamma(1+2\epsilon)} \int_0^1 \!\!dx \,
x^{2\epsilon+1}(1-x)^{\epsilon}
\int_0^{\infty} \!\!dp\,p^3 \frac{ m^{-2(1+3\epsilon)}}{\left( p^2 + m^2 \right)^{1+\epsilon}} 
\nonumber\\
&=
\frac{\epsilon}{(16\pi^2)^2 } \left(\frac{\mu^2}{ k^2}\right)^{4\epsilon}\!\!\!
\frac{\Gamma(1+3\epsilon)}{\Gamma(1+\epsilon)\Gamma(1+2\epsilon)} \int_0^1 \!\!dx \,
x^{2\epsilon+1}(1-x)^{\epsilon} 
\frac{1}{6 \epsilon (1-3 \epsilon) x^2(1-x)^2  }
\nonumber\\
& \hspace{1cm}
 \times\left( 
- \frac{\pi \epsilon \Gamma (4 \epsilon ) x (1-x)  (1+3x (1-x) )   \left(1-\frac{1}{x(1-x)}\right)^{-\epsilon }  }{ \sin (3 \pi  \epsilon )\Gamma (1+\epsilon ) \Gamma (-1+3 \epsilon )\left( x(1-x)-1\right)^{1+3 \epsilon }}
\right.
\nonumber\\
&
\left.\hspace{3cm}
+\frac{(1-3 \epsilon) x (1-x) + \epsilon (1+3x (1-x)  ) \, _2F_1\left[1,1+\epsilon,2-3 \epsilon;\frac{1}{x-x^2}\right] }{1-(1-x) x}
\right)
\nonumber\\
&\ \ \ \ -\frac{\epsilon}{(16\pi^2)^2 } \left(\frac{\mu^2 }{ k^2 }\right)^{4\epsilon}\!\!\!\!\!
\frac{\Gamma(1+3\epsilon)}{\Gamma(1+\epsilon)\Gamma(1+2\epsilon)} \int_0^1 \!\!dx \,
x^{2\epsilon+1}(1-x)^{\epsilon}
\frac{1}{2 (\epsilon -1) \epsilon } \ .
\end{align}
This time however we are not 
allowed to take the $\epsilon\to 0$ limit
before computing the $x$ integral.
In fact, this would result in
the wrong answer
\be
B
= \frac{1}{\left(16\pi^2\right)^2} \int_0^1 dx
\left[ \frac{1}{8 (1-x) } +\frac{x}{2  }
\right] \ .
\ee
In other words, the integral over $x$ does not commute with the  $\epsilon \to 0$ limit, and the latter must be taken
as the last step of the computation.

To perform the integral over $x$ of the $\epsilon$-depedent
expressions, we split also $B$ in four different contributions
\begin{subequations}
\begin{align}
B &= b_1+b_2+b_3+b_4 \ ,\\
b_1 &= \frac{1}{(16\pi^2)^2 } \left(\frac{\mu^2}{ k^2}\right)^{4\epsilon} 
\frac{\pi \Gamma (4 \epsilon )}{2 \sin (3 \pi  \epsilon ) \Gamma (\epsilon )^2 \Gamma (2 \epsilon +2)} 
\int_0^1 dx 
\frac{  (1+3 x(1-x) ) (1-x)^{2 \epsilon -1} x^{3 \epsilon }   }{\left( x(1-x)-1\right)^{1+4 \epsilon }} \ ,
\\
b_2 &= \frac{1}{(16\pi^2)^2 } \left(\frac{\mu^2}{ k^2}\right)^{4\epsilon} 
\frac{\Gamma (3 \epsilon )}{2\Gamma (\epsilon ) \Gamma (2 \epsilon +2)}\int_0^1 dx 
\frac{(1-x)^{\epsilon -1} x^{2 \epsilon } }{ (1-x(1-x) ) } \ ,
\\
b_3 &= \frac{1}{(16\pi^2)^2 } \left(\frac{\mu^2}{ k^2}\right)^{4\epsilon} 
\frac{\pi \epsilon}{2 \sin (3 \pi  \epsilon )\Gamma (2-3 \epsilon ) \Gamma (\epsilon ) \Gamma (2 \epsilon +2)}
\nonumber\\
&\quad\quad \times\int_0^1 dx 
\frac{  (1+3x (1-x) ) (1-x)^{-2+\epsilon } x^{2 \epsilon -1}  \, _2F_1\left(1,\epsilon +1;2-3 \epsilon ;\frac{1}{x-x^2}\right)}{(1-x(1-x) ) } \ ,
\\
b_4 &= \frac{1}{(16\pi^2)^2 } \left(\frac{\mu^2}{ k^2 }\right)^{4\epsilon}
\frac{(1+2\epsilon) }{2 (1-\epsilon ) (2+3\epsilon)(1+3\epsilon) } \ .
\end{align}
\end{subequations}
Now we compute the integrals over $x$
and then expand around $\epsilon =0$,
obtaining
\begin{subequations}
\bea
b_1 &=& \frac{1}{(16\pi^2)^2 } \left[ -\frac{1}{48\epsilon}+\frac{9+18 i\pi -4 \sqrt{3} \pi - 18 \log\left(  \frac{\mu^2}{k^2}\right)}{216} +O(\epsilon) \right] \ ,
\\
b_2 &=& \frac{1}{(16\pi^2)^2 } \left[ \frac{1}{6\epsilon}+\frac{-18+ \sqrt{3} \pi +36 \log\left(  \frac{\mu^2}{k^2 }\right)}{54}  +O(\epsilon) \right] \ ,
\\
b_3 &=& \frac{1}{(16\pi^2)^2 } \left[ -\frac{1}{12\epsilon}+\frac{18-9i \pi - 36 \log\left(  \frac{\mu^2}{k^2 }\right)}{108}  +O(\epsilon) \right] \ ,
\\
b_4 &=& \frac{1}{4(16\pi^2)^2 }  +O(\epsilon) \ .
\eea
\end{subequations}
The sum of these terms leads to
the result
\bea
B &=&
\frac{1}{\left(16\pi^2\right)^2}
\left[ \frac{1}{16 \epsilon}
+ \frac{1+2 \log \left( \frac{\mu^2}{k^2} \right)}{8} 
+O(\epsilon)
\right] \ .
\eea
Also in this case the coefficient of the pole is equal to one fourth of the coefficient in front of $\log \left( \frac{\mu^2}{k^2} \right)$.
As a consequence, the sum $A+B$ which determines the
wave-function renormalization contribution to 
the two-loop beta function is finite, as given
in  \Eqref{eq:AplusB},
and the third line of \Eqref{eq:FRGbetaatorderlambda3}
evaluates to
\bea
&& \frac{N-1}{8\pi^2} \lambda \langle z_1 \rangle^6_{2,0} (0) +\frac{3}{8\pi^2} \lambda \langle \tilde z_1 \rangle^6_{0,2}
\left( 2 \lambda \kappa \right)
 - \frac{N-1}{16\pi^2} \langle z_2 \rangle^6_{1,0}(0) 
 -\frac{1}{16\pi^2} \langle \tilde z_2 \rangle^6_{0,1} \left( 2 \lambda \kappa \right)
 \nonumber\\
&& =-8(5N+22)(A+B) \lambda^3
= -\frac{4(5N+22)}{\left(16\pi^2\right)^2}\lambda^3 \ .
\eea

\subsection{Anomalous dimension contribution}

Within the truncation accounting for a field dependent
wave function renormalization, we define the anomalous
dimension as
\be
\eta = - \frac{d}{dt}\log Z_k(\rho_0) 
= -Z_k^{-1}( \rho_0) \partial_t Z_k(\rho_0) - Z_k( \rho_0)^{-1} Z_k'(\rho_0)\, \frac{d}{dt} \rho_0 \ .
\label{eq:defetaZrho}
\ee
Possible differences between this definition and
a similar one based on $\tilde Z_k(\rho_0)$
are beyond the $O(\lambda^2)$ we are after.
Also, the second term on the r.h.s.~of 
\Eqref{eq:defetaZrho}
would not contribute at this perturbative order.
Hence, the relevant term which can be deduced
from the exact flow equation is
\begin{align}
&\partial_t Z_k( \rho_0)  =
\frac{1}{2} \lim_{Q^2 \to 0} \frac{\partial}{\partial Q^2}
\frac{\delta^2}{\delta \phi(Q) \delta \phi(-Q)}\Tr \left\{ \partial_t R_k(q) \left[ \Gamma_k^{(2)}(q) +R_k(q) \right]^{-1}\right\} \Big|_{ \rho_0}
\nonumber\\
 &=\frac{1}{2} \lim_{Q^2 \to 0} \frac{\partial}{\partial Q^2}
\Tr \left \{  \partial_t R_k(q)
\left( \Gamma_k^{(2)}(q) +R_k(q) \right)^{-1} 
 \left[-\Gamma^{(4)}_k(Q,-Q,q,-q)\left( \Gamma_k^{(2)}(q) +R_k(q) \right)^{-1}
\right. \right.
\nonumber\\
& \left. \left. + 2 \Gamma^{(3)}_k(Q,q,-q-Q)\!\left( \Gamma_k^{(2)}(q) +R_k(q) \right)^{-1} \!\Gamma^{(3)}_k(-Q,q+Q,-q)\!\left( \Gamma_k^{(2)}(q) +R_k(q) \right)^{-1}
\right]
\right \} \Big|_{ \rho_0} \ .
\label{eq:dtZrho}
\end{align}
As described in the main text we have that the anomalous dimension is given by the sum of two terms, $\eta^{(1)}$ and $\eta^{(2)}$: the first one  is the contribution at zero momentum, while in the second one is the momentum contribution.

The flow equation which encodes $\eta^{(1)}$
is the one within the $O(\partial^2)$
derivative expansion, that is:
\begin{align}
\partial_t Z_k(\rho) = &\
-2 v_d k^{d-2} Z_k^{-1}
\left\{\left[ (N-1)Z_k'(\rho)+ Y_k(\rho) \right] l_{1,0}^d(0)
+ \left[ Z_k'(\rho)+ 2 \rho Z''_k(\rho) \right] l_{0,1}^d(w)\right\}
\nonumber\\
& +4 v_d k^{d-6} \rho \left( U''_k(\rho) \right)^2 Q_{2,1}^{d,0}(w)
+4 v_d k^{d-4} \rho Y_k(\rho) U''_k(\rho)  Q_{2,1}^{d,1}(w)
\nonumber\\
& + v_d k^{d-2} \rho \left( Y_k(\rho) \right)^2 Q_{2,1}^{d,2}(w)
+16 v_d k^{d-4} Z_k^{-2} \rho Z'_k(\rho) U''_k(\rho) l_{1,1}^{d}(w)
\nonumber\\
& +\frac{8 v_d}{d} k^{d-2} Z_k^{-2} \rho \left( Z'_k(\rho) \right)^2 l_{1,1}^{d+2}(w)
+8 v_d k^{d-2} Z_k^{-2} \rho Z'_k(\rho) Y_k(\rho) l_{1,1}^{d+2}(w)
\nonumber\\
& +\frac{16 v_d}{d} k^{d-4} \rho Z'_k(\rho) U''_k(\rho) N_{2,1}^{d}(w)
+\frac{8 v_d}{d} k^{d-2} \rho Z'_k(\rho) Y_k(\rho) N_{2,1}^{d+2}(w) \ .
\label{eq:dtZrhoDE}
\end{align}
Following Ref.~\cite{Papenbrock:1994kf} we 
define the threshold functions
\begin{subequations}
\begin{align}
N_{n_1,n_2}^{d}(w) &= 
k^{2(n_1+n_2-1)-d} \int_0^\infty\! dz \, z^{\frac{d}{2}}\partial_t \left\{ \dot P \, P^{-n_1}( \tilde P+w)^{-n_2} \right\} \ ,
\\
Q_{n_1,n_2}^{d,\alpha}(w) &=
k^{2(n_1+n_2-\alpha)-d} \int_0^\infty\! dz \, z^{\frac{d}{2}-1+\alpha}\partial_t \left\{ 
\left[ \dot P + \frac{2z}{d} \ddot P- \frac{4 z}{d} P^{-1} \dot P^2 \right]
P^{-n_1}( \tilde P+w)^{-n_2} \right\} \ ,
\\
M_{n_1,n_2}^{d}(w) &=
k^{2(n_1+n_2-1)-d} \int_0^\infty\! dz \, z^{\frac{d}{2}}\partial_t \left\{
\dot P^2
P^{-n_1}( \tilde P+w)^{-n_2} \right\} \ .
\end{align}
\end{subequations}
These quantities are related in the following way
\begin{align}
Q_{n_1,n_2}^{d,\alpha}(w) =&\,
\frac{2 n_1 -4}{d} M_{n_1+1,n_2}^{d+2\alpha}(w)
+ \frac{2 n_2}{d} M_{n_1,n_2+1}^{d+2\alpha}(w)
\nonumber\\
&+ \frac{2 n_2}{d} \rho Y_k(\rho) N_{n_1,n_2+1}^{d+2\alpha}(w)
- \frac{2 \alpha}{d} N_{n_1,n_2}^{d+2\alpha-2}(w) \ .
\end{align}
Taking $\rho\to\rho_0$ and $w\to w_0$ in \Eqref{eq:dtZrhoDE}
we get the simplified expression
\be
\partial_t Z_k(\rho_0) =
\frac{8}{d} v_d k^{d-6} \bar \lambda^2 \rho_0 M_{4,0}^d(0)
-2v_d k^{d-2} Z_k^{-1} \left( N Z_k'(\rho_0,0) +Y_k(\rho_0,0) \right) l_{1,0}^d(0) \ ,
\ee
which in $d=4$ can be rewritten
\begin{align}
\eta^{(1)} &=
\frac{1}{8\pi^2} m^4_4 \,\kappa \lambda^2 + \frac{Z_k^{-2} k^2}{16\pi^2} l_{1,0}^4(0) \left[ (N-1) Z_k'(\rho_0,0)+ \tilde Z_k'(\rho_0,0) \right]
- Z_k^{-1} Z_k'(\rho_0,0) \frac{d}{dt}\rho_0 \ ,
\\
m_n^d &=
-\frac{Z_k^{n-2}}{2}  M_{n,0}^d(0) \ .
\end{align}
As at the present order and with our pseudo-regulator
both $\kappa$ and $m^4_4 $ vanish,
we are left with \Eqref{eq:preeta1}.
On the other hand, the derivative couplings generated
at one loop are
\begin{subequations}
\bea
Z_k'( \rho_0,0) &=&
-4 \lambda^2 Z_k^4 \lim_{p^2 \to 0} \frac{I_k (p^2)}{p^2} = \frac{1}{3(16\pi^2)} Z_k^{2} m^{-2} \lambda^2 \ ,
\\
\tilde Z_k'(\rho_0, 0 ) &=&
-2(N+8) \lambda^2 Z_k^4 \lim_{p^2 \to 0} \frac{I_k (p^2)}{p^2} = \frac{(N+8)}{6(16\pi^2)} Z_k^{2} m^{-2} \lambda^2 \ ,
\eea
\end{subequations}
which leads to \Eqref{eq:z1sat1loop}.
Nesting the latter in \Eqref{eq:preeta1}
results in the final expression (\ref{eq:theeta1})
for $\eta^{(1)}$.

We then turn to the momentum dependent contribution.
As in Ref.~\cite{Papenbrock:1994kf} we 
define the latter by subtracting
the momentum independent part from the
four-point vertex:
\begin{align}
&\Delta_k (Q,-Q,q,-q) =
\Gamma^{(4)}_k(Q,-Q,q,-q)
\!-\!\Gamma^{(4)}_k(0,0,q,-q)
\!-\!\Gamma^{(4)}_k(Q,-Q,0,0)
\!-\!\Gamma^{(4)}_k(0,0,0,0)
\nonumber\\
&= -\, \lambda^2 \mbox{diag}\left( 2, N+8, \overbrace{2 \ldots 2}^{N-2} \right) 
\frac{1}{2} \int \frac{\dd^4p}{(2\pi)^4} P^{-1}(p) 
\left[ 2 P^{-1}(p) + P^{-1}(p-Q-q)\right.
\nonumber\\
&
\hspace{5cm}\left. +\, P^{-1}(p-Q+q) -2 P^{-1}(p+Q) -2 P^{-1}(p+q)
\right] \ .
\end{align}
For a $\phi^4$ theory at one loop
\begin{align}
&\lim_{Q^2 \to 0} \frac{\partial}{\partial Q^2} \Delta_k (Q,-Q,q,-q) 
= -\lambda^2 \mbox{diag}\left( 2, N+8, \overbrace{2 \ldots 2}^{N-2} \right) 
\nonumber\\
&
\times \frac{1}{2} \lim_{Q^2 \to 0} \frac{\partial}{\partial Q^2} \int \frac{\dd^4p}{(2\pi)^4} P^{-1}(p) \left[ P^{-1}(p-Q-q) + P^{-1}(p-Q+q) -2 P^{-1}(p+Q) 
\right] \ .
\end{align}
To evaluate this expression it is convenient
to define
\be
H(p^2,Q^2) = P^{-1}(p) P^{-1}(p+Q) =
\left( \frac{\mu^2 m^2}{k^2} \right)^{2\epsilon} \frac{\Gamma(2+2\epsilon)}{\Gamma(1+\epsilon)^2}
\int_0^1\! dx\,\frac{x^\epsilon (1-x)^\epsilon }{\left( p^2 +x(1-x) \,Q^2 + m^2 \right)^{2+2\epsilon}} \ .
\label{eq:HformassdepR}
\ee
We then need to expand the following function for small $Q$
\begin{align}
H \left(p^2,(Q \pm q)^2\right) &= H(q^2) + (Q^2 \pm 2 Q \cdot q) H'(q^2) +2 \left(Q \cdot q \right)^2 H''(q^2) + O(Q^3)
\\
&\mathop{\mathrlap{\: =}\int_{q}} H(q^2) + Q^2 H'(q^2) +\frac{1}{2} Q^2 q^2 H''(q^2) + O(Q^4) \ ,
\nonumber
\end{align}
where the second equal sign denotes equivalence
upon integration over $q\in \mathbb{R}^4$, and 
primes denote derivatives with respect to $q^2$.
The anomalous dimension involves the trace
of the four-point vertex, which then reads
\begin{align}
&\Tr \lim_{Q^2 \to 0} \frac{\partial}{\partial Q^2} \Delta_k (Q,-Q,q,-q)
\mathop{\mathrlap{\: =}\int_{q^2}} - 3(N+2)\lambda^2
\nonumber
\\&
\hspace{2cm}\times\int \frac{\dd^4p}{(2\pi)^4}
\left[ H'(p^2,q^2)+\frac{1}{2}q^2 H''(p^2,q^2)-H'(p^2,0)
\right] \ .
\end{align}
This one-loop expression for
the momentum dependence
of the four-point vertex
is to be nested in 
the momentum dependent part of
Eq.~\eqref{eq:dtZrho},
thus obtaining
\bea
\eta^{(2)}= \frac{1}{2}\int \frac{\dd^4q}{(2\pi)^4} \frac{\partial_t R_k(q)}{P(q)^2}
\Tr \lim_{Q^2 \to 0} \frac{\partial}{\partial Q^2} \Delta_k (Q,-Q,q,-q) \ .
\eea
For our pseudo-regulator, we can specify all the
terms in the integrand according to Eqs.~
(\ref{eq:massdepPanddtR}) and  (\ref{eq:HformassdepR}).
Taking the limit for $\epsilon \to 0$ after
all integrals have been performed, we find
the result of \Eqref{eq:theeta2}.

\subsection{Two-loop flow of \texorpdfstring{$m^2$}{TEXT}}
\label{app:2loopsbetam}

In this appendix we show that also
the two-loop beta function of the mass 
can be obtained as the $\epsilon\to0$
limit of the corresponding FRG equation.
We start from \Eqref{eq:betamcomplete},
and neglect higher-loop contributions,
e.g.~by inserting $\kappa=0$, thus obtaining the
simplified result:
\be
\beta_{m^2} -\eta m^2   =
-\frac{k^2}{16 \pi^2}\left[ (N+2) \lambda\, l_{1,0}^4(0) + (N-1) \langle z_1 \rangle _{1,0}^6 (0)
+ \,\langle \tilde z_1 \rangle _{0,1}^6(0) \right] \ .
\label{martina}
\ee
The contribution of the one-loop 
wave function renormalization is
similar to the one discussed in the previous section
\begin{subequations}
\begin{align}
\langle z_1\rangle_{1,0}^6(0) &=
8\pi^2 k^{-2} \int \frac{d^4p}{(2\pi)^4} p^2 Z_k'(0,p^2)  \frac{\partial_t R_k(p^2)}{P(p^2)^2} \ ,
\\
\langle \tilde z_1\rangle_{0,1}^6(0)  &=
8\pi^2  k^{-2} \int \frac{d^4p}{(2\pi)^4} p^2 \tilde Z_k'(0,p^2)  \frac{\partial_t R_k(p^2)}{P(p^2)^2} \ ,
\end{align}
\end{subequations}
where $Z_k'$ and $\tilde Z_k'$ are given in Eqs.~
(\ref{eq:oneloopZ'}) and (\ref{eq:onelooptildeZ'}).
Then the two-loop contributions arise by
replacing in \Eqref{martina}
the following expressions
\begin{align}
l_{1,0}^4(0) &= -\frac{m^2}{k^2} + \left( 1+2f_1 \right) \frac{\beta_{m^2}}{4k^2} \ ,
\\
(N-1) \langle z_1 \rangle_{1,0}^6(0)
+ \,\langle \tilde z_1 \rangle_{0,1}^6(0)  &=
(N+2)\frac{\left(9-\sqrt{3} \pi \right)}{8 \pi ^2} \frac{m^2}{k^2} \lambda^2 \ ,
\\
\eta &= \frac{(N+2)}{2(16\pi^2)^2} \lambda^2 \ ,
\end{align}
where $f_1$ is a free  regularization parameter
as described in Appendix \ref{app:thresholdsmass}.
The combination of these corrections gives
\Eqref{eq:betam2twoloops},
from which it is apparent that the
unique choice
\be
f_1= -\frac{1}{2}+\frac{4 \sqrt{3} \pi - 30 }{N+2} 
\ee
produces the \MSbar two-loop result
\be
\partial_t \log m^2 = \frac{(N+2)}{16\pi^2} \lambda -\frac{5(N+2)}{2(16\pi^2)^2} \lambda^2 \ .
\ee

\end{appendix}

\end{document}